\documentclass{src/DOEproposal}
\usepackage{amsmath, amsthm, amssymb, ulem}
\usepackage{latexsym}
\usepackage{url}
\usepackage{epsfig}
\usepackage{ragged2e}
\usepackage{tabularx}
\usepackage{epstopdf}
\usepackage[range-phrase=\mbox{--},range-units=single,per-mode=symbol]{siunitx}
\usepackage{subcaption}
\usepackage[x11names,dvipsnames,table]{xcolor} 
\usepackage{colortbl}
\usepackage{array}
\usepackage{multirow}
\usepackage[version=4]{mhchem}
\usepackage{graphicx}
\usepackage{xargs}
\usepackage{xfrac}
\usepackage{etoolbox}
\usepackage[total={6.5in,9in},top=1in,headsep=0.1in,headheight=1in]{geometry}
\usepackage{graphbox}
\usepackage[final]{pdfpages}

\makeatletter
\newif\ifAPS@firstauthor
\makeatother

\usepackage{hyperref}
\hypersetup{
    colorlinks=true,        
    linkcolor=blue,         
    citecolor=blue,        
    filecolor=magenta,      
    urlcolor=blue,       
    pdfstartview=           
}                           
\usepackage{enumitem}
\setlist{nolistsep}
\setlist[enumerate]{itemsep=0mm}
\newcolumntype{P}[1]{>{\raggedright\arraybackslash}p{#1}}
\newcolumntype{C}[1]{>{\centering\arraybackslash}p{#1}}
\def\bit{\begin{itemize}
  }
  \def\eit{\end{itemize}
  }
\usepackage[dvipsnames]{xcolor}

\DeclareSIUnit\c{\mbox{$c$}}
\DeclareSIUnit\magn{\mbox{$\times$}}
\DeclareSIUnit\min{min}
\DeclareSIUnit\week{week}
\DeclareSIUnit\month{mo}
\DeclareSIUnit\months{mos}
\DeclareSIUnit\year{yr}
\DeclareSIUnit\years{years}
\DeclareSIUnit\yr{yr}
\DeclareSIUnit\standard{std}
\DeclareSIUnit\str{sr}
\DeclareSIUnit\ppm{ppm}
\DeclareSIUnit\ppb{ppb}
\DeclareSIUnit\ppt{ppt}
\DeclareSIUnit\pe{PE}
\DeclareSIUnit\spe{SPE}
\DeclareSIUnit\pdm{PDM}
\DeclareSIUnit\ev{events}
\DeclareSIUnit\ct{counts}
\DeclareSIUnit\neutron{\mbox{$n$}}
\DeclareSIUnit\smp{samples}
\DeclareSIUnit\Sample{S}
\DeclareSIUnit\ch{ch}
\DeclareSIUnit\hit{hit}
\DeclareSIUnit\hits{hits}
\DeclareSIUnit\bin{(\mbox{5-PE}~bin)}
\DeclareSIUnit\sgm{\mbox{$\sigma$}}
\DeclareSIUnit\rms{RMS}
\DeclareSIUnit\keVee{\mbox{keV$_{e{\rm e}}$}}
\DeclareSIUnit\keVr{\mbox{keV$_{\rm nr}$}}
\DeclareSIUnit\eVee{\mbox{eV$_{\rm ee}$}}
\DeclareSIUnit\eVr{\mbox{eV$_{\rm nr}$}}
\DeclareSIUnit\ph{photon}
\DeclareSIUnit\el{\mbox{$e^-$}}
\DeclareSIUnit\pm{\mbox{PMT}}
\DeclareSIUnit\pixel{\mbox{pixel}}
\DeclareSIUnit\inch{''}
\DeclareSIUnit\foot{'}
\DeclareSIUnit\ft{\mbox{ft}}
\DeclareSIUnit\bit{bit}
\DeclareSIUnit\sample{samples}
\DeclareSIUnit\barn{barn}
\DeclareSIUnit\bara{bar}
\DeclareSIUnit\barg{barg}
\DeclareSIUnit\mlardepth{\mbox(meter~of~\LAr~depth)}
\DeclareSIUnit\Curie{Ci}
\DeclareSIUnit\psf{psf}
\DeclareSIUnit\pcf{pcf}
\DeclareSIUnit\parsec{pc}
\DeclareSIUnit\liveday{\mbox{live-days}}
\DeclareSIUnit\days{\mbox{days}}
\DeclareSIUnit\miles{\mbox{miles}}
\DeclareSIUnit\lumens{\mbox{lm}}
\DeclareSIUnit\degreeC{\mbox{$^{\circ}$C}}
\DeclareSIUnit\degreeF{\mbox{$^{\circ}$F}}
\DeclareSIUnit\electron{\mbox{$e^-$}}
\DeclareSIUnit\Euro{\mbox{\euro}}
\DeclareSIUnit\cph{cph}
\DeclareSIUnit\neq{neq}
\DeclareSIUnit\normal{\mbox{N}}

\newcommand{\gr}{$\gamma$-ray}

\newcommand{\alphan}{$(\alpha,n)$}

\newcommand{\ngamma}{$(n,\gamma)$}

\let\OLDthebibliography\thebibliography
\renewcommand\thebibliography[1]{%
  \OLDthebibliography{#1}%
  \raggedright       
  \sloppy            
}

\usepackage{orcidlink}
\usepackage{fontawesome}

\usepackage{titlesec}
\titlespacing*{\subsection}
  {0pt}               
  {12pt plus 4pt minus 2pt}   
  {6pt plus 2pt minus 1pt}    
\titlespacing*{\section}
  {0pt}               
  {12pt plus 4pt minus 2pt}   
  {6pt plus 2pt minus 1pt}    

\investigator{Jaehoon Yu}
\institute{University of Texas, Arlington}
\phone{(817) 272-2814}
\email{jaehoon@uta.edu}

\title{DAMSA Experiment Conceptual Design White Paper}

\affiliation{uta}{University of Texas at Arlington, Arlington,~TX~76019,~USA}
\affiliation{fnal}{Fermi National Accelerator Laboratory,~Batavia,~Illinois,~USA}
\affiliation{wau}{Washington University, St. Louis,~MO~63130,~USA}
\affiliation{tamu}{Texas~A\&M~University,~College~Station,~TX~77843,~USA}
\affiliation{ucr}{Department of Physics \& Astronomy, University of California, Riverside,~CA~92521,~USA}
\affiliation{bnl}{Brookhaven National Laboratory, Upton, NY 11973,~USA}
\affiliation{itpnn}{Department of Physics and Institute of Theoretical Physics Nanjing Normal University, Nanjing, 210023, China}
\affiliation{ibs}{Particle Theory and Cosmology Group, Center for Theoretical Physics of the Universe, \\ Institute for Basic Science (IBS), Daejeon 34126,~Republic~of~Korea}
\affiliation{usd}{Department of Physics, University of South Dakota, Vermillion,~SD~57069,~USA
}
\affiliation{snu}{Department of Physics \& Astronomy, Seoul National University, 1~Gwanak-ro,~Gwanak-gu,~Seoul~08826,~Republic~of~Korea}
\affiliation{knu}{Department of Physics, Kyungpook~National~University,~Daegu 41566,~Republic~of~Korea}
\affiliation{uchicago}{Department of Physics, University of Chicago, Chicago,~IL~60637,~USA}
\affiliation{pitt}{University of Pittsburgh,~Pittsburgh,~PA~15260,~USA}
\affiliation{kus}{Department of Accelerator Science, Korea University Sejong Campus, 2511~Sejong-ro,~Sejong~30019,~Republic~of~Korea}
\affiliation{chungnam}{Department of Physics and Institute for Sciences of the Universe, Chungnam National University, Daejeon~34134,~Republic~of~Korea}
\affiliation{sdm}{South Dakota School of Mines and Technology, Rapid City,~SD~57701,~USA}
\affiliation{umd}{Department of Physics, University of Maryland, College Park,~MD~20742,~USA}
\affiliation{jnu}{Laboratory for Symmetry and Structure of the Universe, Department of Physics, Jeonbuk~National~University,~Jeonju,~Jeonbuk~54896,~Republic~of~Korea}
\affiliation{nu}{Northwestern~University,~Evanston,~IL~60208,~USA}

\author[uta]{Prithak~Bhattarai}
\author[uta]{Andrew~Brandt}
\author[fnal]{Alan~Bross}
\author[uta]{Bradley~Brown}
\author[uta]{Samriddha~Chakraborty}
\author[wau]{Haohui~Che}
\author[wau]{Bhupal~Dev}
\author[tamu]{Bhaskar~Dutta}
\author[fnal]{Juan~V.~Estrada}
\author[uta]{Eric~Garcia}
\author[ucr]{Anthony~Gomez}
\author[uta]{Gajendra~Gurung}
\author[uta]{Brian~Joshua~Gomez~Hernandez}
\author[uta]{Wooyoung~Jang}
\author[bnl]{Jay~Hyun~Jo}
\author[itpnn,ibs]{Krzysztof~Jod\l{}owski}
\author[usd]{Doojin~Kim}
\author[snu]{Eunsu~Kim}
\author[snu]{Hyunyong~Kim}
\author[knu]{Shin~Hyung~Kim}
\author[uchicago]{Young-Kee~Kim}
\author[usd]{Jing~Liu}
\author[knu]{Chang-Seong Moon}
\author[pitt]{Donna~Naples}
\author[uta]{David~Nygren}
\author[snu]{Minseok~Oh}
\author[pitt]{Vittorio~Paolone}
\author[kus]{Hyangkyu~Park}
\author[chungnam]{Jong-Chul~Park}
\author[fnal]{Nathaniel~J.~Pastika}
\author[uta]{Rohit~Raut}
\author[sdm]{Juergen~Reichenbacher}
\author[fnal]{Paul~Rubinov}
\author[umd]{Eunsuk~Seo}
\author[ucr]{Veronika~Shalamova}
\author[jnu]{Seodong~Shin}
\author[uchicago]{Melvin~Shochet}
\author[nu]{Adrian~Thompson}
\author[uchicago]{Yau~Wah}
\author[ucr]{Shawn~Westerdale}
\author[bnl]{Guang~Yang}
\author[snu]{Un-Ki~Yang}
\author[snu]{Inseok~Yoon}
\author[uta]{Jaehoon~Yu}

\begin{abstract}
    DAMSA (DArk Messenger Searches at an Accelerator) is a novel short-baseline accelerator experiment aimed at probing short-lived physics processes, including searches for evidence of a dark sector of particle physics and well-motivated Standard Model signals. Motivated by open questions in neutrino physics and the absence of conclusive evidence for conventional weakly interacting massive particles, DAMSA targets MeV-to-sub-GeV dark-sector messengers with feeble couplings that can be produced in abundance at the PIP-II LINAC. By employing an ultra-short baseline of order one meter, DAMSA is uniquely positioned to overcome the beam-dump “ceiling” that limits sensitivity to promptly decaying particles in longer-baseline experiments. The conceptual design emphasizes a beam-dump production scheme combined with a compact detector optimized for rare decays while mitigating intense neutron-induced backgrounds inherent to high-power proton beams. To validate the experimental strategy and detector technologies, the Little DAMSA Path-Finder (LDPF) proof-of-concept experiment is proposed, focusing on axion-like particles decaying to two photons and operating with 300 MeV electron beams at FAST. Successful realization of LDPF will establish the feasibility of the DAMSA approach, enabling a broad and powerful program to explore short-lived new physics and precision Standard Model processes in a previously inaccessible regime. This conceptual design document outlines the technical details of DAMSA's physics goals, the beam facility proposals, key experimental challenges and how to overcome them, and the proposed experimental staging campaigns.
\end{abstract}

\begin{document}

    \maketitle
    \tableofcontents

    \phantomsection
    \section{Introduction}
\label{sec:Introduction}

Neutrinos make up a quarter of the elementary particle landscape in the Standard Model (SM) of particle physics. However, they do not behave as initially predicted by the SM, as they possess non-zero mass. Consequently, the model must be modified to retain its predictive power. Future neutrino experiments, such as the Deep Underground Neutrino Experiment (DUNE)~\cite{DUNE:2020lwj} at Fermilab, aim to precisely measure neutrino properties to address this gap. These experiments utilize high-flux neutrino beams generated from proton interactions with a target, combined with a suite of detectors for precision measurements. Fermilab’s ``Proton Improvement Plan II'' linear accelerator program (PIP-II LINAC)~\cite{pip2-linac} is a crucial component, delivering the required high-intensity proton beams with a total proton current of 2~mA to support DUNE.

Another compelling motivation for new physics comes from the existence of dark matter comprising about 25\% of the Universe's energy budget, which is strongly supported by numerous astrophysical and cosmological observations through its gravitational effects. However, its particle properties, including the mass scale, remain unknown. Scenarios involving GeV-scale weakly interacting massive particles (WIMPs) have garnered significant attention, as these candidates are thermally produced, independent of specific model details. Moreover, since WIMPs predict non-gravitational interactions between dark matter and SM particles, extensive experimental and theoretical efforts have been devoted to their study over the past few decades (see, e.g., Ref.~\cite{Bertone:2004pz}). However, no conclusive evidence has been found, prompting the exploration of alternative ideas which are now being actively investigated.

Among these alternatives, MeV-scale (light) dark matter is a promising candidate, as it can be thermally produced and its parameter space remains largely unexplored. Additionally, portal scenarios suggest that other dark-sector particles--such as messenger particles mediating the interactions between dark matter and SM particles--of similar mass should exist, with feeble interactions with SM particles (see, e.g., Refs.~\cite{Holdom:1985ag,Patt:2006fw,Pospelov:2007mp,Pospelov:2008jd,Pospelov:2008zw,Falkowski:2009yz,Arcadi:2019lka,Batell:2022xau}). The predicted mass scale is within reach of existing beam facilities, which can produce these dark-sector particles, including MeV-scale dark matter, while the very weak coupling strengths motivate the need for experiments with intensified beams to accommodate the rare production of such particles. In addition to scenarios involving light dark matter, visibly-decaying messenger particles are often proposed to explain various experimental anomalies, such as the MiniBooNE low-energy excess~\cite{MiniBooNE:2008yuf,MiniBooNE:2018esg,MiniBooNE:2020pnu}, the muon $g-2$~\cite{Muong-2:2006rrc,Muong-2:2021ojo}, and the LSND anomaly~\cite{LSND:2001aii}.

With its powerful beams, the PIP-II facility is expected to produce the aforementioned feebly-interacting messenger particles in abundance, shedding light on dark-sector physics. The discovery of dark-sector particles at an accelerator would pave the way for a deeper understanding of the nature of dark matter. In particular, high-intensity beam experiments have provided leading constraints on messenger particles in the MeV-to-sub-GeV range. Schematically, messenger particles produced at the target or dump must survive long enough to reach the detector and decay within its fiducial volume. The sensitivity of an experiment depends on factors such as baseline distance, effective decay volume, beam intensity, and beam energy. A key challenge is probing messenger particles that decay relatively early, as longer baselines reduce detection sensitivity, which is referred to as beam-dump ``ceiling''~\cite{Dutta:2023abe,Kim:2024vxg}. To address this, experiments with very short baselines are better suited for exploring these regions of parameter space~\cite{Dutta:2023abe,Kim:2024vxg}.

In this regard, a novel experiment for searching for dark-sector particles, DAMSA ({\bf{\underline {DA}}}rk {\bf{\underline {M}}}essenger {\bf{\underline {S}}}earches at an {\bf{\underline {A}}}ccelerator - pronounced ``da-m-sa'' and means rumination) has been proposed, with a baseline scale of approximately 1 meter, inspired by the experimental scheme proposed in Ref.~\cite{Jang:2022tsp}. An essential element of DAMSA is the beam-dump facility, which can produce dark-sector particles using high-intensity proton beams from the PIP-II LINAC. Given the potential of DAMSA to explore the dark sector, it is crucial to place this effort within the broader scope of dark matter research. 

The use of high-intensity proton beams and the close proximity of the detector inevitably lead to a high flux of low-energy neutrons, which may induce unwanted backgrounds. In this proposal, we aim to overcome this challenge by selecting specific final states, optimizing beam configurations, and employing a detector system designed to mitigate neutron backgrounds. To this end, we propose constructing the {\bf Little DAMSA Path-Finder (LDPF)} proof-of-concept experiment, which will focus on the two-photon final state of the axion-like particles and operate using \SI{300}{\MeV} electron beams---potentially offering a more controlled environment regarding neutron-induced backgrounds---at Fermilab’s Facility for Accelerator Science and Technology (FAST) with the detector system placed $\sim10$~cm away from the ALP production point. Successful completion of the proposed tasks of LDPF will allow us not only to discover the messenger particle but also to build confidence in the experimental techniques, paving the way for many more discoveries through the full-scale DAMSA experiment and other future efforts in the field.

\vspace{0.5em}
\section{Physics Goals}  
\label{sec:PhysicsGoals}
This section outlines the potential physics topics that DAMSA can explore, as well as the experimental capabilities required to access them. To achieve all four goals, the full-scale DAMSA with precision tracking capability and total absorption 4D electromagnetic calorimeter (ECAL) is necessary. However, Goals 1 and 2 can be readily accomplished with the LDPF experiment to be performed under the scope of this proposal. The remaining goals can also be achieved with a potential upgrade of the LDPF's ECAL system.  

\subsection{Axion-like Particles Coupling to Photons} 
\paragraph{Motivations}
Axion-like particles (ALPs) provide a well-motivated and economical extension of the SM in which a new pseudoscalar field \(a\) interacts through the leading electromagnetic portal,
\begin{equation}
    \mathcal{L}_{\rm{ALP}} \supset -\frac{1}{4}\, g_{a\gamma\gamma}\, a \, F_{\mu\nu}\,\tilde{F}^{\mu\nu},
    \label{eq:alagamma}
\end{equation}
where \(g_{a\gamma\gamma}\) (with mass dimension \(-1\)) parameterizes the ALP-photon coupling, \(F_{\mu\nu}\) is the electromagnetic field-strength tensor, and \(\tilde{F}^{\mu\nu}\equiv \tfrac{1}{2}\epsilon^{\mu\nu\rho\sigma}F_{\rho\sigma}\) is its dual. This dimension-5 operator is often the dominant low-energy interaction in theories where \(a\) arises as a pseudo-Nambu-Goldstone boson of an approximate global symmetry. The associated (approximate) shift symmetry makes light ALPs technically natural---radiative corrections to the ALP mass \(m_a\) are symmetry suppressed---so sub-eV to GeV-scale masses with feeble couplings can arise without fine-tuning. The QCD axion~\cite{Peccei:1977hh,Wilczek:1977pj,Weinberg:1977ma}, introduced to resolve the strong-CP problem, provides a canonical benchmark within this broader framework; more general ALPs relax the strict mass-coupling relation and motivate an experimentally driven exploration of the \((m_a, g_{a\gamma\gamma})\) parameter space.

From a UV perspective, an ALP coupling to photons is generically expected. If the ALP couples to heavy electrically charged states, integrating them out typically induces the low-energy effective interaction in Eq.~\eqref{eq:alagamma} through anomaly-like matching and loop effects, making \(g_{a\gamma\gamma}\) a robust infrared imprint even when other SM couplings are absent or subdominant. Moreover, many well-studied UV settings---ranging from composite sectors to extra-dimensional and string-motivated constructions---naturally contain multiple axion-like degrees of freedom (often termed an ``axiverse''), for which electromagnetic couplings are common and phenomenologically consequential. In cosmology and astrophysics, photon-coupled ALPs are equally compelling: depending on their mass and coupling, they can contribute to dark matter~\cite{Duffy:2009ig,Marsh:2015xka,Battaglieri:2017aum,Adams:2022pbo} (e.g., via misalignment~\cite{Preskill:1982cy,Abbott:1982af,Dine:1982ah}), act as dark radiation, and be efficiently produced or converted in stellar and plasma environments, linking laboratory searches to complementary astrophysical sensitivity (see, e.g., Ref.~\cite{Fortin:2021cog} and references therein).

From a phenomenological standpoint, the \(a\gamma\gamma\) portal is especially attractive for a conceptual design study because it yields clean, versatile experimental signatures that map directly onto detector and infrastructure choices. The same operator governs (i) Primakoff-like production~\cite{Primakoff:1951iae} in electromagnetic fields or on charged targets, (ii) coherent \(\gamma \leftrightarrow a\) conversion~\cite{Graham:2015ouw} in macroscopic magnetic fields, and (iii) visible decays \(a\to\gamma\gamma\) when kinematically allowed, including prompt and displaced topologies depending on the ALP lifetime. Together, these channels provide multiple, complementary discovery handles with controllable systematics, enabling DAMSA to define a clear and broadly interpretable physics program in the standard \((m_a, g_{a\gamma\gamma})\) plane while retaining sensitivity across wide ranges of lifetime and mass.

\paragraph{Sensitivity Prospects}
As described above, in the presence of the coupling in Eq.~\eqref{eq:alagamma}, an incoming photon can convert into an ALP via the Primakoff process on a nuclear target,
\(\gamma N \to a N\).
The corresponding differential production cross section may be written as
\begin{equation}
    \frac{d\sigma_P}{d\theta_a}
    =\frac{1}{4}\,g_{a\gamma\gamma}^2\,\alpha\,Z^2\,\left|F(t)\right|^2\,
    \frac{p_a^{\,2}\,\sin^3\theta_a}{t^2}\,,
\end{equation}
where \(\alpha\) is the electromagnetic fine-structure constant, \(Z\) is the atomic number of the target nucleus, \(p_a\) is the magnitude of the outgoing ALP three-momentum, and \(\theta_a\) is the ALP polar angle measured with respect to the incident photon direction. The squared momentum transfer to the target is
\begin{equation}
    t \equiv (p_\gamma-p_a)^2
    = m_a^2 - 2E_\gamma\!\left(E_a - p_a\cos\theta_a\right),
\end{equation}
with \(E_\gamma\) (\(E_a\)) the incident photon (outgoing ALP) energy. The factor \(F(t)\) encodes the finite nuclear size and atomic screening effects; for small momentum transfer (i.e., \(|t|\) sufficiently small) the process is strongly forward-peaked and one typically has \(E_a\simeq E_\gamma\). In this regime, \(F(t)\) is commonly modeled using the Helm form factor (or an equivalent parameterization appropriate to the target material). 

A realistic estimate of the ALP yield further requires an accurate determination of the photon flux and energy-angle distribution within the target, including showering and secondary production. To this end, we model the photon transport and interactions in the target using the \textsc{GEANT4}~\cite{GEANT4:2002zbu} simulation framework.

Once produced in the target, an ALP must propagate toward the detector and decay within the detector fiducial region (including the dedicated decay volume). The relevant timescale is set by the ALP partial width to two photons,
\begin{equation}
    \Gamma(a\to\gamma\gamma)=\frac{g_{a\gamma\gamma}^2\,m_a^3}{64\pi}\,,
\end{equation}
which determines the mean decay length in the laboratory frame,
\begin{equation}
    \ell_a^{\rm lab} \equiv \beta_a\gamma_a\,c\tau_a
    = \frac{\beta_a\gamma_a\,c}{\Gamma_a}\,,
\end{equation}
where \(\Gamma_a\) is the total ALP width [equal to \(\Gamma(a\to\gamma\gamma)\) in the minimal scenario considered here], \(\beta_a\) is the ALP speed in units of \(c\), and \(\gamma_a\) is the Lorentz boost factor. The probability for an ALP produced in (and emerging from) the target region to survive a distance \(\ell_t\) and then decay within a downstream decay volume of length \(L_D\) is therefore
\begin{equation}
    P_D
    = \exp\!\left(-\frac{\ell_t}{\ell_a^{\rm lab}}\right)
      \left[1-\exp\!\left(-\frac{L_D}{\ell_a^{\rm lab}}\right)\right].
\end{equation}
In the DAMSA geometry used for our sensitivity study, \(\ell_t\) corresponds to the effective target length along the ALP flight path, and \(L_D\) denotes the length of the downstream fiducial region over which the \(a\to\gamma\gamma\) decay can be efficiently reconstructed. In our sensitivity estimates, we adopt a conservative definition of this acceptance by taking \(L_D\) to be the physical length of the dedicated decay volume only.

\begin{figure}[htb]
    \centering
    \includegraphics[width=0.495\linewidth]{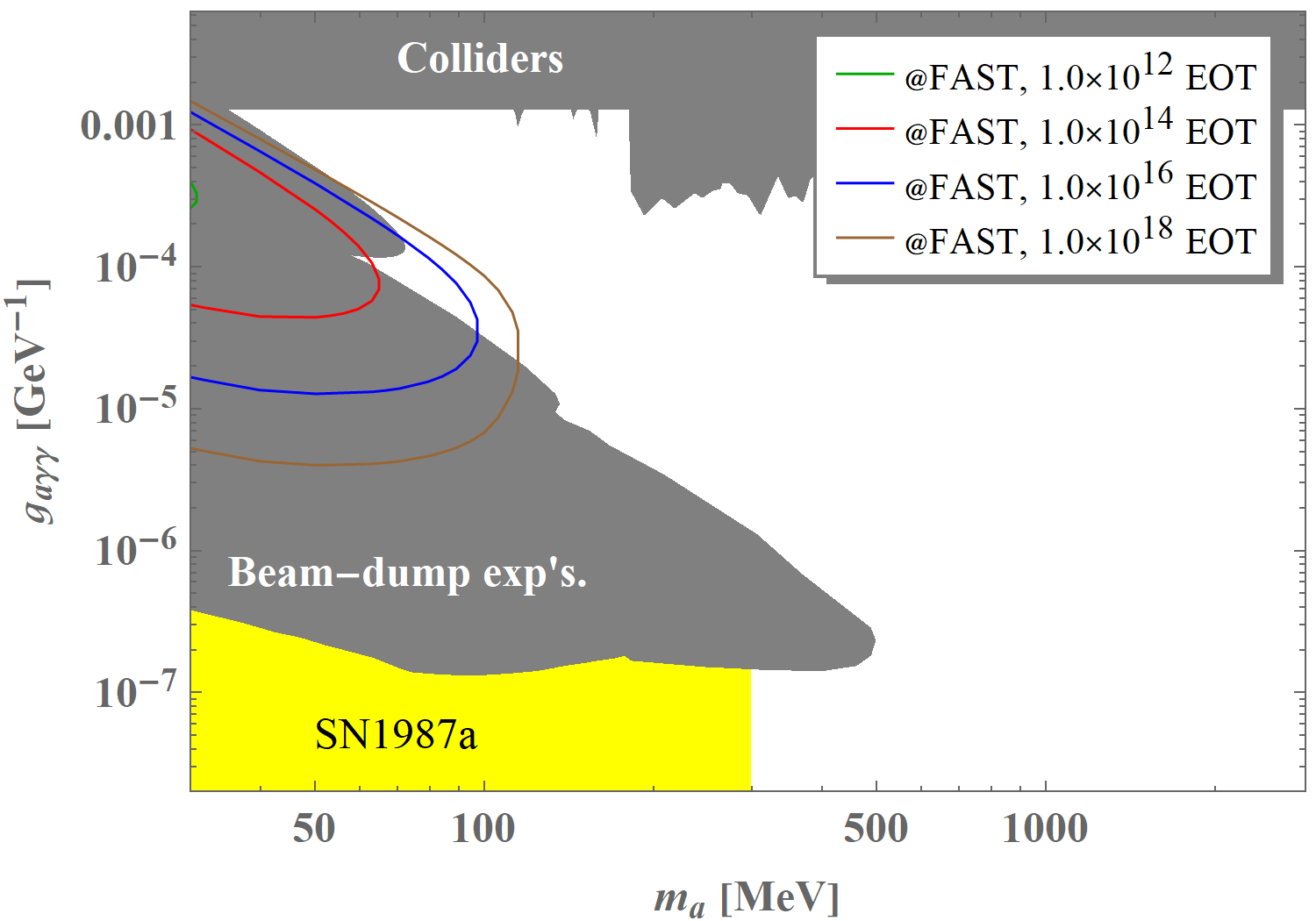}
    \includegraphics[width=0.495\linewidth]{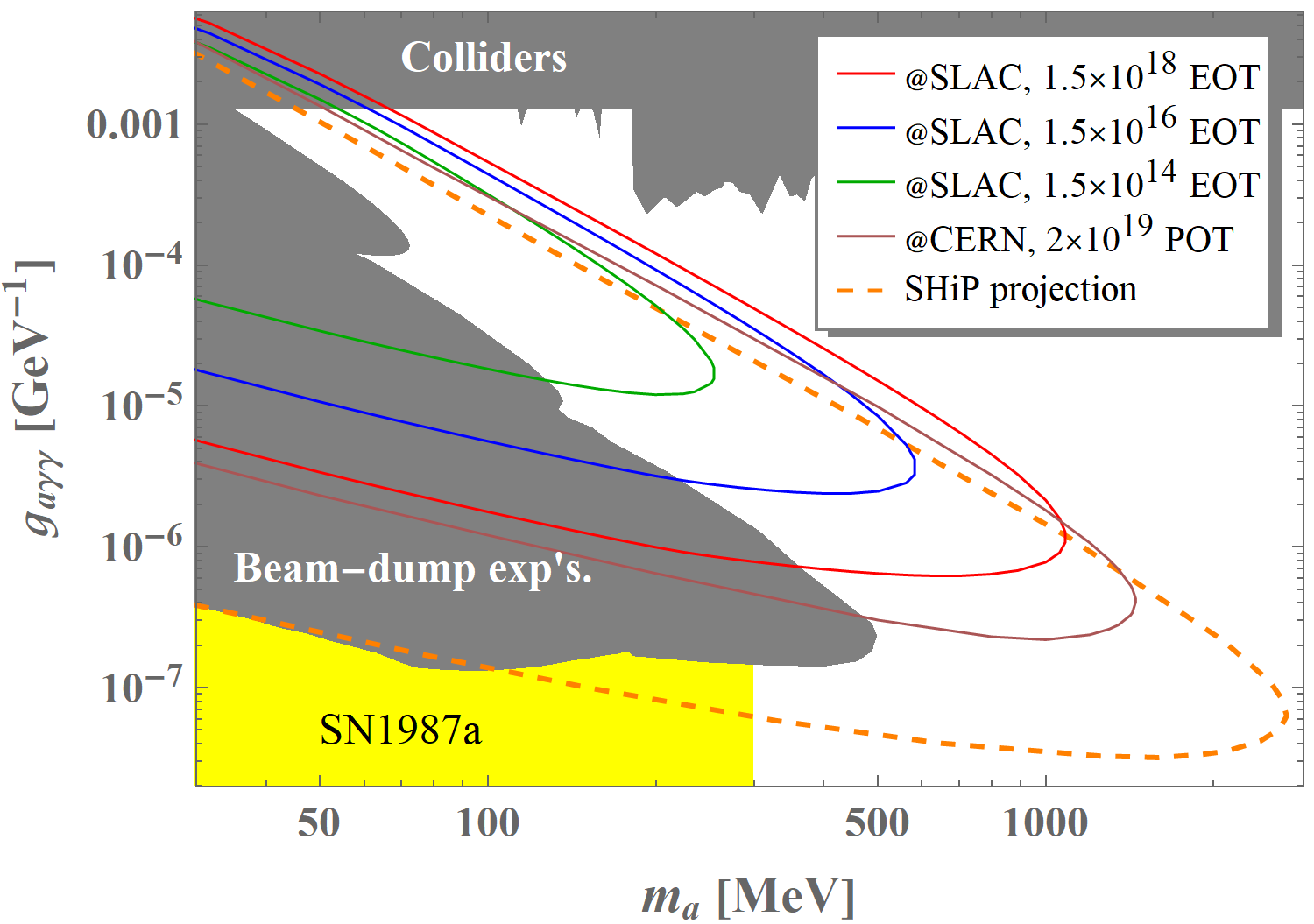}
    \caption{
        Expected sensitivity coverages of the DAMSA experiment using the 300~MeV electron beam at Fermilab's FAST facility (Stage 0, Left) and the 8~GeV electron beam at SLAC's LESA facility (Stage 1, Right). The sensitivity estimate for CERN’s BDF facility (Stage 3) is shown by the brown curve in the right-hand panel.
    }
    \label{fig:alp-slac}
\end{figure}

The DAMSA experimental staging, discussed further in \S~\ref{sec:stages}, provides a set of benchmarks to compare the experimental sensitivity over the ALP model parameter space. The left-hand panel of Fig.~\ref{fig:alp-slac} shows the projected sensitivity of DAMSA using a 10~cm-long tungsten target and Fermilab’s FAST beam, corresponding to a 300~MeV electron beam. The longitudinal length of the decay volume is taken to be 30~cm, and 
the instantaneous beam intensity is adjusted such that potential backgrounds are reduced to a negligible level.
We consider four representative values of the total integrated electrons on target: $1.0\times 10^{12}$ (green), $1.0\times 10^{14}$ (red), $1.0\times 10^{16}$ (blue), and $1.0\times 10^{18}$ (brown). Our simulation study indicates that a new region of parameter space can already be explored with a beam exposure of $10^{14}$ electrons on target. 

By contrast, the right-hand panel of Fig.~\ref{fig:alp-slac} shows the projected sensitivity of DAMSA for a $15~\mathrm{cm},(L)\times 12~\mathrm{cm},(W)\times 12~\mathrm{cm},(H)$ tungsten target, using SLAC's LESA beam facility. The longitudinal length of the decay volume is taken to be 30~cm, and we assume an instantaneous beam intensity of $10^{4}$ electrons per pulse with a repetition rate of 1~kHz. Under this beam-target configuration, our \textsc{GEANT} simulation indicates that backgrounds can be reduced to a negligible level. We consider three reference values of the total integrated electrons on target: $1.5\times 10^{14}$ (green curve), $1.5\times 10^{16}$ (blue curve), and $1.5\times 10^{18}$ (red curve). Given the assumed beam intensity and repetition rate, an exposure of $1.5\times 10^{14}$ electrons on target can be achieved within approximately 3 -- 4 months of data taking. These results clearly demonstrate that a substantially extended region of previously unexplored parameter space can be probed even with a modest 3 -- 4 month beam exposure. 
Following the successful completion of DAMSA at SLAC’s LESA beam facility, we plan to deploy DAMSA at CERN’s BDF facility. In this configuration, we assume a tungsten target with a length of order 100~cm, followed by an additional shielding region of order 2{,}000~cm to suppress potential backgrounds to a negligible level. The corresponding projected sensitivity is shown by the brown curve in the right-hand panel of Fig.~\ref{fig:alp-slac}.

\subsection{Axion-like Particles Coupling to Electrons}
\paragraph{Motivations}
We also consider ALPs with dominant couplings to electrons, which can arise by the mediation of an extended Higgs sector that lifts the fermion mass terms to pseudoscalar currents, as in the DFSZ (Dine-Fischler-Srednicki-Zhitnitsky) variants of the QCD axion~\cite{Dine:1982ah,Zhitnitsky:1980tq,Sun:2020iim,DiLuzio:2020wdo}, and others~\cite{Han:2020dwo,Ganguly:2022imo}. Their interactions can be parameterized by an effective Lagrangian with a pseudoscalar coupling,
\begin{equation}
    \mathcal{L}_{\rm{ALP}} \supset -i g_{ae} a \bar{e} \gamma^5 e \, .
\end{equation}
This phenomenological coupling allows for production of ALPs in the DAMSA targets in several ways. In the case of an electron beam impinging on the target, they can be emitted in the electron bremsstrahlung process, $e^\pm Z \to e^\pm Z a$, where $Z$ is the target atom. In addition, secondary electrons, positrons, and photons produced in the beam target can support ALP emission through resonant and non-resonant annihilation of positrons with target electrons, $e^+ e^- \to a$ and $e^+ e^- \to a \gamma$, through Compton scattering ($e^- \gamma \to a e^-$) and through further bremsstrahlung emission from secondary positrons and electrons. These production processes have been studied in the contexts of other accelerator beam target experiments, and their associated cross sections can be found, for example, in refs.~\cite{CCM:2021jmk,AristizabalSierra:2020rom,PhysRevD.34.1326,Avignone:1988bv,Gondolo:2008dd,Brodsky:1986mi}. For the moment we neglect production in the 3-body decays of charged pions and kaons produced in the beam target, which may offer additional enhanced sources of ALP flux in this scenario~\cite{PhysRevLett.130.241801}.

The propagation and subsequent decay of the ALP could then take place, where in this case we consider primarily their decay to $e^+ e^-$ pairs in the decay pipe. The decay width in this case is
\begin{equation}
    \Gamma(a \to e^+ e^-) = \frac{g_{ae}^2 m_a}{8\pi}\sqrt{1 - \frac{4 m_e^2}{m_a^2}} \, ,
\end{equation}
valid for ALP masses satisfying $m_a > 2 m_e \simeq 1$~MeV. The theoretical parameter space for ALPs coupled to photons can be probed in a similar way to photons as discussed in the previous section, where the access to short path lengths for the ALP propagation and decay at DAMSA allow us to probe the short-lifetime limit in the parameter space.

\begin{figure}
    \centering
    \includegraphics[width=0.485\linewidth]{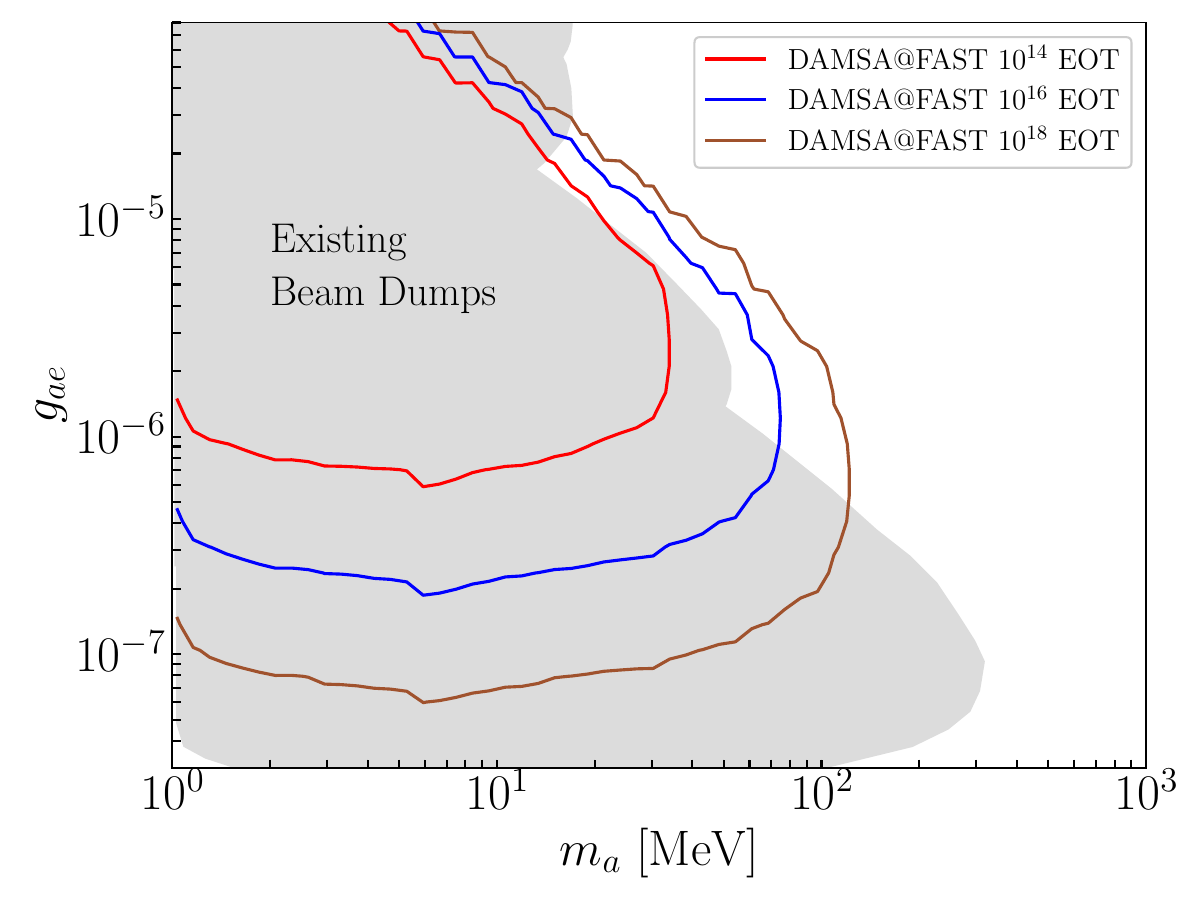}
    \includegraphics[width=0.485\linewidth]{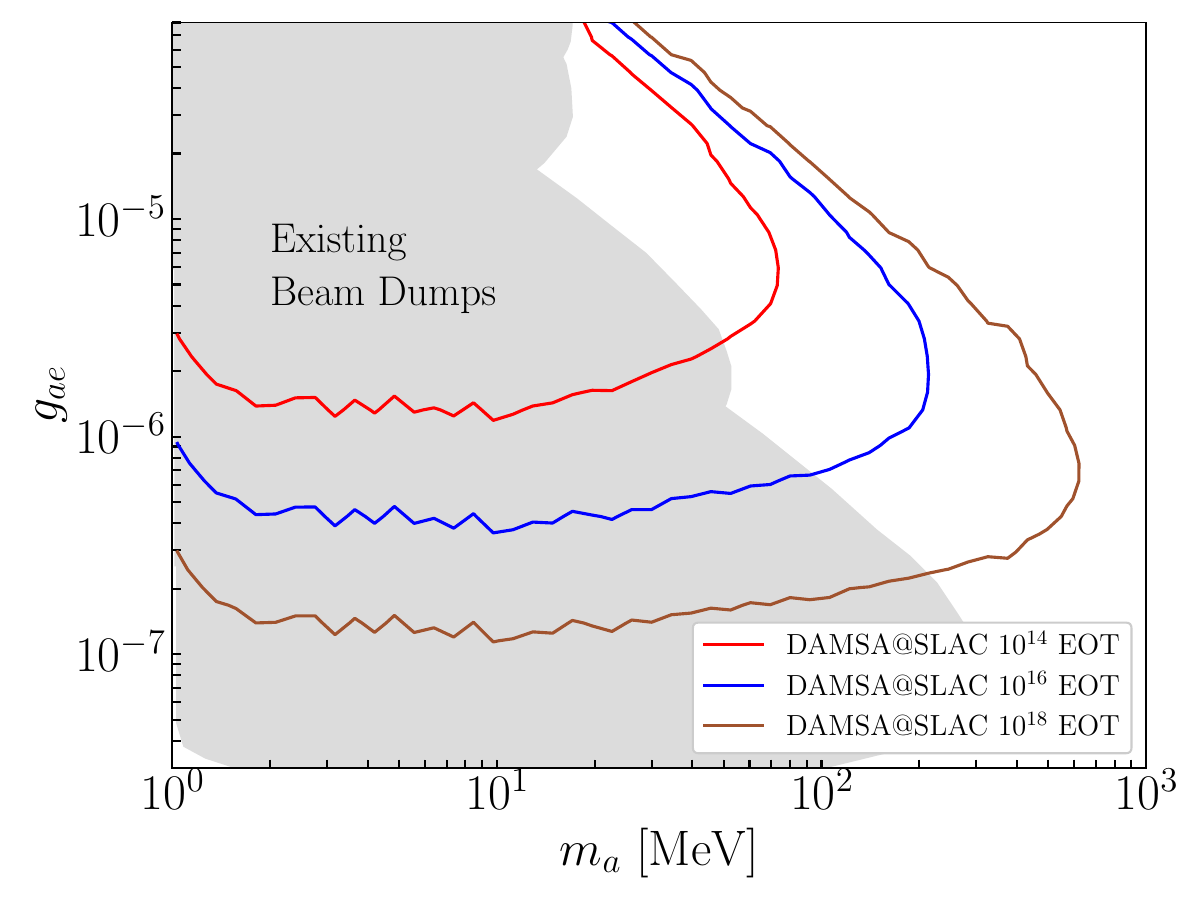}
    \caption{Expected sensitivity coverages over the ALP-electron coupling parameter space in the ($m_a$, $g_{ae}$) plane of the DAMSA experiment, using the 300 MeV electron beam at Fermilab's FAST facility (Stage 0, left) and the 8 GeV electron beam at SLAC's LESA facility (Stage 1, right). }
    \label{fig:alp_electron_sens}
\end{figure}

\paragraph{Sensitivity Prospects}
We utilize the \textsc{GEANT4}-simulated electron, positron and photon fluxes in conjunction with the \textsc{alplib} library \href{https://github.com/athompson-git/alplib}{{\large\color{violet}\faGithub}}~\cite{alplib} to simulate the ALP production channels in the DAMSA target, in addition to ALP bremsstrahlung production from primary electron beams, discussed above. The resulting sensitivity to electron-coupled ALPs is shown in Fig.~\ref{fig:alp_electron_sens} for the FAST and LESA facility benchmark electron beam experimental setups used in the previous section. The projected reach at DAMSA expands upon existing beam dump limits (shown in gray, combined from E137~\cite{PhysRevD.38.3375,Andreas:2010ms}, Orsay~\cite{Bechis:1979kp}, E141~\cite{Riordan:1987aw}, E774~\cite{Bross:1989mp}, NA64~\cite{NA64:2021ked,Andreev:2021fzd,Gninenko:2017yus}, and the CCM120 engineering run~\cite{CCM:2021jmk}) and gain access to heavier ALPs approximately in the $g_{ae} \in [10^{-7}, 10^{-5}]$ coupling range. Pushing the sensitivity envelope in this direction complements collider probes~\cite{Bauer:2017ris, Eberhart:2025lyu, Alimena:2025kjv}, which test regions of parameter space at higher couplings and larger masses.

\subsection{Dark Photons}
We consider a model in which a dark photon interacts via a kinetic mixing term:
\begin{equation}
\mathcal{L}_{kin} = -\frac{\epsilon}{2}F_{\mu \nu} F^{'\mu \nu} , 
\end{equation}
where $F$ is the electromagnetic field strength, $F'$ is the dark photon field strength, and $\epsilon$ is a mixing parameter. After diagonalizing and unitarizing the kinetic term, the dark photon field, $A^\prime$, has an effective interaction with the electromagnetic current as follows:
\begin{equation}
\mathcal{L}_{eff} = \epsilon e J_{\mu} A^{'\mu} .
\end{equation}
Any interactions involving the photon is also allowed for the dark photon but will be scaled by a factor of $\epsilon^{2}$, as well as kinematic factors due to the mass of the dark photon $m_{\gamma^\prime}$. 

\paragraph{Dark Photon Simulation} In order to estimate dark photon production at DAMSA, 800 MeV protons striking a tungsten target are simulated using Geant4. For every photon that is produced in the collision, the production process, the production position, and the four-momentum of the primary particle that produced the photon is recorded in addition to the four-momentum of the produced photon. The dominant production processes for photons above 10 MeV are pion decay and bremsstrahlung. We consider the two cases. A neutral pion may decay as $\pi^0 \rightarrow \gamma\gamma$. For this decay:
\begin{equation}
\frac{\Gamma(\pi^0 \rightarrow \gamma \gamma^\prime)} {\Gamma(\pi^0 \rightarrow \gamma\gamma)} = 2\epsilon^{2} \bigg(1-\frac{m^2_{\gamma^\prime}}{m^2_{\pi}} \bigg)^3
\end{equation}
The factor of 2 appears from the fact that the photons in the usual pion decay are identical, while for the dark photon decay
they are distinguishable. 
There are two ways one can proceed: Simply take pions that produce a photon, re-weight by above and resample the decay of the pion. Alternately one can simply re-weight each photon by $\epsilon^{2}$ and impose a cut that checks if the decay is kinematically allowed. We therefore assign a weight of 0 for $m_{\gamma^\prime} > m_{\pi}$ and require that $E_{\gamma} > m_{\gamma^\prime}$. By allowing for either photon from the pion decay to be reweighted into a dark photon, we therefore correctly incorporate the factor of 2. The photon momentum can then be redefined such that it is on shell. 

The other major production process considered is dark bremsstrahlung, $e^\pm + Z \rightarrow e^\pm + Z + \gamma^\prime$, where, like in the ALP case, we consider radiative production of $\gamma^\prime$ off target atoms $Z$. The resulting $\gamma^\prime$ flux is dominated by this radiative process in the presence of a primary accelerator electron beam, but it can also take place efficiently from secondary $e^\pm$ produced in the electromagnetic showers of electron or proton beams impinging on the target. The cross section can be computed exactly at tree-level~\cite{Gninenko:2017yus}, and has been implemented in several codes and other analyses~\cite{Blinov:2024pza,Zhou:2024aeu,Ge:2025aui}. In addition, secondary positrons produced in the electromagnetic cascades can also lead to $\gamma^\prime$ production through $e^+ e^- \to \gamma^\prime$ resonant production and through the associated production of a photon, $e^+ e^- \to \gamma \gamma^\prime$. In the case of electron beam facilities, these channels are expected to be subdominant to bremsstrahlung production of $\gamma^\prime$ by the primary beam. Lastly, vector meson mixing with $\gamma^\prime$ could also lead to enhanced $\gamma^\prime$ production in the target, see for example ref.~\cite{Schuster:2021mlr}.

\paragraph{Decay}
Given the dark photon four-vectors obtained from above, we can obtain a probability for the dark photon to decay in the DAMSA decay volume. The dark photon can decay into $e^{+} e^{-}$, $\mu^{+} \mu^{-}$, or hadrons.
The decays into leptons can be calculated at tree level and are:
\begin{equation}
\Gamma_{\ell^{+} \ell^{-}} = \frac{\epsilon^{2} \alpha m_{\gamma~'} }{3}\sqrt{1-\frac{4m^{2}_{\ell}}{m^2_{\gamma^\prime}}} \bigg(1+\frac{2m^{2}_{\ell}}{m^2_{\gamma^\prime}} \bigg)
\end{equation}
The hadronic decay rate can be obtained from the ratio R of the cross section for $e^{+} e^{-} \rightarrow$ hadrons to that
of $e^{+} e^{-} \rightarrow \mu^{+} \mu{-}$ 
\paragraph{Results}
Like the sensitivity to electron-coupled ALPs, we expect di-lepton final states to drive the sensitivity to this dark photon search, with the possibility of hadronic decay modes opening up for higher energy beams. Preliminary results point to extended sensitivity both at large mixing and small mixing angles, pushing beyond existing constraints from accelerator searches at NA62, NA48/2, and other efforts like those at JLab~\cite{Adrian:2022nkt}. A dedicated analysis of sensitivity projections, with the inclusion of the $\gamma^\prime$ production channels and di-lepton final states, discussed above, for the DAMSA experimental stages (see \S~\ref{sec:stages}) is to be conducted in the near future. This study would aim to investigate the sensitivity reach of DAMSA especially for kinetic mixings of the order $\epsilon \simeq 10^{-7}$ and dark photon masses $m_{\gamma^\prime} \gtrsim 20$~MeV, complementing NA62 and NA48/2 reach and the reach of other future beam dumps.


\subsection{Large Extra Dimensions}
\paragraph{Motivation}
Massive spin-2 fields are a common feature of many extensions of the Standard Model of particle physics. They naturally appear in extra-dimensional models, such as the Randall-Sundrum model~\cite{Randall:1999ee}, Linear Dilaton (LD)~\cite{Antoniadis:2001sw}, and the ADD model \cite{Arkani-Hamed:1998jmv}, among others, which aim to solve the electroweak hierarchy problem. In these models, the SM fields are confined to a 3-brane, while gravity propagates in the bulk. As a result of propagating in the extra dimension, the 4D Lagrangian contains an infinite tower of massive spin-2 fields, the Kaluza-Klein (KK) modes of the metric. The lightest of these modes, the graviton, is massless and mediates the gravitational interaction between the SM fields with interactions suppressed by the Planck mass. The heavier KK modes, on the other hand, are massive and can mediate new interactions between the SM fields, whose strength can be much larger than for the massless graviton, leading to a rich phenomenology.
The LD model is particularly interesting due to its interpolating behavior between Randall-Sundrum and ADD models -- the LD spectrum contains a mass gap and continuum of narrowly spaced KK resonances, which interact with the SM by sizable couplings. Moreover, LD is characterized by an approximate shift symmetry of the dilaton field, meaning in turn that the $k\ll M_5$ region is technically natural, which means that light, long-lived KK-gravitons are its rather natural prediction; $k$ is the curvature of the extra spatial dimension and $M_5$ is fundamental scale of gravity.
On the other hand, massive spin-2 field $G$ can also appear in the context of thermal dark matter (DM) models, where $G$ mediates the interaction between the DM and the SM~\cite{Lee:2013bua,Kang:2020huh}.
In such case, the origin of the mass of the spin-2 field $G$ is typically left unspecified, but it can be generated by the St\"ukelberg formalism, among other possibilities~\cite{Hinterbichler:2011tt}.

In both cases, the effective Lagrangian for the massive spin-2 field $G$ can be schematically written as:
\begin{equation}
  \mathcal{L} \supset g_{\gamma}\, G^{\mu \nu}\left(\frac{1}{4} \eta_{\mu \nu} F_{\lambda \rho} F^{\lambda \rho}+F_{\mu \lambda} F_\nu^{\ \lambda}\right)  -i \sum_l \frac{g_\ell}{2} G^{\mu \nu}\left(\bar{l} \gamma_\mu D_\nu l-\eta_{\mu \nu} \bar{l} \gamma_\rho D^\rho l\right)\,,
\end{equation}
where $G^{\mu \nu}$ is the symmetric tensor describing spin-2 field, $F_{\lambda \rho}$ is EM field strength tensor, $D$ is covariant derivative, and $l$ goes over the electrically charged SM fermions.
Note that for perturbative unitarity of the interactions, the couplings need to be equal $g_{\gamma}=g_{\ell}$, and that similar interactions can be written for the other gauge fields - for the complete Lagrangian, see~\cite{Lee:2013bua,Kang:2020huh}.

\paragraph{Results}
The leading decay channel of a light massive spin-2 field $G$ is to two photons~\cite{Lee:2013bua,Kang:2020huh}:
\begin{equation}
  \Gamma_G=\frac{g_{\gamma}^2 m_G^3}{80\pi}\,,
\end{equation}
which allows for the possibility of observing $G$ decays in DAMSA.

The leading production mechanism of producing sub-GeV spin-2 field $G$ is through the Primakoff-like process of conversion of on-shell photons into $G$: $\gamma N \to G N$~\cite{Jodlowski:2023yne}, while bremsstrahlung is typically subdominant.
The cross section for this process can be estimated as:
\begin{equation}
  \sigma_{\gamma N \to G N} \simeq \frac{\alpha_{em} g_{G\gamma \gamma}^2 Z^2}{2} \left(\log \left(\frac{d}{1/a^2 + t_{max}}\right)-2\right)\,,
\end{equation}
where $a$, $d$ are atomic form factor parameters - for more details, see~\cite{Jodlowski:2023yne}.

\begin{figure}[tb]
    \centering
    \includegraphics[scale=0.42]{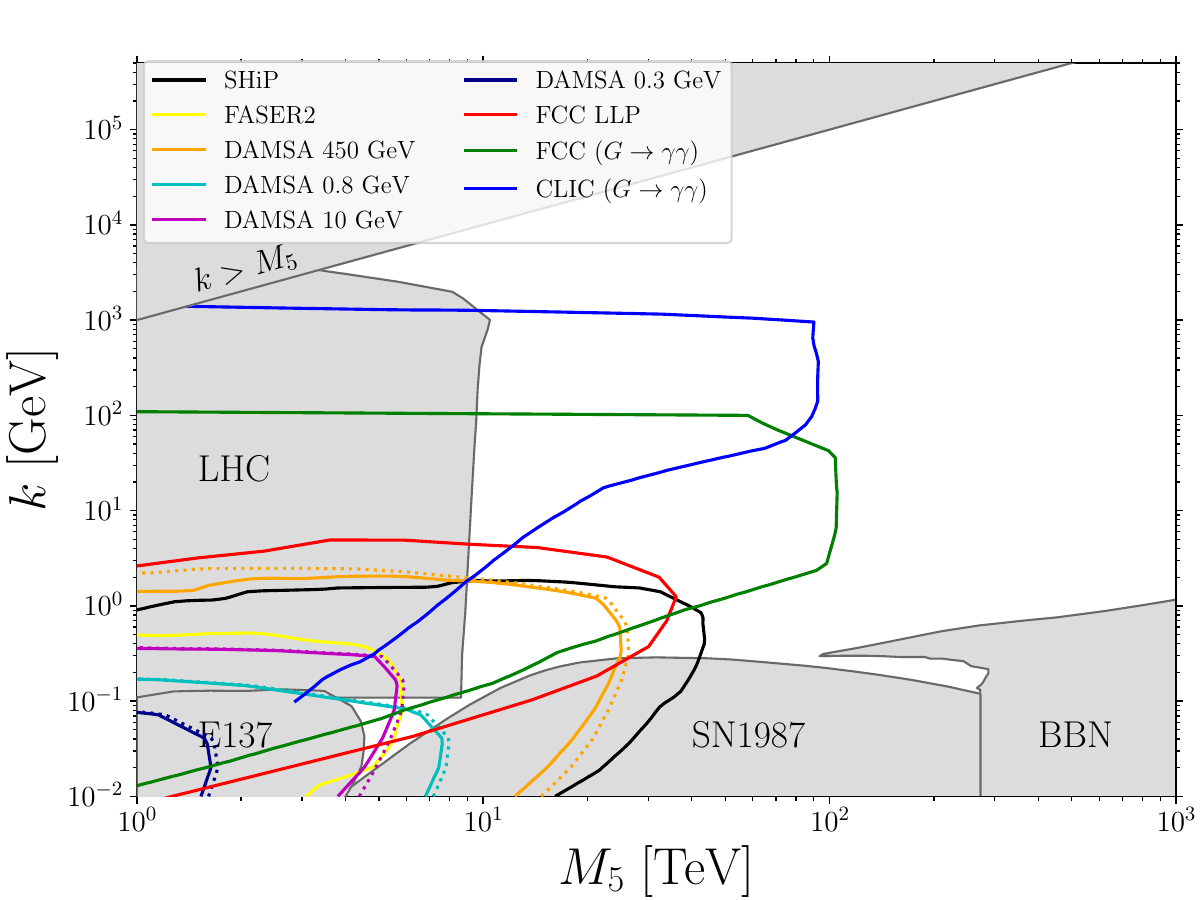}\vspace{0.35cm}
    \includegraphics[scale=0.3]{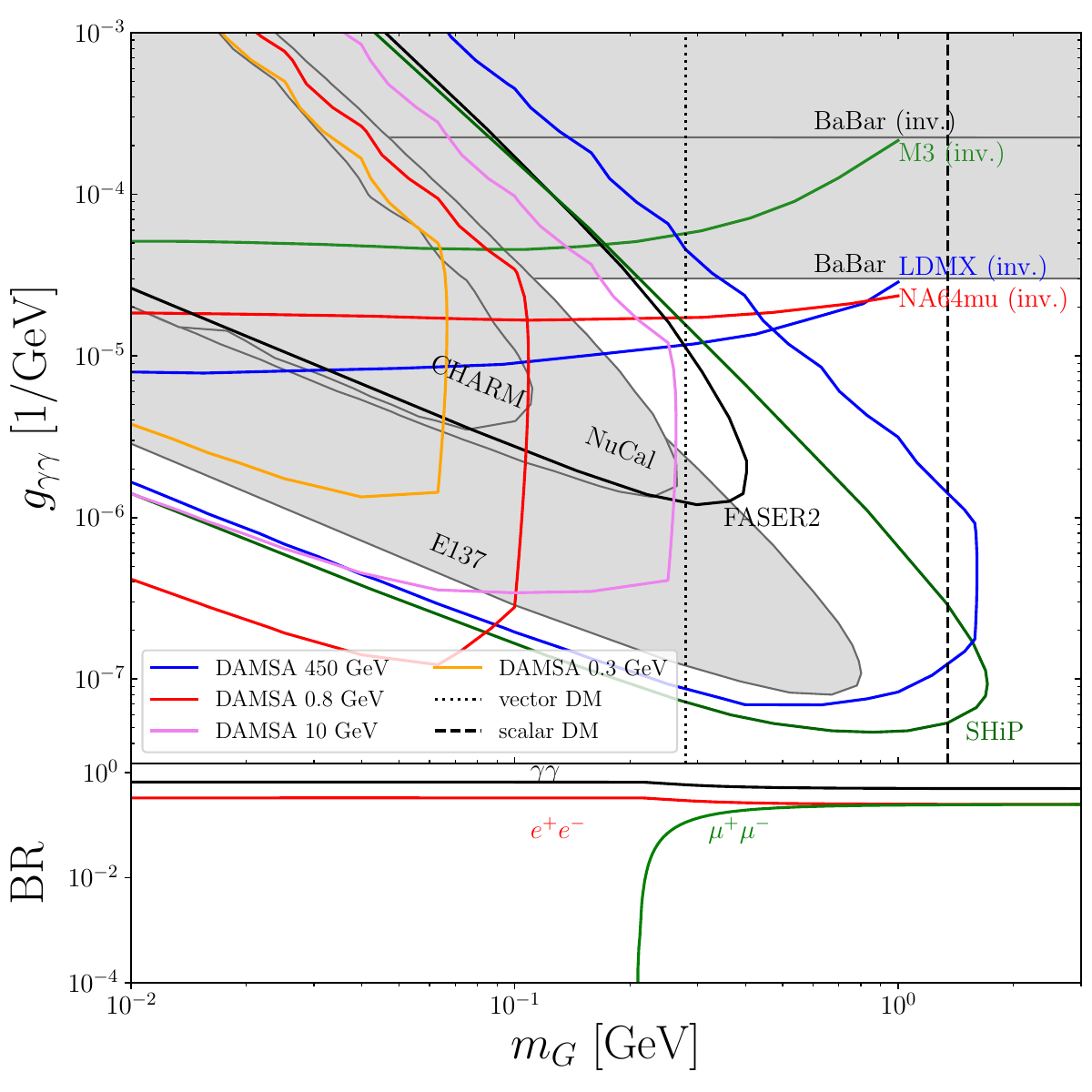}
    \caption{
      DAMSA sensitivity to the spin-2 field $G$ in the context of the Linear Dilaton model (left) and the massive spin-2 mediator to DM (right).
      The other limits shown in these panels have been determined in \cite{Im:2024jqx} and~\cite{Jodlowski:2023yne}, respectively. In addition to the baseline configuration of DAMSA, see text for details, we also show prospects of DAMSA with smaller separation between the target and the decay vessel, $L_{min}=0.1\,m$, which is denoted with dotted lines. As a result, shorter-lived KK gravitons can be probed.
      In the right panel, the DM relic density is shown for a benchmark considered in~\cite{Kang:2020huh}.
      }
    \label{fig:result_1}
\end{figure}

In Fig.~\ref{fig:result_1}, we show the results of our simulation, where we used the baseline DAMSA setup with $\theta=0.5\,\mathrm{rad}$, $L_{min}=1\,m$ and decay vessel length of $10\,m$. In both cases, DAMSA could probe previously unexplored parameter space, improving over E137, NuCal, and colliders. We note that the default DAMSA setup, characterized by beam $E_p=0.8\,\mathrm{GeV}$, can probe most of the gap between the LHC and supernovae limits. This is due to its main characteristics, which allow to probe the intermediate long-lived regime - the one between collider-stable species and species decaying on astrophysical length scales.

\paragraph{Conclusions}
Long-lived spin-2 particles can be naturally realized as KK-gravitons or mediators between the SM and DM.
The sub-GeV mass region is motivated by shift symmetry of the Linear Dilaton (technical naturalness) or in context of light thermal DM.
In both cases, various scenarios for DAMSA will have sensitivity to open parameter space, improving over E137, NuCal, and colliders. 
The signature is decays of $G$ not only into two photons, but also into a pair of charged leptons, since the branching ratios to matter are fixed by unitarity.
We have determined the sensitivity of the baseline DAMSA setup, which is comparable to other beam dump experiments (for example, FASER and SHiP). As a result, DAMSA could complement collider and astrophysical searches for extra spatial dimensions and massive spin-2 mediators to dark matter.

\subsection{Light Dark Matter}

\paragraph{Motivation}

Identification of dark matter remains a long-standing challenge in the particle physics community.
DAMSA can contribute to this effort by probing light dark matter (LDM) in the sub-GeV mass range, as motivated by theories with light mediators such as dark photons, dark scalars, axion-like particles, or neutrino-philic mediators.
At low-energy accelerator facilities, LDM can be produced through processes such as meson decays and electromagnetic radiation, depending on the underlying dark matter scenario~\cite{Bjorken:2009mm,Batell:2014mga,Batell:2022xau,Choi:2025wbw}.

From the experimental perspective, two broad categories of signals can be considered: elastic and inelastic dark matter scattering.
The first category corresponds to the elastic scattering of dark matter off electrons, nucleons, or nuclei (coherently), producing visible recoil signals in the detector.
If the transferred energy is sufficiently large, deep inelastic scattering (DIS) may also occur, leading to hadronic activity in the final state.
Such elastic and DIS-induced events are generic signatures expected for LDM produced at accelerator-based experiments.
The second category involves inelastic scattering of dark matter, in which either an excited dark sector state~\cite{Tucker-Smith:2001myb, Izaguirre:2014dua, Kim:2016zjx} or an excited target nucleus~\cite{Dutta:2024kuj, Choi:2024ism} is produced.
Depending on the model parameters and the amount of energy transferred, these excited states can give rise to secondary signatures, providing distinctive experimental features compared to elastic scattering processes.~\footnote{Although deep inelastic scattering is formally an inelastic process, it is treated separately here due to the absence of secondary signatures associated with excited dark sector states.}

\paragraph{Prospects}

Although no dedicated LDM analysis has yet been developed for DAMSA, several well-defined search avenues are already implicit in its design. 
Light dark matter particles produced at the target may scatter elastically or inelastically within the detector volume, generating low-energy electromagnetic or hadronic activity. 
The extremely short source–detector distance of DAMSA enhances the expected event rates compared to conventional beam-dump or long-baseline configurations, allowing competitive sensitivity even for very small couplings. Moreover, DAMSA’s proximity to the target opens opportunities to probe regions of parameter space where the dark sector particle is moderately long-lived, but would decay before reaching far detectors.
For a facility with high intensity but pulse type beam, utilization of the timing information can be important~\cite{Dutta:2019nbn,Dutta:2020vop}.
From a broader perspective, DAMSA can provide complementary coverage to existing beam-dump and collider experiments by probing LDM scenarios in which production is dominated by meson decays or photon-induced processes at low momentum transfer. 
Even in the absence of a signal, DAMSA can place meaningful constraints on dark matter–electron and dark matter–nucleon scattering cross sections, as well as on the couplings of light mediators, in parameter regions that are otherwise weakly constrained. 
With modest extensions to simulation and background modeling, DAMSA has the potential to evolve into a versatile probe of light dark sectors, leveraging its near-target geometry as a decisive experimental advantage.

\subsection{The Bread-and-Butter Standard Model Physics: Mesons and Physics Validation}
\begin{figure}
    \centering
    \includegraphics[width=0.8\linewidth]{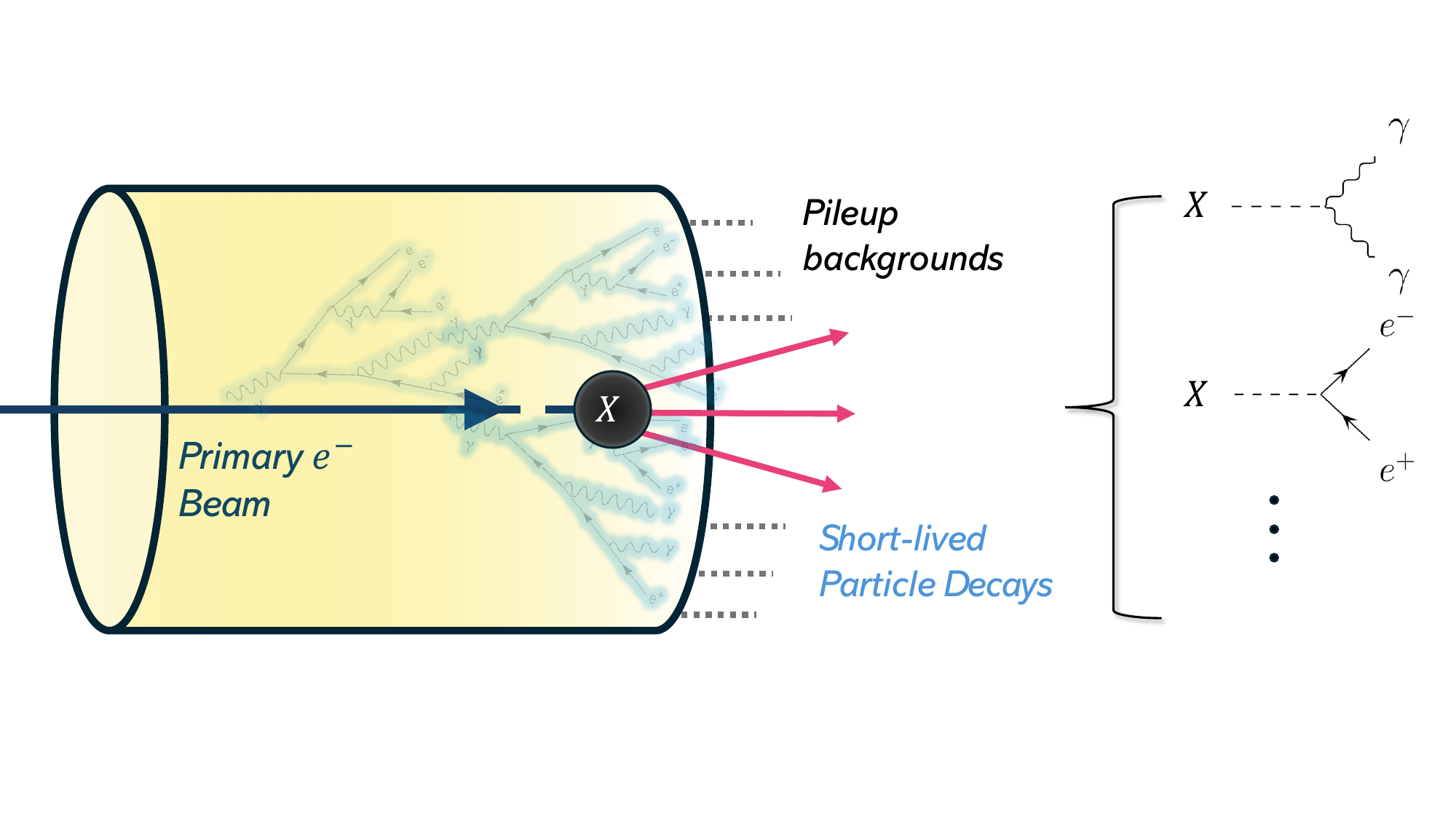}
    \caption{Searching for short-lived particle decays emerging from the DAMSA target.}
    \label{fig:damsa_diagram}
\end{figure}

The intense electron beam source impinging on the tungsten target also leads to meson production. Depending on the mass, lifetime, boost factor, and closeness of the production site to the decay pipe of these mesons when they decay, their final states can act as backgrounds to new physics signatures described in the previous sections. Quality vertex resolution for decay signatures (e.g. $\gamma\gamma$, $e^+ e^-$) will be very important for discriminating \textit{short-lived} particles decaying within the target from those decaying within the decay pipe. This also presents an opportunity to reconstruct mass resonances of promptly decaying mesonic states, both for the purposes of physics validation as well as studies of rarer decay channels predicted within the SM.

A key example is the production of neutral $\pi^0$ from the beam, in addition to possible production of $\eta$, and $J/\psi$ in the beam target. Their decays to $\gamma\gamma$ and $e^+ e^-$ final states will serve as benchmark decays to test the function and calibration of the detectors and selection efficiency to demonstrate the sensitivity to BSM particle searches with similar final states, and can complement searches for invisible final states~\cite{Schuster:2021mlr}. The lifetimes of these particles are incredibly short, so they need to be produced close to the end of the target geometry so that their decay products can escape into the decay pipe without absorption or significant energy loss before detection. In addition, the primary challenge will be to differentiate these signals, e.g. $\pi^0 \to \gamma \gamma$ or $\eta \to \gamma \gamma$, from the background leakage flux of electromagnetic showers through the end of the beam target.

Estimations of the background pileup flux with $10^4$ EOT per pulse indicate a background $\gamma$ flux multiplicity entering the detection volume of $N_\gamma \sim \mathcal{O}(100)$, though their energy spectrum is softer than the typical $\pi^0 \to \gamma \gamma$ decay at rest energy $\langle E_\gamma\rangle \sim m_\pi / 2$, with additional boost from the typical pion kinetic energies of $\mathcal{O}(100)$ MeV. A lower energy cut to reduce the high multiplicity can be optimized for $\pi^0 \to \gamma \gamma$ selection on a pulse-by-pulse basis, in addition to vertex finding for $\gamma\gamma$ pairs. Acceptance of both photons from neutral pion decay into the decay pipe will also require that the parent pion carries forward momentum. For the remaining $N_\gamma$ photons in the pulse window after selection cuts, a pair-wise invariant mass distribution can be defined,
\begin{equation}
    M_{i,j}^2 \equiv 2  E_i \cdot E_j (1 - \cos\theta_{ij}) \, ,
\end{equation}
for $\gamma$ energies $E_i$, $E_j$, $i\neq j$ and separation polar angle $\theta_{ij}$. Resonances in the large $M_{i,j}^2 \gtrsim m_\pi^2$ region should stand out over a falling pileup background from the softer electromagnetic shower.

Other candidate particle decays for physics validation on $\gamma\gamma$, $e^+ e^-$, $\pi^+ \pi^-$, and $\mu^+ \mu^-$ final state resonances and Dalitz 3-body final states could include~\cite{ParticleDataGroup:2024cfk}
\begin{align}
    \pi^0 &\to \gamma \gamma & BR_{\rm exp} \simeq 100\% \nonumber \\
    \pi^0 &\to e^+ e^- \gamma & BR_{\rm exp} \simeq 1\% \nonumber \\
    \eta &\to \gamma \gamma & BR_{\rm exp} \simeq 100\% \nonumber \\
    K_S &\to \pi^0 \pi^0 (4\gamma) & BR_{\rm exp}= 30.69 \pm 0.05\% \nonumber \\
    K_S &\to \pi^+ \pi^- & BR_{\rm exp} = 69.20 \pm 0.05 \% \nonumber
\end{align}
In addition, searches for rare meson decays and their final states emerging from the DAMSA target can be investigated. Namely, $\pi^0 \to e^+ e^- e^+ e^-$ ($\pi^0 \to 4e$, $BR \simeq 10^{-5}$), which originally helped determine the neutral pion's pseudoscalar nature, may be possible with its unique final state given an enriched sample of forward-boosted neutral pions. Eta meson decays such as $\eta \to \mu^+ \mu^- \gamma$ Dalitz supply a dimuon final state ($BR \simeq 10^{-4}$) and even rarer decays with limited experimental bounds could be improved such as $\eta \to e^+ e^- \ell^+ \ell^-$~\cite{CELSIUSWASA:2007ifz}. Depending on the beam energy, $J/\psi$ production near threshold may also be accessible with characteristic tests of $e^+ e^-$ and $\mu^+ \mu^-$ decays ($BR \simeq 5\%$) and offer tests of non-perturbative QCD~\cite{JeffersonLabSoLID:2022iod,JointPhysicsAnalysisCenter:2023qgg}.

The intense electron beam incident on the DAMSA targets produces a copious flux of Standard Model mesons whose prompt decays provide a powerful in situ calibration and validation program, beginning in Stage 1 of the experiment at the SLAC LESA 8 GeV electron beam and extending to later stages. The primary calibration benchmark is neutral pion production, with $\pi^{0} \to \gamma\gamma$ used to calibrate the electromagnetic energy scale, vertexing performance, and photon reconstruction efficiency, while a secondary meson mass resonance (such as $\eta \to \gamma\gamma$ or $K_{S} \to \pi^{+}\pi^{-}$) is employed to validate mass resolution and overall normalization. By exploiting pulse-by-pulse energy thresholds, vertex reconstruction, and invariant mass spectra, meson decay resonances are expected to emerge above the softer electromagnetic pileup background. The reconstructed yields, kinematic distributions, and invariant-mass peaks are to be compared with detailed beam-target and detector simulations, providing a quantitative cross-check of background modeling and detector response and establishing sensitivity to beyond-the-Standard-Model searches with similar final states across the DAMSA program.



\vspace{0.5em}
\section{The Beam Facility}      
\label{sec:Facility}

\subsection{Electron beam facilities}
\subsubsection{SLAC/LESA Test Beam Facility}

\begin{figure}[htb]
\centering
\includegraphics[width=0.68\linewidth]{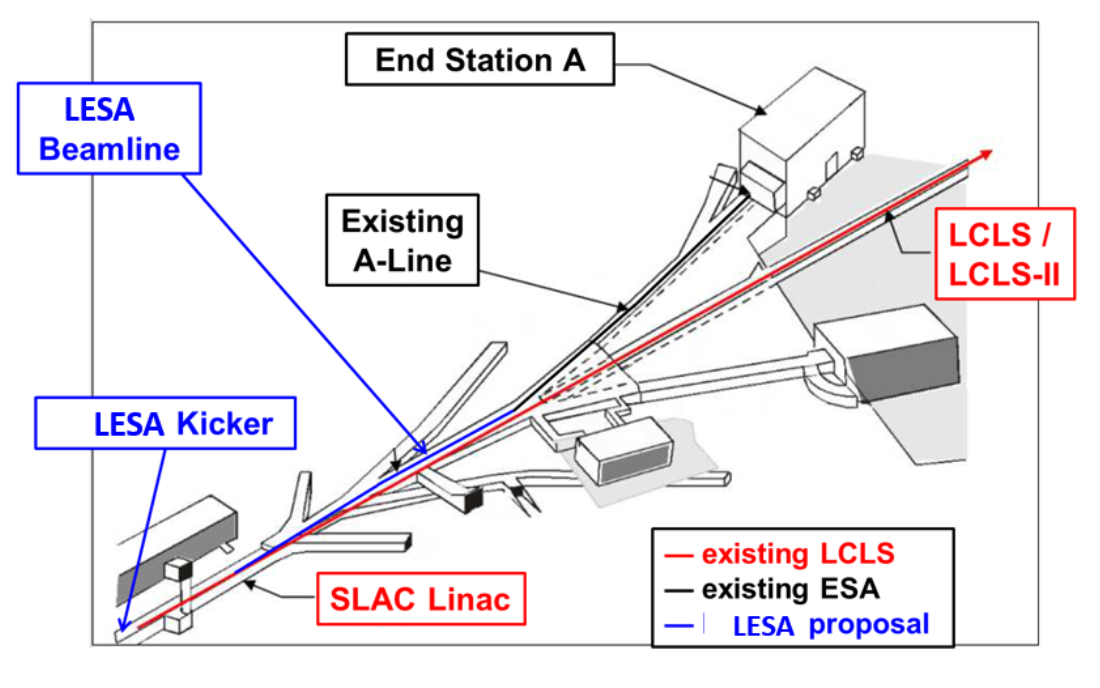}
\caption{Layout of the LESA Beam-line at SLAC.} 
\label{fig:LESA_layout}
\end{figure}

The LINAC-to-ESA (LESA) beam-line at SLAC is a revamping of the A-Line’s beam transport capability to reestablish a test beam program at End Station A (ESA). It is expected to be available in 2027. LESA is capable of high repetition rates, precision timed short bunches, and variable operating modes. It is ideal for advanced detector R\&D for HEP and NP programs. A layout of LESA is shown in Fig.~\ref{fig:LESA_layout}. The maximum beam energy available at LESA is 8 GeV with flexibility
to select lower energies down to about 2 GeV (limited by stable operation of the A-Line magnets. See Fig.~\ref{fig:LESA_layout}). However the maximum beam intensity at lowest energies will be orders of magnitude less than at 8 GeV. The accelerator beam parameters of the LESA system are listed in Table~\ref{tab:LESA_para} for the 4 expected modes of LESA operation.

\begin{table}[]
\begin{tabular}{|l|c|c|c|c|}
\hline
\textbf{Accelerator Parameters} & \multicolumn{1}{l|}{\textbf{Interleaved}} & \multicolumn{1}{l|}{\textbf{X-LEAP Laser}} & \multicolumn{1}{l|}{\textbf{LESA Laser}} & \multicolumn{1}{l|}{\textbf{Dark current}} \\ \hline
SRF LINAC Energy                & 8 GeV                                     & 8 GeV                                      & 8 GeV                                    & 8 GeV                                      \\ \hline
Max Current in LESA             & \textless{}10 nA                          & 25 nA                                      & 25 nA                                    & \textless{}$\sim$1 pA                          \\ \hline
Maximum Kicker Rate             & 100 Hz                                    & 929 kHz                                    & 929 kHz                                  & 929 kHz                                    \\ \hline
\# bunches per Kick             & 1                                         & 1                                          & 18                                       & 100                                        \\ \hline
Avg. Bunch Charge               & 10-100 pC                                 & \textgreater{}167ke-(27 fC)                & \textgreater{}4.2ke-(0.7 fC)             & 0.07 e-                                    \\ \hline
Bunch spacing                   & \textgreater 10 ms                        & \textgreater 1.08 µs                       & 26.9 ns                                  & 5.4 ns                                     \\ \hline
Max beam power                  & 8 - 80 W                                  & 200 W                                      & 100 W                                    & 8 mW                                       \\ \hline
\end{tabular}
\caption{LESA Beam Parameters for 4 Operation Modes.}
\label{tab:LESA_para}
\end{table}

\subsubsection{Facility for Accelerator Science and Technology, FAST}

\begin{figure}[htb]
\centering
\includegraphics[width=0.98\linewidth]{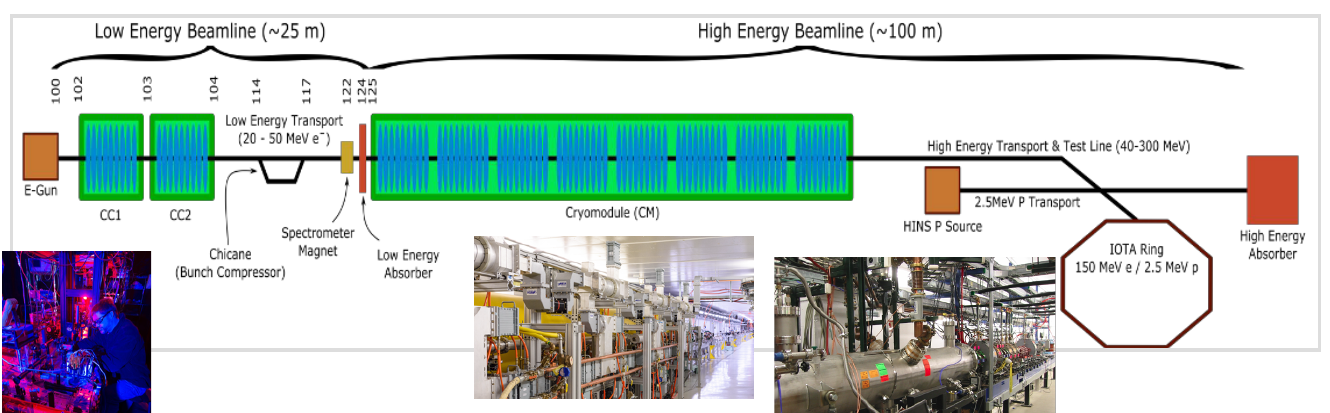}
\caption{Layout of the FAST Beam-line at Fermilab (From M. Wallbank/FNAL).} 
\label{fig:FAST_layout}
\end{figure}

The Fermilab Accelerator Science and Technology (FAST) facility is a R\&D accelerator chain to primarily support research and development of accelerator technology for the next generation of particle accelerators but in addition the electron beam could be used for future experiments. The FAST facility is located at NML, formerly known as the New Muon Lab. FAST is a superconducting RF LINAC with the electron injector running through the NML building and a tunnel extension to provide electrons to the IOTA ring (IOTA is an electron-proton storage ring) as shown in Fig.~\ref{fig:FAST_layout}. The electron injector includes a 5 MeV electron RF photo-injector, a 25 meter long low energy ($\le$40 MeV) beam-line followed by a $\sim$100 meter long high energy ($\le$300 MeV) beam-line. The beam parameters for the FAST facility are listed in Table~\ref{tab:FAST_para}. The CM (High Energy Beam-line) is commissioned and reached more than 250 MeV for its beam energy. In addition, 500 bunches at 3~MHz has been achieved.
Several beam line locations and spaces are available for R\&D including Fixed Target experiments.

\begin{table}[]
\begin{tabular}{|l|c|}
\hline
\multicolumn{1}{|c|}{\textbf{Parameter}} & \textbf{Value}                                           \\ \hline
Beam Energy                              & 20 MeV – 300 MeV                                         \\ \hline
Bunch Charge                             & \textless 10 fC – 3.2 nC per pulse                       \\ \hline
Bunch Train (Macropulse)                  & 0.5 – 9 MHz for up to 1 ms (3000 bunches, 3 MHz nominal) \\ \hline
Bunch Train Frequency                    & 1 – 5 Hz                                                 \\ \hline
Bunch Length                             & Range: 0.9 – 70 ps (Nominal: 5 ps)                       \\ \hline
Bunch Emittance (50 MeV, 50 pC/pulse)    & Horz: 1.6 ± 0.2 $\mu$m, Vert: 3.4 ± 0.1 $\mu$m                   \\ \hline
\end{tabular}
\caption{FAST Beam Parameters}
\label{tab:FAST_para}
\end{table}

\subsubsection{Pohang Accelerator Laboratory}
Pohang University of Science and Technology operates the Pohang Radiation Accelerator (PAL) and is also known as the Pohang Light Source (PLS). PAL is the largest particle accelerator facility operating in Korea and was launched in 1988. Multiple electron beams serve as light sources, including the X-ray free electron laser (XFEL), a fourth-generation synchrotron accelerator based on the design of SwissFEL and was completed in 2017. It has a total length of 1.1 km, a beam energy of 10 GeV, and a bunch length of 5 ps. In addition, PAL houses a compact synchrotron accelerator with a storage ring circumference of 36 m and an electron beam energy of 400 MeV, which specializes in providing Extreme Ultraviolet (EUV) light sources.

\subsubsection{Jefferson Laboratory}
\begin{figure}[htb]
\centering
\includegraphics[width=0.68\linewidth]{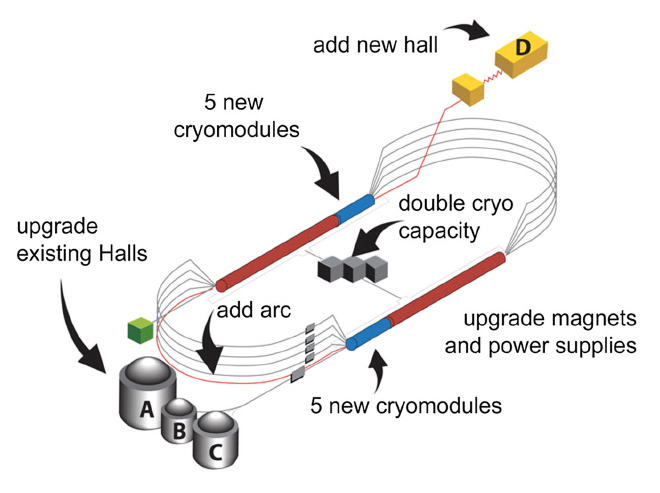}
\caption{Schematic of the 12 GeV CEBAF accelerator at JLAB } 
\label{fig:CEBAF_12GeV}
\end{figure}

The laboratory houses the CEBAF accelerator which includes a polarized electron source and injector and a pair of superconducting RF linear accelerators that are 1400 m (7/8-mile) in length connected to each other by two arc sections containing steering magnets. With each revolution around the accelerator, the beam passes through the two LINAC accelerators, but through a different set of bending magnets in semi-circular arcs at the ends of the LINACs. The electron beam makes five successive orbits, increasing the energy to a maximum of 12 GeV. CEBAF is fundamentally a linear accelerator, but because of the unique stacked design the length is reduced to a tenth of its expected size as shown in Fig.~\ref{fig:CEBAF_12GeV}. The design of CEBAF allows the electron beam to be roughly continuous. There is however beam structure, but the pulses are very much shorter and closer together compared to other electron beam facilities. The pulse width is 250 $\mu$s at 60 Hz (16.67 ms pulse spacing) resulting in a 1.5\% duty cycle. Nominal pulse height is 4 mA with a beam power of 720W at 12 GeV. Three 499 MHz bunch trains are interleaved producing a 1497 MHz bunch train injected into the LINACs with bunch widths of 1 picosecond. The electron beam can be directed to four potential targets halls. The accelerator is located 8 meters underground with 2 foot thick tunnel walls.

\subsection{Proton beam facilities}
\subsubsection{CERN Beam Dump Facility/SHiP}
\begin{figure}[htb]
\centering
\includegraphics[width=0.88\linewidth]{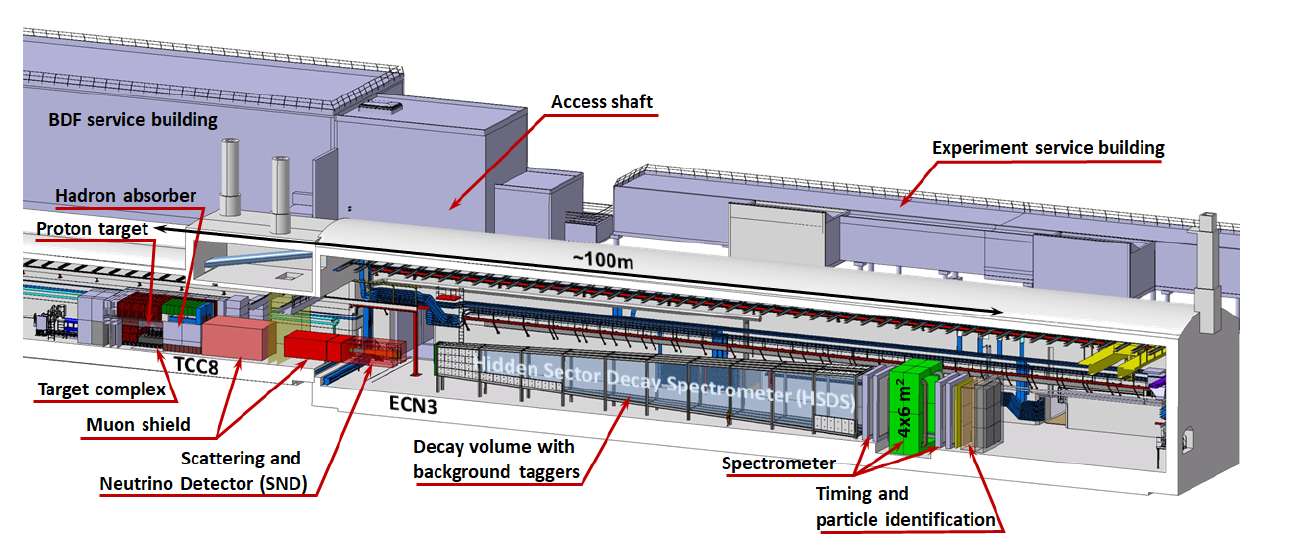}
\caption{The beam dump facility and the SHiP detector.} 
\label{fig:BDF_SHIP}
\end{figure}

The CERN Beam Dump Facility (BDF) is a an upgrade to the existing ECN3 experimental facility at CERN's SPS accelerator and is designed for high-intensity beam-dump experiments. ECN3 is located along with the TCC8 target area at the end of the 750m transfer tunnel downstream of TCC2. Both ECN3 and TCC8 are located below ground with an 8 m thick layer of dirt for shielding placed on top. The 400 GeV/c proton beam extracted from the SPS is dumped on a heavy target. The SPS is expected to deliver on average 5000 spills ($4\times 10^{13}$ protons per spill) per day. The BDF/SHiP target is therefore designed to fully absorb 400 GeV 2.6 MJ/pulse every 7.2 seconds (with a 1.2~s flat-top), \textit{i.e.} roughly 350 kW of average beam power. The facility implements a large proton target/dump, shielding, and space for large detectors. The space in front of SHiP is too small 
for a large-scale experiment, but DAMSA, because of its small size, could be placed immediately behind the muon magnetic shield. BDF operation is planned to operate for at least 15 years. The beam dump facility and the SHiP detector are shown in Fig.~\ref{fig:BDF_SHIP}.

\subsubsection{Fermilab Facility for Dark sector particle Discovery}
\begin{figure}[htb]
\centering
\includegraphics[width=0.88\linewidth]{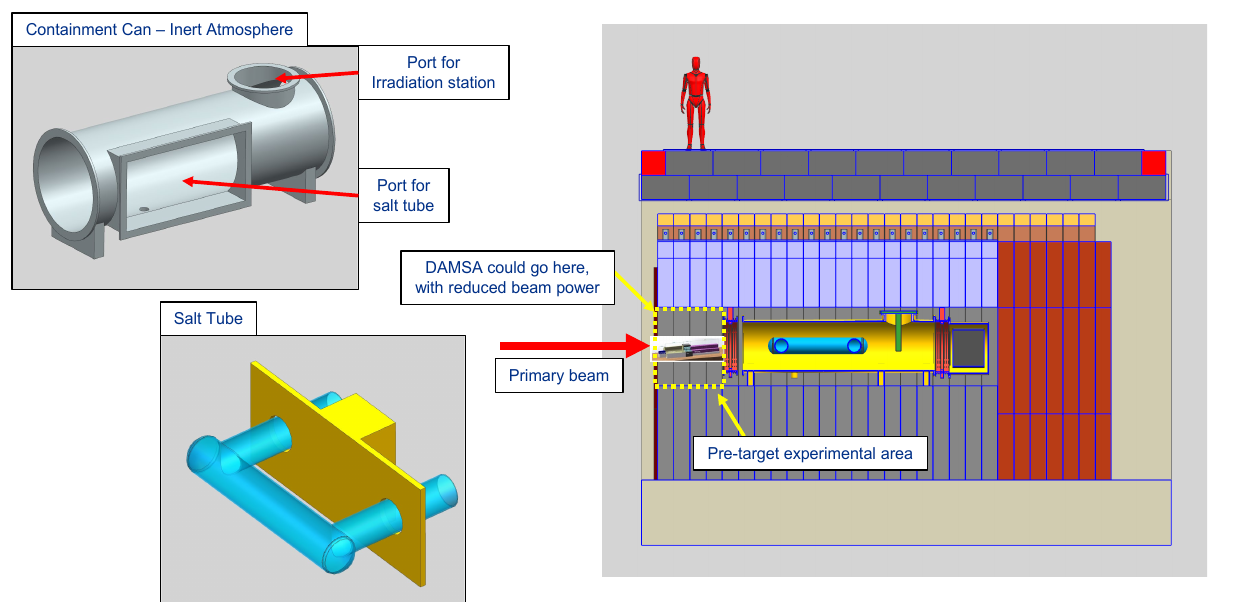}
\caption{The F2D2 target region and possible location for the DASMA detector.} 
\label{fig:F2D2_DAMSA}
\end{figure}

Several workshops have been held at Fermilab resulting in the commissioning of a task force for the study of the feasibility  of a beam dump facility, tentatively called Fermilab Facility for Dark sector Discovery (F2D2). The task force recently submitted a report to Fermilab leadership with DAMSA as one of the proposed experiments. The primary requirements for F2D2 is to implement a high-power, low-beam energy target/beam dump facility utilizing the 2.5 MW power 1-GeV PIP-II beam to produce dark sector interactions. The expected number of Protons on Target (POT) per second is $1.565\times 10^{16}$. Experiments in general would be located downstream in a separate hall from the target/beam area but also would allow for additional experimental space upstream of the main target/dump where DAMSA could be located. DAMSA could be placed in front of a shielding block with its own small target to produce a short baseline as shown in Fig.~\ref{fig:F2D2_DAMSA}.

\vspace{0.5em}
\section{Experiment and The Detector}      
\label{sec:Detector}
DAMSA~\cite{Jang:2022tsp} is a very short baseline beam dump experiment, aiming to probe the parameter space inaccessible in previous beam dump experiments. Given the proximity to the beam dump/target, the primary background comes from the large number of beam-related neutrons (BRN), resulting from the beam interactions in the dump. In order to overcome these backgrounds, DAMSA aims to detect two-photon or $e^{+}e^{-}$ final states of dark sector particles.

Our extensive simulation-based studies~\cite{Kim:2024vxg} show that DAMSA can access the targeted parameter space with a tabletop scale experiment equipped with a vacuum decay chamber, a precision tracking detector under a magnetic field and a fine granular 4D total absorption electromagnetic calorimeter.
This section presents the DAMSA experiment components in detail.

\begin{figure}[htb]
    \centering
    \includegraphics[scale=0.2]{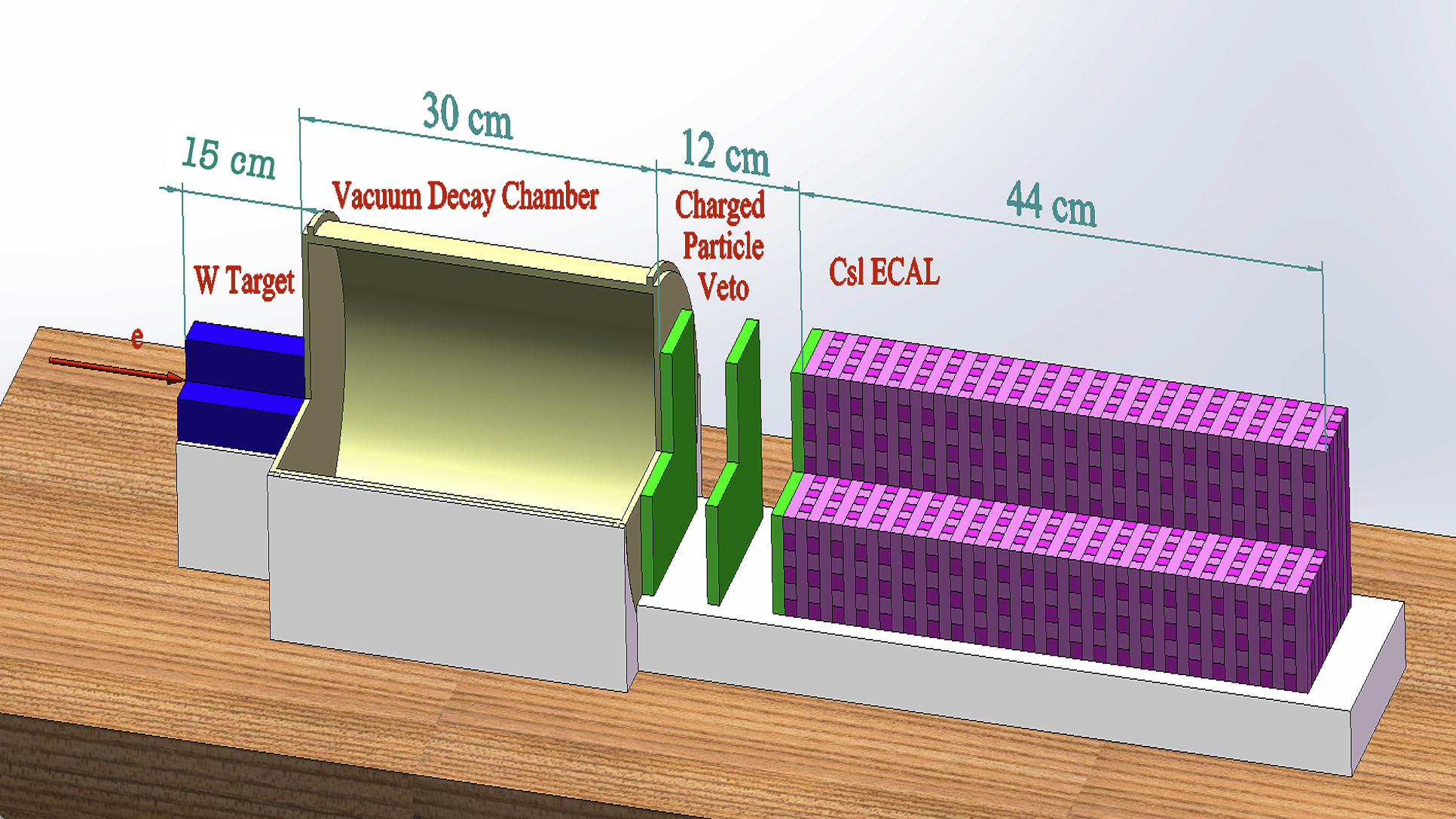}
    \caption{
        Stage 1 DAMSA experiment which consists of the 15~cm tungsten target, the vacuum decay chamber, a three scintillator charged particle ID system and the 4D total absorption ECAL.
    }
    \label{fig:stage1}
\end{figure}

\subsection{The Target}
The target of DAMSA detector will be made of a block of tungsten of $5~{\rm cm} \times 5~{\rm cm}$ in the transverse direction and $10~{\rm cm}$ long along the beam direction, providing $28.5X_{0}$ radiation length, sufficient to absorb most of the 300~MeV electrons in the beam.
Given the expected electron beam intensity, in the range of $10^{7}\sim 10^{8}/{\rm pulse}$, with the pulse width 10~ns and repetition rate of about 100~Hz to minimize both the fast neutron component from the target and the remaining neutrons in the experimental hall, the expected peak beam power on the target for the 300~MeV electron beam at FAST is $ 5\times 10^{4} \sim 5\times 10^{5} {\rm (W)}$.  
This represents $ 0.05 \sim 0.5 {\rm (W)}$ of the beam power averaged over one second, a negligible number.

The configuration and shape of the target will be studied to further reduce the flux of the electromagnetic particles, such as X-rays.
For example, we may add a 1~cm thick lead sheet at the end of the 10~cm W target to absorb a large fraction of X-rays that could be produced in the target.
We may also explore a rotating, cylindrical shape target that is positioned slightly off center with respect to the incident beam position to help further dissipate heat from the beam interactions.
This is a poor-man's way of rastering the beam to further disperse the heat.  
This kind of target geometry may be necessary if the beam pulse intensity at a future experimental configuration increases.

\subsubsection{Target Composition of Tungsten and Lead}
\begin{figure}[htb]
\centering
\includegraphics[width=0.48\linewidth]{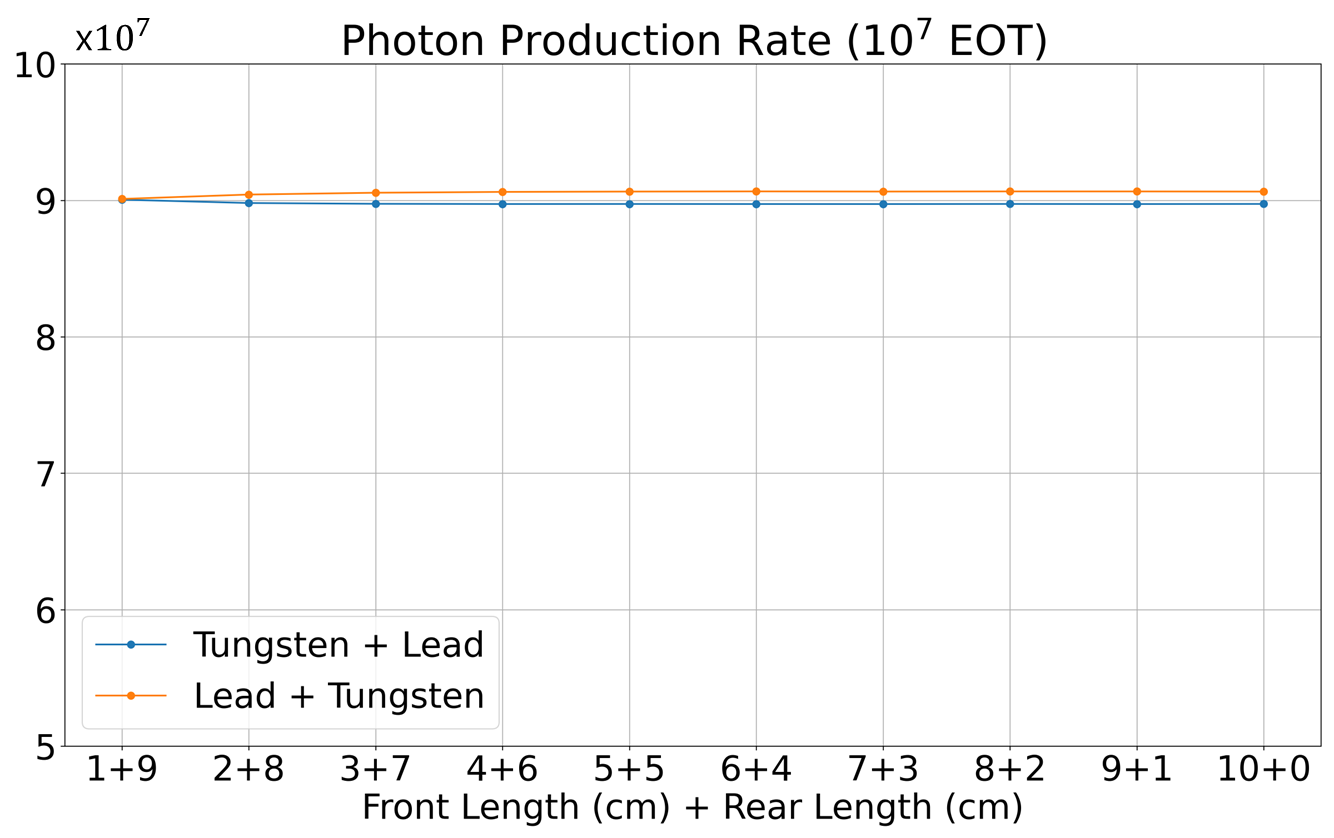}
\includegraphics[width=0.48\linewidth]{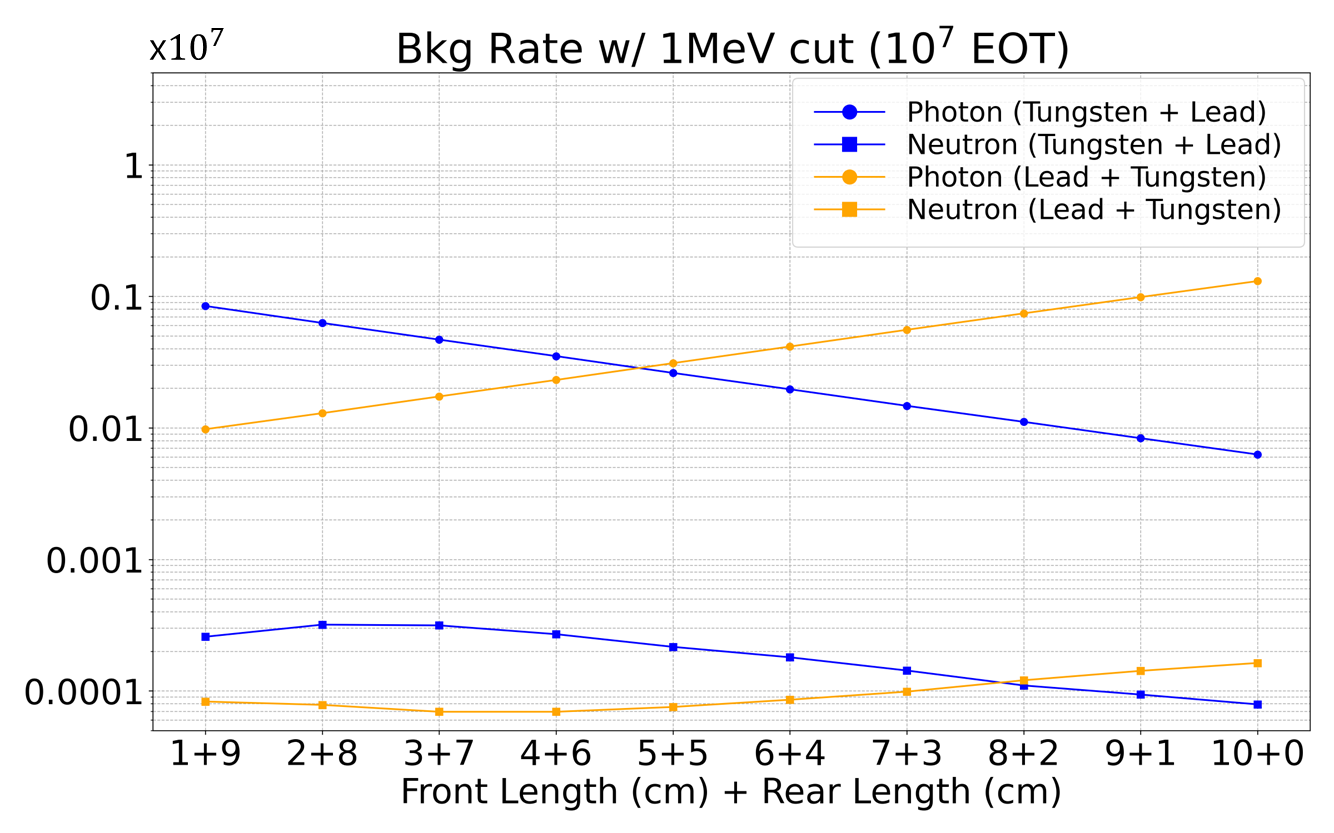}
\caption{Photon production rate(left) and Background rate(right)} 
\label{fig:Target}
\end{figure}

The optimal target composition of tungsten and lead was studied using Geant4 simulations, considering both photon production and backgrounds. For photon production, we calculated the number of photons produced with energy above 10 MeV and positive P$_z$, which are potential candidates for ALP generation. As shown in the left plot of Fig.~\ref{fig:Target}, the difference is at most 1\%. For background, electrons and positrons are omitted in right plot of Fig.~\ref{fig:Target} because they exhibit the same trend as photons. Considering both photon production and background, the full tungsten target is optimal when only tungsten and lead are considered. 

\subsection{The Vacuum Decay Chamber}
The vacuum decay chamber is an essential element of the LDPF experiment.
It is located immediately downstream of the target, as show in Fig.~\ref{fig:stage1} and, in particular, allows unstable dark-sector particles like ALPs to decay inside.
The decay vertex will then be reconstructed from the fine granular electromagnetic calorimeter, reducing the background from random overlap of the photons from neutron interactions which require materials to interact.
The diameter of the vacuum chamber is 20~cm to provide sufficient angular coverage to capture a large fraction of the unstable particles produced in the target.
The length of the chamber is 30~cm to allow sufficient separation between the two decay products in the ECAL downstream.
The small size of the vacuum decay chamber makes fabrication easy.  
We plan to utilize a commercial vendor for the fabrication.  
A chamber with convex heads and side pump out is the baseline.
Maintaining a high level of vacuum is important in minimizing neutron interactions inside the vacuum volume.
One of the outcomes of the LDPF is to measure the vacuum level dependence of the backgrounds and determine the optimal level of vacuum for the subsequent data taking.

\subsection{The Magnet}
The magnet provides a strong magnetic field in the 12 cm space allocated for the tracking detectors, located between the vacuum decay chamber and the electromagnetic calorimeter. This enables the separation of charged particles from photons. It also allows for the identification of charge polarity, such as distinguishing between electrons and positrons, and can support momentum reconstruction.
Due to its compactness and maintenance-free characteristics, a permanent magnet is preferred over an electromagnet. Among permanent magnets, two options are considered: NdFe magnets, which provide the strongest magnetic field strength, and SmCo magnets, which are highly resistant to radiation.

Due to the narrow space limitation, a 12 cm window is maintained to accommodate the trackers. By placing two NdFe magnet blocks with dimensions of 30 cm(x) $\times$ 25 cm (y) $\times$ 12 cm (z), separated by a 12 cm gap, the maximum central magnetic field strength achieved is 0.55 T. When a magnet yoke structure is added, the central magnetic field strength increased to 0.65 T.

\subsection{The Tracking Detector}

The tracking detector aims to reconstruct electrons and positrons with high efficiency, separating from photons deposited in the calorimeter. \\

The tracking technology we consider is the LGAD sensor developed for the upgrade of the CMS timing detector.
Other technology is a micro resistive well ($\mu$RWELL) detector.
In particular, $\mu$RWELL will play a crucial role in the early stages of the experiment.

The tracker will be located right after  the vacuum decay chamber. The transverse size of the tracker is 10 cm~$\times$~10 cm with a length of 12 cm in the beam direction. This chamber will be covered  by 0.5 T of magnetic field of the dipole magnet.  The tracker will have four layers along the beam line, aiming for a tracking efficiency above 95\%
for momenta above 30 MeV/c. With this tracking detector, a vertex resolution for a short-lived particle decaying into an electron-positron pair is better than \SI{1}{mm}. In addition, the tracker with LGAD sensor
provides a timing resolution of 50 ps, and  around a level \SI{5}{\nano\second} with $\mu$RWELL technology.
\newline

\subsubsection{LGAD Si Pixel Detector}

The DAMSA silicon tracker is designed to identify the electron and positron decaying from the ALP or Dark Photon. The silicon tracker will be installed in 4 to 6 layers depending on the available detector space allowed. The total amount of space required for two-pixel layers along the z-axis for the detector is about 45 mm. An additional space of 20 mm for each side of the silicon tracker is needed for the thermal screen. Each pixel layer will use Low Gain Avalanche Diodes (LGAD) sensors, which enable precision timing and position reconstruction for charged particles. The sensors with 50 $\mu m$ active region, within normal 300 $\mu m$ thick silicon wafer, and a thin implanted gain layer are expected to provide the desired performance. Former studies demonstrated a resolution of timing reconstruction about 30 to 50 ps. The LGAD sensor is a 16×16 pixel array consisting of square pads, each pixel size being 1.3 × 1.3 $mm^2$, resulting in a total sensor size of 21 × 21 $mm^2$. Each module is read out by an ASIC, of approximately 20 × 20 $mm^2$, which processes signals from a 16×16 submatrix. The size of the sensor will be larger than the chip size to allow the ASIC to be bump-bonded to it for electrical connection. A set of readout chips interfaces with an on-detector board, known as the service hybrid, which supplies DC power, bias voltage, communication (both slow and fast control), and monitoring for the ASICs. The service hybrid consists of two components: a readout board and a power board. The pixel layer will be operated with the cooling system to provide temperatures of at least -30 $^{\circ}$C.
\newline

\subsubsection {Other technologies}

\begin{figure}[htb]
    \centering
    \includegraphics[scale=0.40]{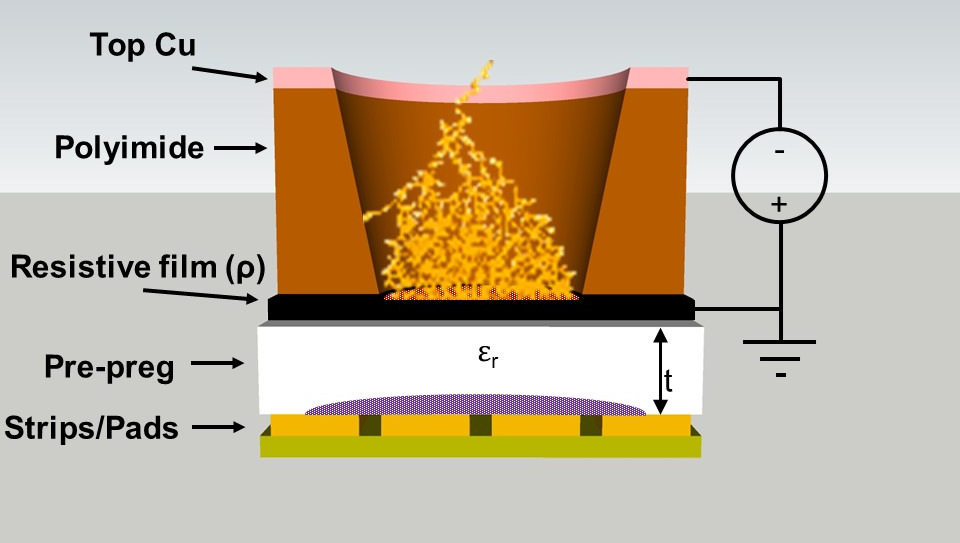}
    \caption{
        Structure of unit cell of $\mu$RWELL \cite{Bencivenni_2019}.
    }
    \label{fig:detector:uRWELL_Unit_Cell}
\end{figure}

Another technology to consider is $\mu$RWELL, the resistive variant of gas electron multiplier (GEM) \cite{POLILENER2016565}.
Figure~\ref{fig:detector:uRWELL_Unit_Cell} shows the unit cell structure of $\mu$RWELL.
The strong electric field provided by the GEM microelectrodes amplifies electrons generated by primary ionization, producing a measurable signal.
The resistive film composed of diamond-like carbon (DLC) suppresses discharges by inducing local voltage drop when  streamers form.
Through this mechanism, $\mu$RWELL suppresses discharge occurrence, and even if a discharge occurs, the discharge current is highly quenched.
Therefore, a higher voltage can be applied to the GEM foil, enabling a single GEM foil to achieve a high gain.

Such discharge suppression offers various advantages.
First, it mitigates channel loss in the readout electronics caused by discharges.
Additionally, it eliminates the need for spatial separation between the avalanche region and the induction region.
the GEM foil can be directly stacked onto the readout PCB.
As a result, the GEM foil obtains  self rigidity, eliminating the need for foil stretching.
This enables a simpler detector structure and significantly simplifies the assembly process.
This makes the $\mu$RWELL detector more cost-effective.

Depending on the readout configuration, $\mu$RWELL can achieve an excellent position resolution of \SI{70}{\micro\meter} or slightly better.
Time resolution varies depending on the drift gap spacing and gas selection, but it can reach around \SI{5}{\nano\second}.
Other advantages of the $\mu$RWELL detector in this experiment include its thin structure, allowing it to fit within limited space, and its low material budget.

For more stable detector operation, a GEM + $\mu$RWELL hybrid (or GRWELL), where GEM serves as a preamplifier for $\mu$RWELL can be considered.
In particular, for 2D tracking, the $\mu$RWELL detector must be operated close to its maximum gain to achieve full efficiency.
From the perspective of stable detector operation, this may introduce a risk.
The added structural complexity from introducing an additional GEM is a drawback, but it is not a significant issue for small detectors of around $\SI{10}{\centi\meter}\times\SI{10}{\centi\meter}$ area.

The readout structure is being designed with an XY strip readout to maintain an appropriate number of readout channels.
Given that at least four $\mu$RWELL or GEM + $\mu$RWELL detectors will be used to construct the tracker and that the charged particle flux in the experimental environment is not expected to be high, the occurrence of ``ghost cluster'' is anticipated to remain low.

Additionally, capacitive sharing readout is being considered to keep the number of readout channels low.
This method allows signal sharing between adjacent readout channels through capacitive coupling, ensuring high strip multiplicity even with a large strip pitch \cite{GNANVO2023167782}.
As a result, it enables high position resolution while maintaining a reduced number of readout channels. 
By using the capacitive sharing readout with an XY strip structure, a $\SI{10}{\centi\meter}\times\SI{10}{\centi\meter}$ area can be covered with only 2$\times$128 channels, achieving a position resolution of around \SI{60}{\micro\meter}.

For data acquisition (DAQ), the use of VMM and the Scalable Readout System (SRS) \cite{LUPBERGER201891} is being considered.
VMM and SRS are commonly used options for MPGD detector DAQ, with their performance already well demonstrated.
If four $\mu$RWELL or GEM + $\mu$RWELL detectors are used to construct the tracker and capacitive sharing readout is applied, resulting in 2$\times$128 channels per detector, then only a DVMM adapter card will be required to operate eight VMM hybrids.

One concern is the highly constrained space imposed by the magnet.
Due to this limitation, the commonly used readout PCB for $\SI{10}{\centi\meter}\times\SI{10}{\centi\meter}$ detectors cannot be utilized.
Instead, a compact readout PCB must be designed and used, minimizing space for signal routing and services.
Additionally, it is important to find a balance between minimizing potential damage to the VMM hybrids from the expected high neutron flux and the space constraints that limit their optimal placement for radiation protection.

\subsection{The Total Absorption 4D Electromagnetic Calorimeter}
\begin{figure}[htb]
    \centering
    \begin{minipage}{0.37\linewidth}
    \includegraphics[width=\linewidth]{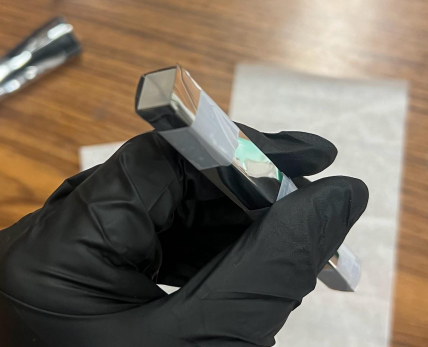}
    \end{minipage}
    \begin{minipage}{0.62\linewidth}
    \includegraphics[width=\linewidth]{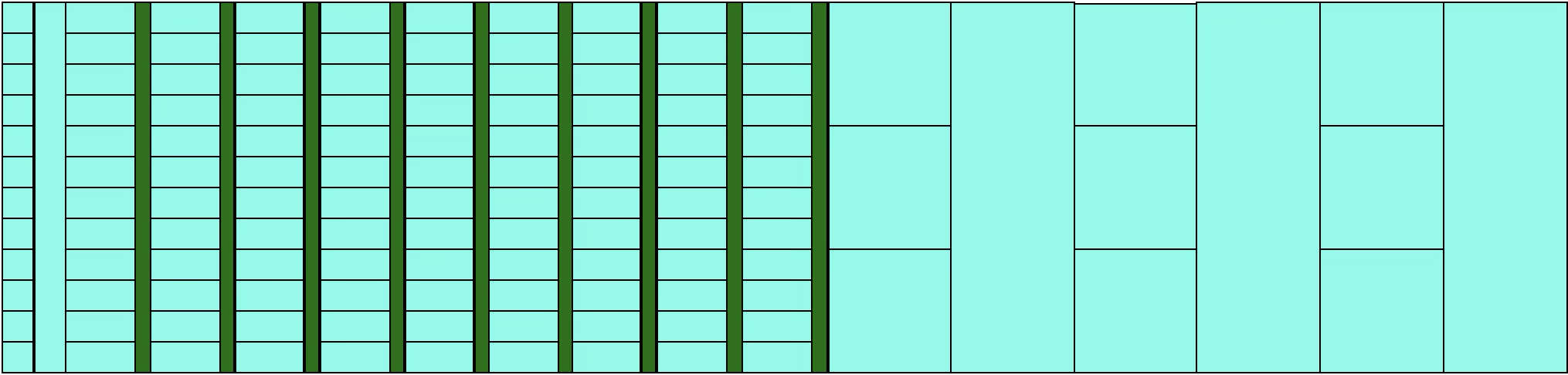}
    \includegraphics[width=\linewidth]{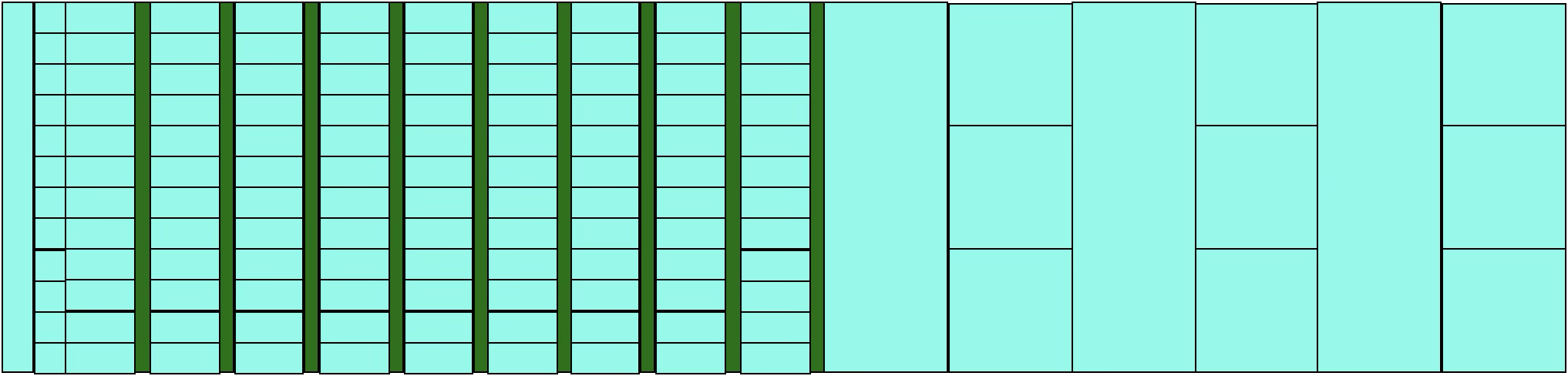}
    \end{minipage}
    \caption{(Left) Photograph of a CsI crystal wrapped in aluminized mylar. (Right) 3D ECal layout. (Top) An x--z side view and (Bottom) y--z side view). The first two layers consist of $1\times1\times12~\rm{cm}^3$ CsI crystal bars stacked along the x and y directions. The next nine layers are cubic \numproduct{12x12} array composed of $1\times1\times2~\rm{cm}^3$ CsI bars. The final six layers are $4\times4\times12~\rm{cm}^3$ CsI bars in the same configuration as the first two layers.}
    \label{fig:ecal_design}
\end{figure}

DAMSA will use an electromagnetic calorimeter (ECal) comprised of an array of undoped CsI scintillating crystals, each viewed from both ends by silicon photomultipliers (SiPMS) with nanosecond timing resolution mounted on a PCB. 
This design is illustrated in Fig.~\ref{fig:ecal_design}.
Each crystal in the CsI array will be \qtyproduct[product-units=single]{1x1x12}{\cubic\cm} and wrapped in reflector to increase light collection efficiency.
Undoped CsI scintillates at \SI{310}{\nm} with a \SI{\sim10}{\ns} time constant at room temperature and is radiation hard.
Aluminized mylar, 3M$^\mathrm{TM}$ Enhanced Specular Reflector (ESR), among other materials, are being considered for the reflector, while wavelength-shifting layers are under consideration to improve the detection efficiency of scintillation light.

The ECal design shown in Fig.~\ref{fig:ecal_design} consists of three types of CsI crystal sections, each optimized for a specific function. 
The first section uses an array of \qtyproduct[product-units=single]{1x1x12} crystal bars to detect pile-up; since most signal photons begin to interact after approximately \SI{2}{\cm}, this section is not critical for signal reconstruction. 
The middle section is designed to measure signal photons with high energy and spatial resolution. The final section is optimized to measure electromagnetic shower tails, providing high energy resolution with reduced spatial resolution.

\subsubsection{Validation tests}
\begin{figure}
    \centering
    \begin{minipage}{0.1\linewidth}
    \includegraphics[width=\linewidth]{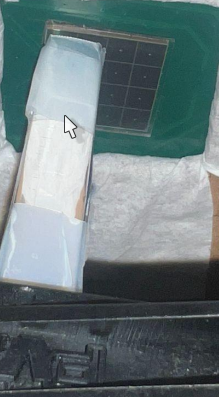}
    \end{minipage}
    \begin{minipage}{0.4\linewidth}
    \includegraphics[width=\linewidth]{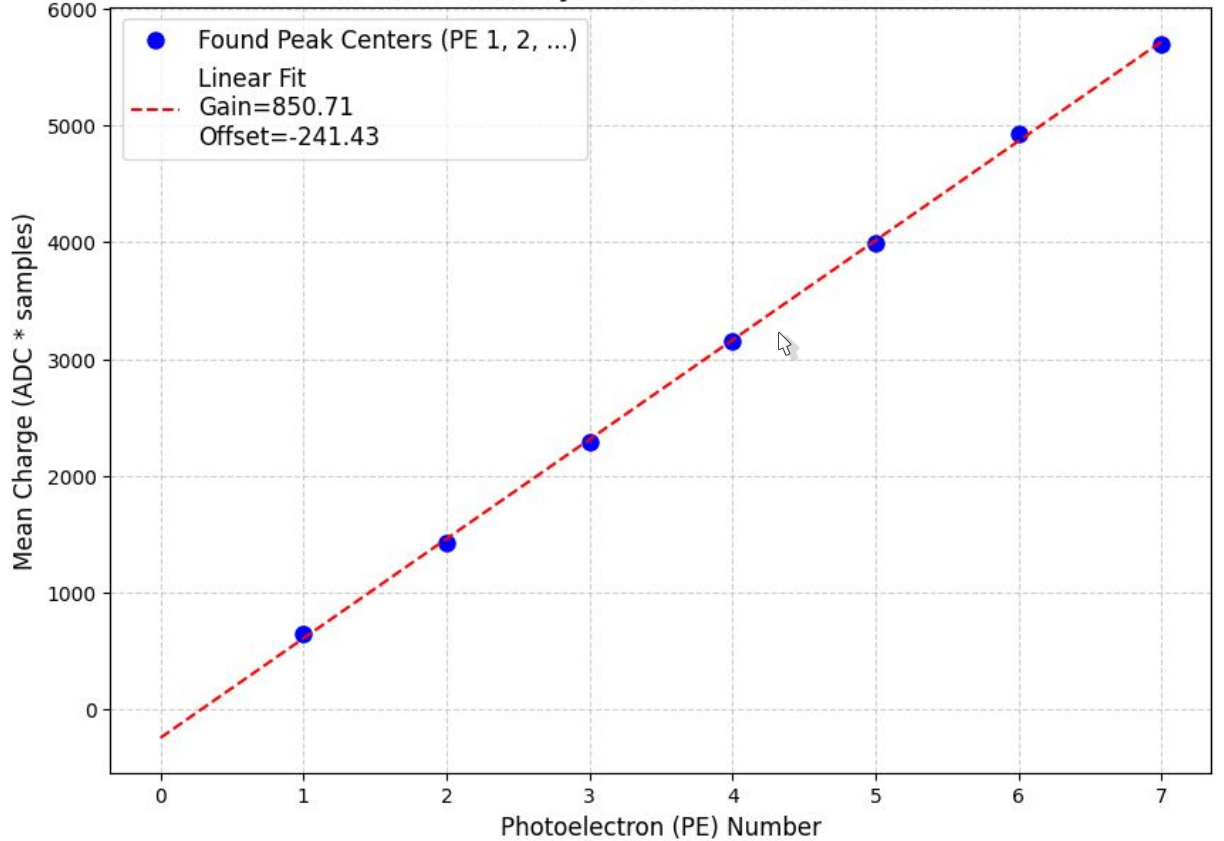}
    \end{minipage}
    \begin{minipage}{0.42\linewidth}
    \includegraphics[width=\linewidth]{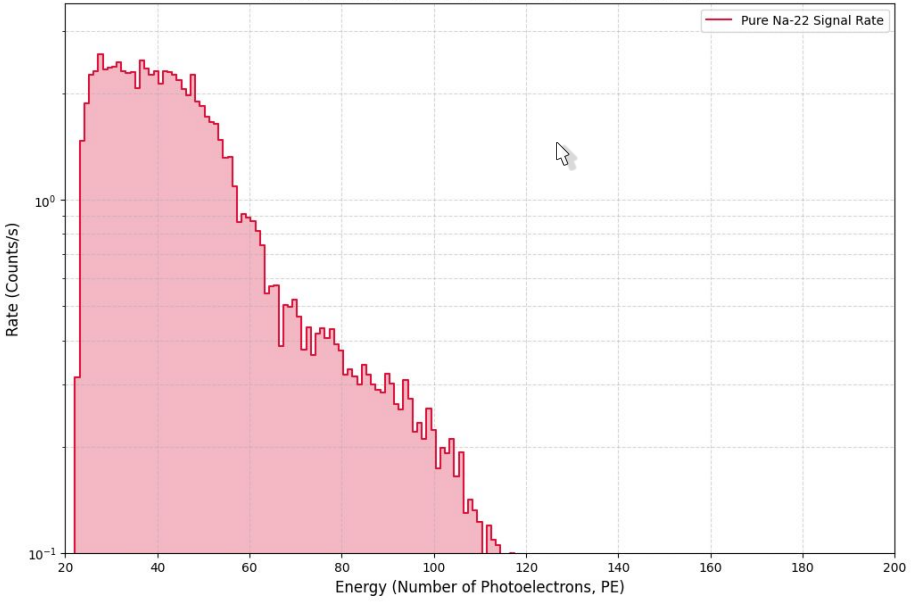}
    \end{minipage}
    \caption{(Left) Illustration of the CsI crystal calibration measurements at UC\,Riverside, using a Hamamatsu S14161 SiPM array coupled to the crystal using a silicone pad. (Middle) Single photoelectron calibration measurement at room temperature. (Right) Measured photoelectron spectrum with a \ce{^22Na} calibration source, after background subtraction, showing Compton spectra with edges at \SI{511}{\keV} and \SI{1.27}{\MeV}.}
    \label{fig:ucr_calibration}
\end{figure}

Ongoing studies at Brookhaven National Laboratory, University of California--Riverside, Kyungpook National University, and Seoul National University are performing prototype tests to optimize and validate the ECal design. These studies aim to test various ECal designs to determine the configuration that gives the best energy and timing resolution and to ensure the spatial uniformity of each crystal.

While tests are still underway, early results using a \ce{^22Na} calibration source, with \SI{511}{\keV} and \SI{1.27}{\MeV} $\gamma$-lines, are shown in Fig.~\ref{fig:ucr_calibration}.
After subtracting backgrounds and correcting for the single photonelectron (PE) gain of the SiPM, the observed spectrum from these $\gamma$-rays Compton scattering in the crystal are estimated to be in the range of \SIrange{100}{150}{PE\per\MeV} with a SiPM only on one side of the crystal. 

Similar studies at Brookhaven National Laboratory used a \ce{^90Sr} $\beta$-source (\SI{546}{\keV} endpoint) at secular equilibrium with a \ce{^90Y} $\beta$-sourcE (\SI{2.278}{\MeV} endpoint) to compare the SiPM performance with visible and VUV-sensitive Hamamatsu SiPMs. 
These studies found a higher light yield with visible SiPMs, consistent with their higher quantum efficiency for \SI{310}{\nano\meter} photons.

\subsubsection{Optical simulations}
\begin{figure}[htb]
    \centering
    \includegraphics[width=0.325\linewidth]{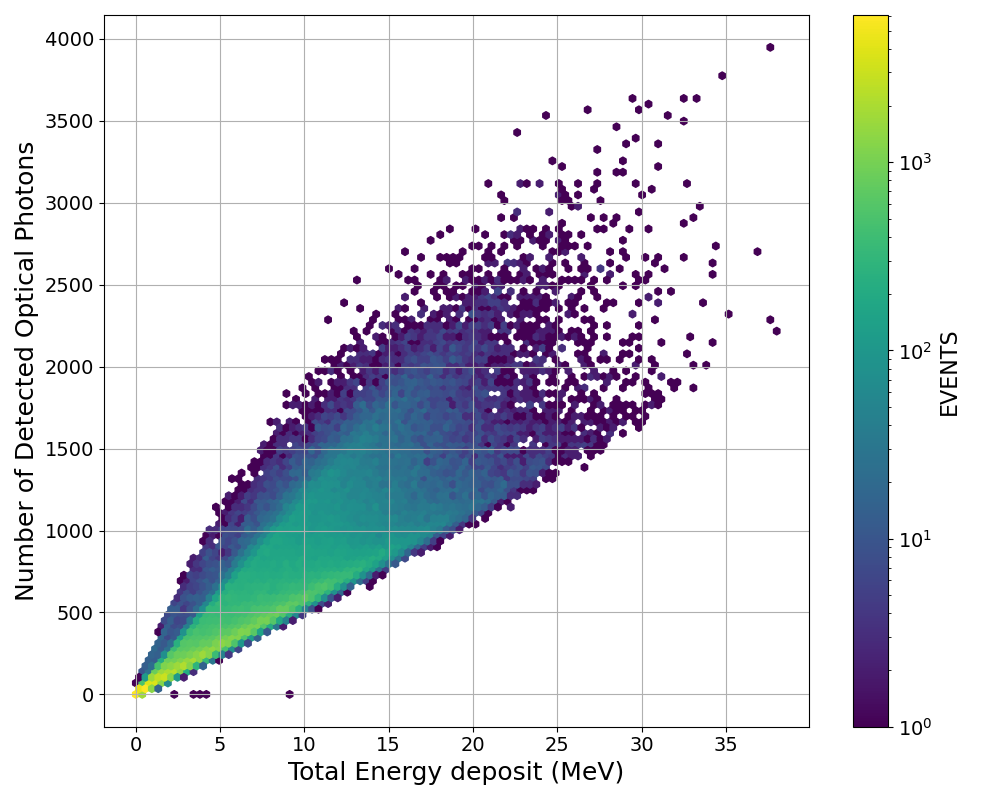}
    \includegraphics[width=0.325\linewidth]{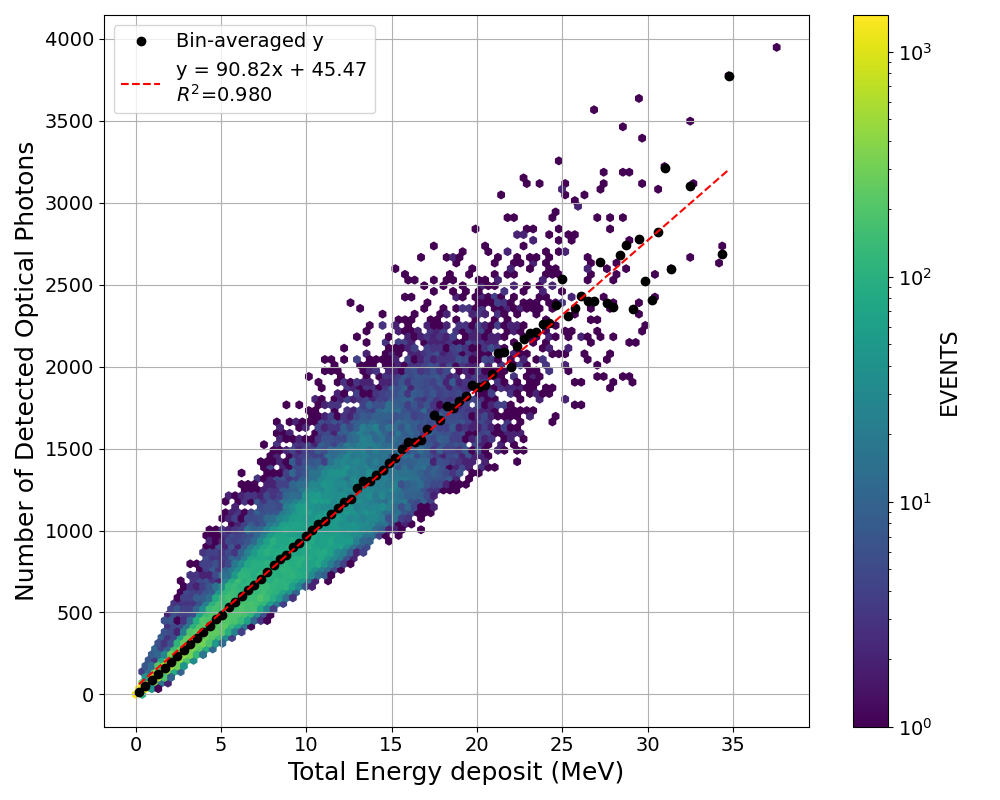}
    \includegraphics[width=0.325\linewidth]{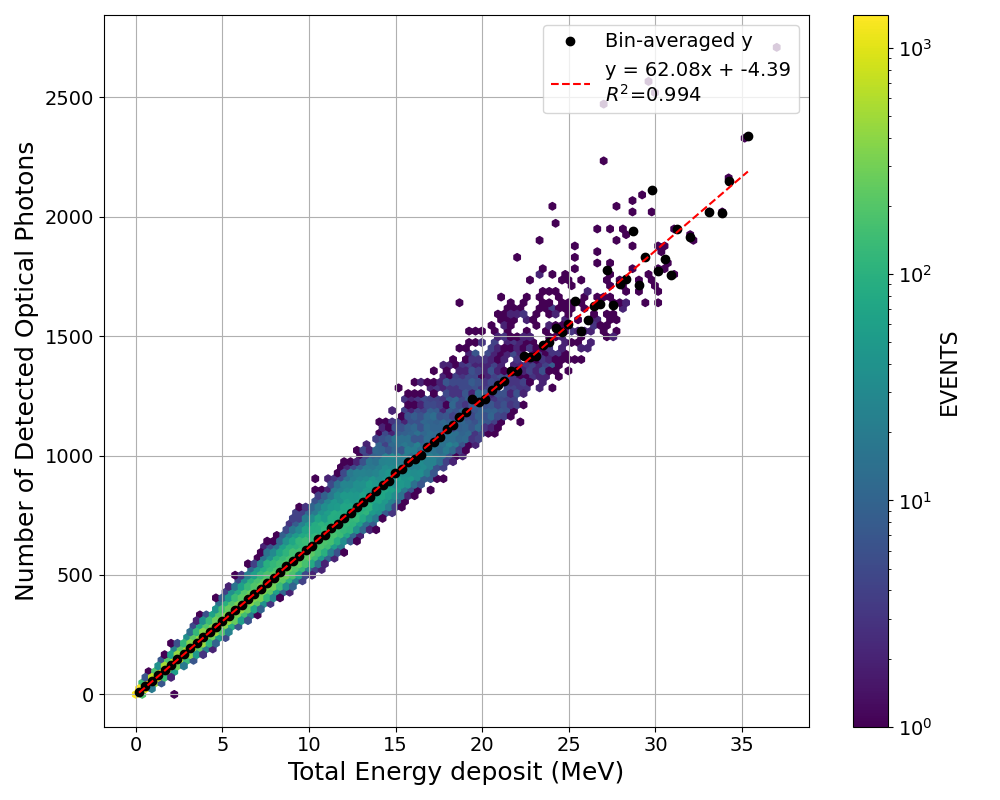}
    \caption{\textsc{Geant4} optical simulations a CsI detector showing the measured scintillation light versus deposited energy for \SIrange{0}{35}{\MeV} depositions in the CsI (Left) for all events, (middle) for energy deposited within \SI{3}{\cm} of the SiPM at either end of the CsI bar, and (right) for energy deposited in the central \SI{6}{\cm}.}
    \label{fig:CsI_Geant4_conf}
\end{figure}

A \textsc{Geant4} simulation was performed to estimate the energy resolution of a single CsI crystal bar, in order to calculate the scintillation light yield by propagating optical photons through the crystal. The result is shown in Fig.~\ref{fig:CsI_Geant4_conf}.
Each plot shows the number of photons detected as a function of the energy deposited in the detector.
The leftmost plot shows this distribution for all simulated energy depositions and has a very broad spread: an event for which \num{15000} photons is detected could be anywhere from \SI{10}{\MeV} to \SI{25}{\MeV}.
This spread is driven by spatial dependence of the light yield: energy deposited near the SiPMs (middle plot of Fig.~\ref{fig:CsI_Geant4_conf}) produce more detected light than energy deposited near the edges (rightmost plot), due to the large number of photons required for photons starting far from the SiPMs.
Better energy resolution is therefore achieved with shorter or wider bars. 
Reconstructing the transverse position may allow corrections to improve the resolution.

\begin{figure}[htb]
    \centering
    \begin{minipage}[c]{0.45\linewidth}
    \includegraphics[width=\linewidth]{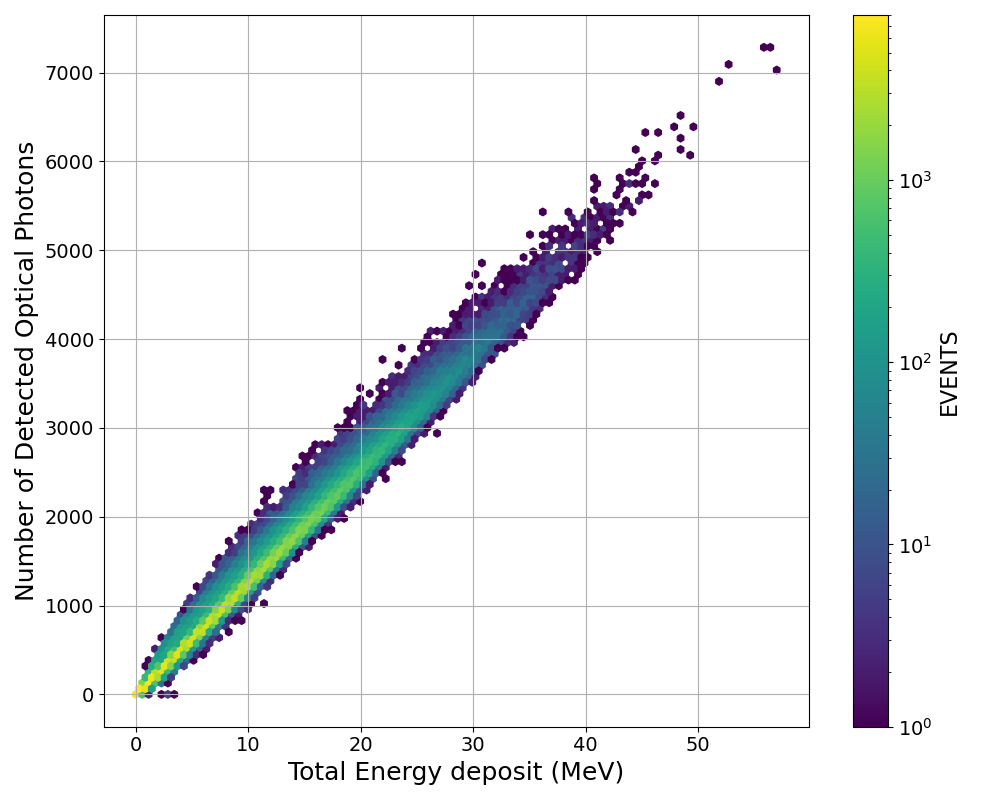}
    \end{minipage}
    \begin{minipage}[c]{0.45\linewidth}
    \includegraphics[width=\linewidth]{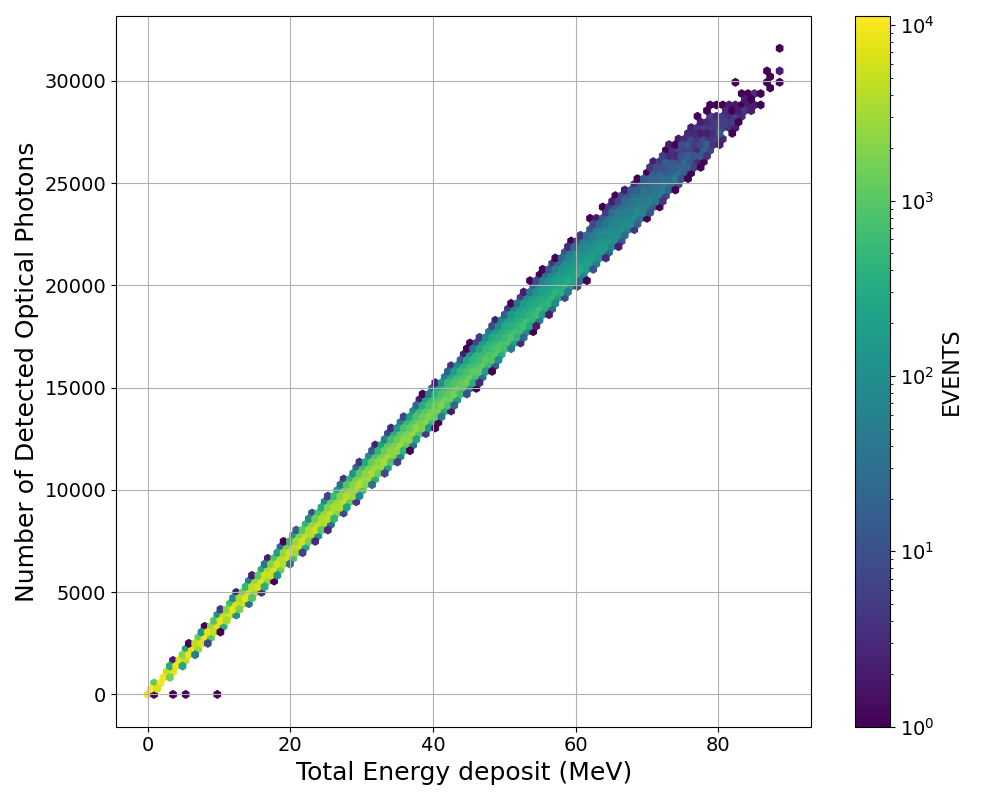}
    \end{minipage}
    \caption{Detection photons as a function of energy deposited in the CsI for (Left) \qtyproduct[product-units=single]{1x1x2}{\cubic\cm} cube-like crystals being considered for an alternative design and (Right) \qtyproduct[product-units=single]{4x4x12}{\cubic\cm} crystals forming the final six layers of the nominal design.}
    \label{fig:cube_like}
\end{figure}

An alternative design using \qtyproduct[product-units=single]{1x1x2}{\cubic\cm} is also being considered.
The improved energy resolution with this design is shown in Fig.~\ref{fig:cube_like} (left).
An ECal composed of such cube-like crystals arranged in a two-dimensional array over multiple layers enhances both the energy resolution and the suppression of ghost hits. In the current design using \qtyproduct[product-units=single]{1x1x12}{\cubic\cm} crystal bars, the bars are stacked along the x- and y-axes, and these two layers are used to reconstruct the hit position. However, when multiple hits occur simultaneously, the fired x- and y-axis bars can produce ghost hits at each of their crossing points.

A thicker \qtyproduct[product-units=single]{4x4x12}{\cubic\cm} crystal improves the energy resolution by reducing internal reflections, as shown in Fig.~\ref{fig:cube_like} (right). Although this geometry degrades the spatial resolution, the 3D ECal employs it for energy measurements of the electromagnetic shower produced by signal photons. High spatial resolution is not required at this stage, since after the incident energetic photons initiate an electromagnetic shower, the transverse spread of secondary particles increases and the total deposited energy becomes the primary observable. Consequently, the \qtyproduct[product-units=single]{4x4x12}{\cubic\cm} crystal bar reduces the number of DAQ channels while improving the energy measurement.

\subsubsection{Radiation hardness}
\begin{figure}[htb]
    \centering
    \includegraphics[width=\linewidth]{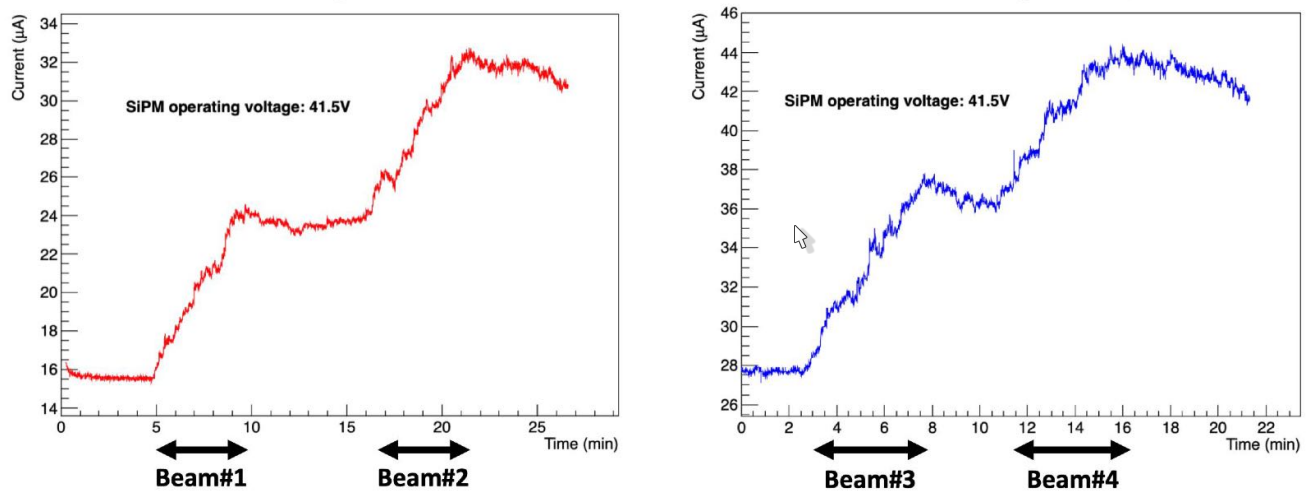}
    \caption{Hamamatsu S14161 SiPM array leakage current with intermittently irradiation by a \SI{<90}{\MeV} neutron beam at a \SI[per-mode=reciprocal]{e4}{\per\square\cm} fluence in a \SI{100}{\micro\second} bunch pulsed at \SI{1}{\hertz} at the KOMAC Neutron Facility.}
    \label{fig:sipm_hardness}
\end{figure}
Figure~\ref{fig:sipm_hardness} shows the results of exposing a Hamamatsu S14161 \numproduct{4x4} SiPM array to a \SI[per-mode=reciprocal]{e4}{\per\square\cm} fluence of \SI{<90}{\MeV} neutrons at the KOMAC Neutron Facility, using \SI{100}{\micro\second} bunches pulsed at \SI{1}{\hertz}.
After successive \SI{5}{\minute} exposures, the dark count rate in the SiPMs was found to increase by approximately \SI{8}{\micro\ampere}.
This leakage current was seen to slowly recover over time.

\subsection{The Data Acquisition System}
There are two components for the DAQ system, one is the digitization and the other is the trigger. As discussed in the trigger, the existing Chicago pipeline system will work for both tasks. Currently, there are two pipeline FADC modules, each with 16 channels, 14-bits dynamic range, and 125 or 500 Mega-Sampling-per-second. Due to both readout and trigger pipelines, there will be no delay cables (for external triggers) for the calorimeter SiPM analog outputs.

The final readout to the computer will be optical-fiber-based 40~G Ethernet. This will allow total flexibility for the data collection computer configuration.

\subsection{The Trigger}

Depending on the final DAQ equipment and setup, for example, if the Chicago pipeline ADC system is used, the trigger will also be pipelined without the need of an external trigger, i.e., various triggers could be constructed internally within the DAQ system. This will be similar to the J-PARC KOTO system. If a commercial FADC system is used, then likely an external trigger system is needed and the associated trigger hardware sophistication will depend on the content. For example, the total energy deposited in the electromagnetic calorimeter is one of the key components of the trigger. Within the Chicago DAQ system, it is performed digitally inline without much issue. However, a different hardware system is needed to perform the analog summation of all calorimeter channels. Similar arguments apply to the veto content of the trigger.

\vspace{0.5em}
\section{Background Mitigation Strategy}~\label{sec:BackgroundMitigation}
As stated in Section~\ref{sec:PhysicsGoals}, DAMSA is a short baseline beam dump experiment that aims to greatly increase the accessible parameter space by minimizing the distance between the target and the detector. In this process, as the distance between the detector and the target decreases, managing the beam-induced background becomes important. In this section, we will describe what we plan on discriminating and mitigating the beam-induced backgrounds through actual test beam measurements and Monte Carlo simulations.

\subsection {Environmental Neutrons}
\label{subsec:neutron-flux}
Since neutrons can be captured by materials surrounding the detector to produce \SIrange{2}{11}{\MeV} \gr\ cascades, they potentially present a significant amount of background to the ALP search. Neutrons are produced by natural processes in the surrounding environment as well as by the beam itself.

\begin{figure}
    \centering
    \includegraphics[width=0.5\linewidth]{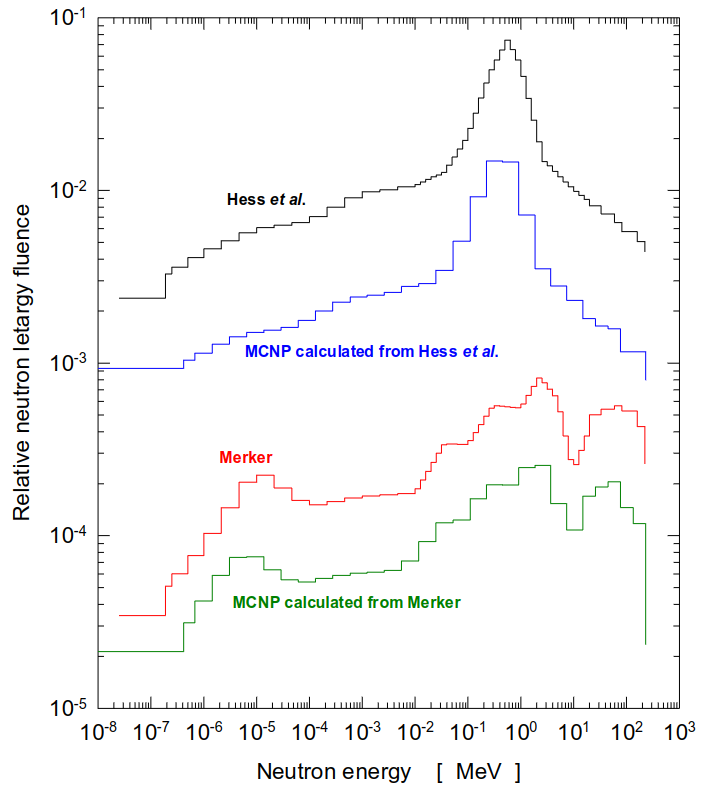}
    \caption{Various ambient neutron fluence calculations at Zacatecas City, from~\cite{cStudyEnvironmentalNeutron2003}}
    \label{fig:ambient_neutrons}
\end{figure}
In above-ground laboratories, the ambient neutron flux is dominated by cosmic rays and the products of their interactions.
Although cosmogenic neutron fluxes reported in the literature vary between different measurements and studies, typical values range from \SIrange{e-4}{e-2}{n\per\square\cm\per\second}, with an energy spectrum spanning from thermal to fast energies, as shown in Fig.~\ref{fig:ambient_neutrons}.
At an altitude of \SI{340}{\meter}, Ref.~\cite{korunMeasurementAmbientNeutron1996} measures a thermal-neutron flux of \SI{1.9\pm0.7e-3}{n\per\square\cm\per\second} and a fast-neutron flux of \SI{16\pm3e-3}{n\per\square\cm\per\second}, consistent with other measurements.
This creates a steady-state background that will need to be mitigated with tight beam-timing cuts.

\subsection {Beam-Related Neutron Flux}
\begin{figure}
    \centering
    \includegraphics[width=0.7\linewidth]{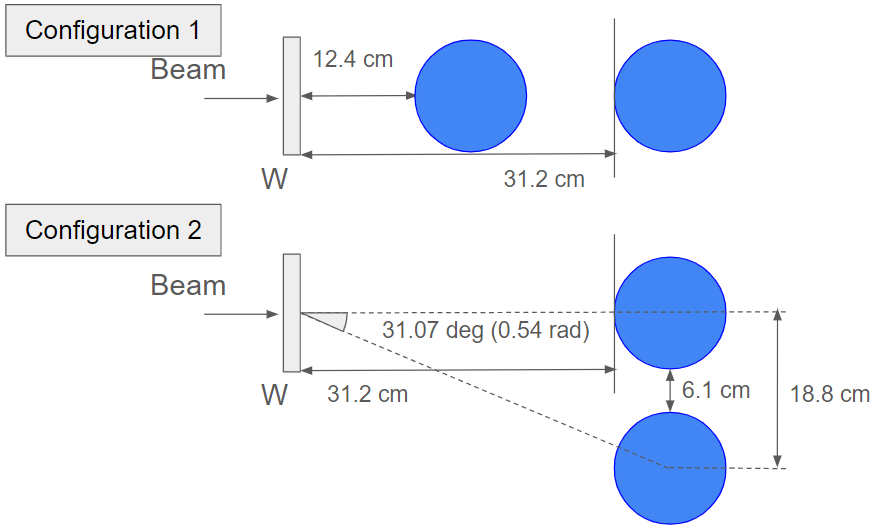}
    \caption{Detector setups for observing variations in background signals based on distance and angle. Detectors are shown in blue, targets in gray, and the beam direction is indicated by arrows. {\it Top} -- A configuration where two detectors are placed at different distances from the target along the beam on-axis to measure the effect of distance. {\it Bottom} -- A configuration where one detector is placed on-axis and the other off-axis to measure the effect of angle.}
    \label{fig:detector-configurations}
\end{figure}

To understand the beam-induced background, we previously measured the neutron and photon production when an electron beam interacts with a tungsten target, using actual tungsten targets and detectors at the Fermilab Test Beam Facility (FTBF). FTBF provides a 120 GeV primary proton beam and secondary mixed particle beams, consisting of pions and electrons with energies as low as 1 GeV. In our previous measurement, similar to the plan proposed for the FAST facility, we used a 2 GeV mixed beam to obtain electron beams. FTBF is also equipped with differential Cherenkov counters for particle identification. We have already conducted one measurement and plan to carry out several mFigore in the coming year. We used two liquid scintillation detectors filled with EJ-301~\cite{Eljen:2021eljen} and multiple tungsten targets, each with a thickness of 4 mm, varying the tungsten thickness to observe how the beam-induced background changes. Additionally, we set up two different detector configurations to compare particle populations on-axis, where the beam is directed, and off-axis. In the on-axis configuration, the two detectors were placed along the beam axis at distances of 12.4 cm and 31.2 cm from the target, respectively. In the off-axis configuration, one detector was positioned 31.2 cm from the target along the beam axis, while the other was located 18.8 cm perpendicular to the beam axis. Figure~\ref{fig:detector-configurations} shows these two different configurations.

\begin{figure}
    \centering
    \includegraphics[width=0.7\linewidth]{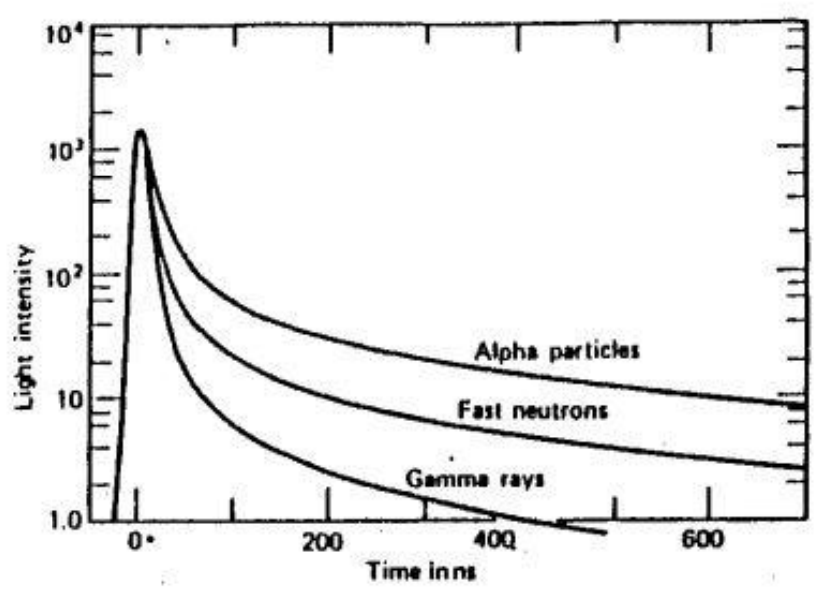}
    \caption{Particle identification in a typical liquid scintillator detector based on the differences in signal falling time that depends on the type of particle. Typically, photons exhibit a shorter falling time than neutrons.}
    \label{fig:liquidscint-particleid}
\end{figure}

Using EJ-301, we can distinguish between photons and neutrons by employing the Pulse Shape Discrimination (PSD) method. This method is widely used for identifying neutrons and photons, with numerous studies supporting its application to EJ-301 liquid scintillators~\cite{DAS2022167405, LANG201726, doi:10.14407/jrpr.2019.44.2.53}. The PSD technique relies on differences in the fall times of pulse signals generated by photons and neutrons. Typically, neutron signals have a longer fall time compared to photon signals, resulting in a longer tail in the pulse. Figure~\ref{fig:liquidscint-particleid} shows the typical pulse shape differences of photons and neutrons.

By applying the PSD method to real beam data, we will compare the results with \textsc{GEANT4} simulations to validate the accuracy of the simulation. This validation will then be used to inform the experimental design. See Sections~\ref{subsec:300-MeV-geant4-simulation} and \ref{sec:2-GeV-electron-beam-geant4-simulation} for the details of the simulations.

\subsection {Electron and Photon Flux Out of Target}
ALPs can be generated through Primakoff reactions from energetic photons, making it crucial to understand the characteristics of photons produced in the interaction between the accelerator's beam and the target material. As discussed in Section~\ref{subsec:neutron-flux}, the PSD method enables the identification of neutrons and photons entering the liquid scintillator detector. We plan to use this method to compare the number of photons generated by accelerator beams of known energy interacting with the target material to simulated results, which will assist in planning the experiments.

\subsection {Mitigation Strategy}
\begin{figure}
    \centering
    \includegraphics[width=0.7\linewidth]{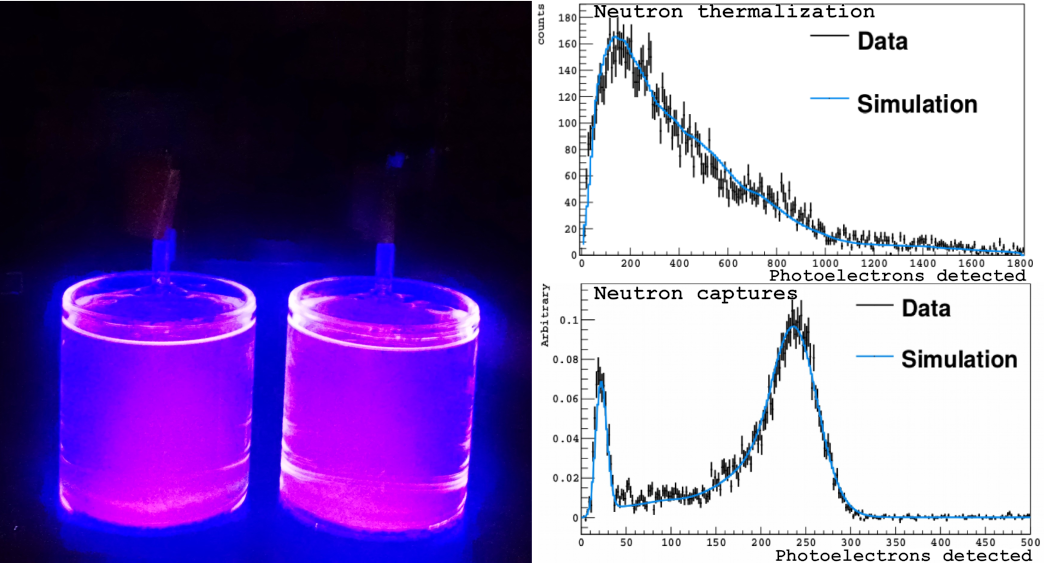}
    \caption{{\it Left} -- Neutron detectors, similar to those for this project, exposed to UV light. {\it Right} -- Prompt thermalization and delayed capture signals of calibration source neutrons in DarkSide-50's neutron veto.}
    \label{fig:neutron-detectors}
\end{figure}
LDPF will develop techniques for modeling and mitigating neutron backgrounds, using techniques the University of California, Riverside (UCR) PI developed for dark matter direct detection. 
We aim to (1) design  techniques for characterizing and modelling environmental neutrons and BRNs; (2) define shielding and coincidence timing requirements to mitigate neutrons.
Doing so requires measuring the rate and energies of thermal, fast, and high-energy neutrons.

The neutron detectors for this project, like those in Fig.~\ref{fig:neutron-detectors}, are based on the borated liquid scintillator neutron veto used for the neutron veto of the DarkSide-50 dark matter detector, which the UCR PI designed and lead the analysis of~\cite{westerdalePrototypeNeutronVeto2016,agnesVetoSystemDarkSide502016}.
Commercial detectors typically choose between energy resolution and efficiency~\cite{mukhopadhyayReviewDirectNeutron2022}.
Liquid scintillator detectors produce signals proportional to the energy lost by neutrons but typically have low efficiency, especially for low-energy neutrons. 
On the other hand, \ce{^3He} detectors efficiently count thermal neutron captures, but are insensitive to neutron energies.
Borated liquid scintillator detectors capture the best of both techniques: the hydrogenous scintillator quickly and efficiently thermalizes neutrons, producing a prompt signal proportional to energy, and thermal neutrons capture on \ce{^10B}, via,
\begin{gather}
    \begin{aligned}
        n+\ce{^10B} \rightarrow \begin{cases}
            \ce{^7Li}\text{ (\SI{1015}{\keV})} + \alpha\text{ (\SI{1775}{\keV})} & (\SI{6.4}{\percent}) \\
            \ce{^7Li^*} + \alpha\text{ (\SI{1471}{keV})}, & (\SI{93.6}{\percent}) \\
            \qquad \ce{^7Li^*} \rightarrow \ce{^7Li}\text{ (\SI{839}{\keV})} + \gamma\text{ (\SI{478}{\keV})}
        \end{cases}
    \end{aligned}
\end{gather}
PI Westerdale found that the $\alpha$ and \ce{^7Li} are quenched to \SI{30}{~\keVee} (\si{\keV} electron equivalent)~\cite{westerdaleQuenchingMeasurementsModeling2017}. 
Since they cannot escape the detectors, sensitivity to them allows for high tagging efficiency, regardless of the energy a neutron deposited.
While, nuclear recoils in the thermalization signal are quenched to \SI{10}{\percent}, Westerdale developed a quenching model able to reconstruct their energies~\cite{westerdaleQuenchingMeasurementsModeling2017}, as demonstrated by Fig.~\ref{fig:neutron-detectors} (top right), which shows accurate reproduction of the neutron energy spectrum in DarkSide-50's neutron veto, using an \ce{^241Am}-\ce{^13C} \alphan\ calibration source.
As such, these detectors can achieve both high efficiency and energy resolution, with $\sim1$\,ns timing resolution for fast and high-energy neutron thermalization signals.
Figure~\ref{fig:neutron-detectors} shows these signals in the DarkSide-50 neutron veto, which achieved $>99.5\%$ tagging efficiency.

These detectors will use equal volumes of pseudocumene (PC: the primary scintillator) and trimethyl borate (TMB: the boron-loading agent), with \SI{3}{\gram\per\liter} of 2,5-Diphenyloxazole (PPO) and \SI{25}{\milli\gram\per\liter} 1,4-Bis(2-methylstyryl)benzene (bis-MSB) as primary and secondary wavelength shifters, to enhance the light yield and convert UV PC sicntillation light to the visible spectrum, where photons will be efficiently detected by arrays of silicon photomultipliers (SiPM) coupled to the detectors' face.
Based on prior experience, we expect them to have a light yield of \SI{2}{\pe\per\keV}, high enough to efficiently measure neutron captures.

Thermal neutrons will produce isolated capture signals that will be detected with \SI{>99}{\percent} efficiency.
About \SI{50}{\percent} of ambient fast neutrons will thermalize in the detector, creating a signal proportional to the neutron's energy, while the capture signal, which we will detect down to low energies by triggering on the capture signal that follows by \SI{2.2}{~\micro\second}.
High-energy neutrons are less likely to thermalize in the detector and will instead produce nuclear recoils without subsequent capture signals; their spectrum can be unfolded from the nuclear recoil energy spectrum.
Since PC scintillates with nanosecond timing, fast and high-energy BRNs can be measured in coincidence with the beam to infer the timing structure of neutrons produced.

We request funding to build an array of \num{8} detectors, which will be placed around the lab and near the target detector and beam, to measure  neutrons.
These measurements will be benchmarked against simulations, using the \alphan~\cite{westerdaleRadiogenicNeutronYield2017,gromovCalculationNeutronGamma2023a} and \ngamma~\cite{weimerG4CASCADEDatadrivenImplementation2024} modeling tools developed by the UCR PI and building upon the experience of the UCR PI~\cite{cano-ottWhitePaperAlpha2024}.
These comparisons will validate our technique for characterizing background neutrons and define shielding and timing requirements needed to reduce neutron backgrounds.
In year~1, the UCR PI and graduate student will build the detectors and characterize them at UCR, establishing the full analysis chain, and in years~2 and~3 these detectors will be transferred to the FAST facility for tests with LDPF.
The equipment cost budgeted for the UCR PI covers the cost of instrumenting the borated liquid scintillator detectors with Hamamstu SiPM arrays needed to provide adequate SiPM coverage for each detector to achieve the target light yield.

\subsection{300 MeV Electron Beam GEANT4 Simulation}
\label{subsec:300-MeV-geant4-simulation}
The interaction of the 300 MeV electron beam with the 10 cm tungsten target was simulated using \textsc{GEANT4} to estimate the production of background and signal sources at the target. Electronuclear reactions assume the photon approximation, where real $\gamma$'s are generated from virtual ones at the electromagnetic vertex. This cross section is valid for incident electrons and positrons across all energy ranges. The background estimation considers all sources of electromagnetic interactions within the electromagnetic calorimeter detector volume, with particular emphasis on neutron-induced background.

\begin{figure}[t]
\centering
\includegraphics[width=0.49\linewidth]{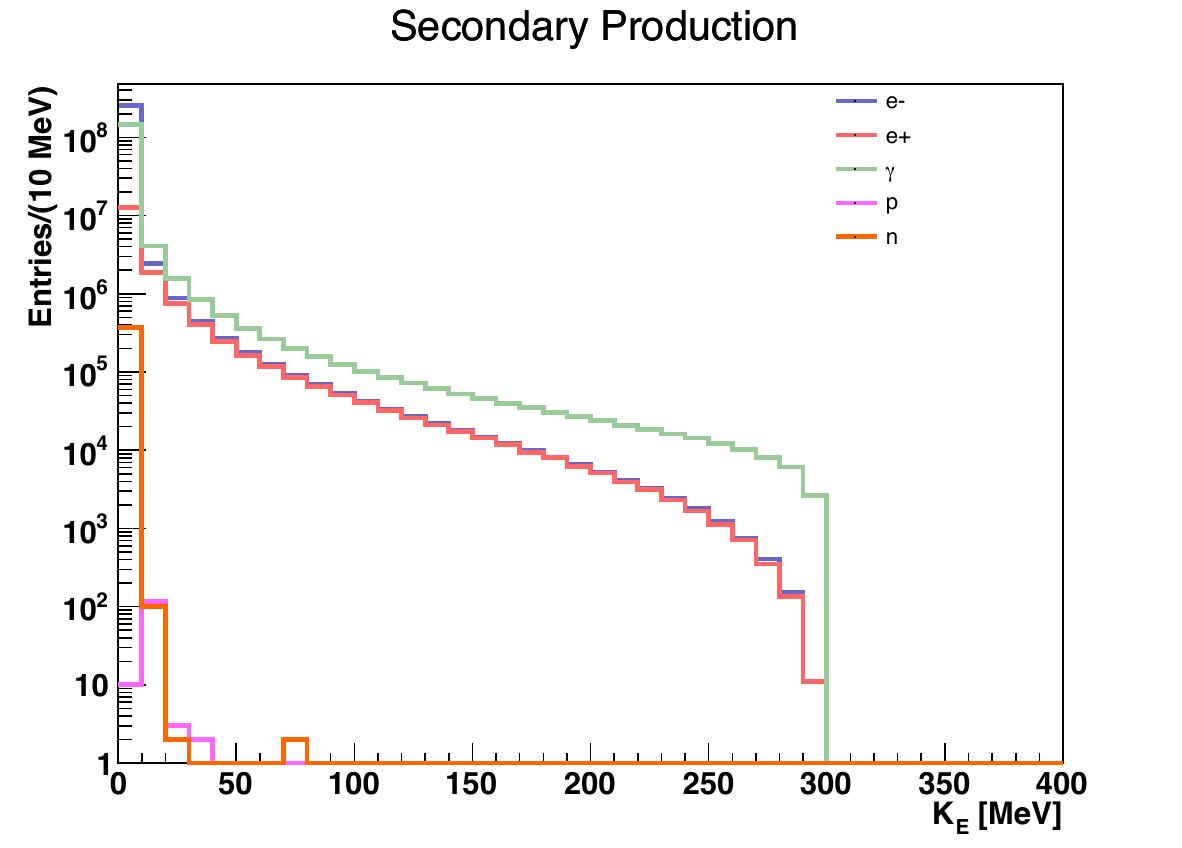}
\includegraphics[width=0.49\linewidth]{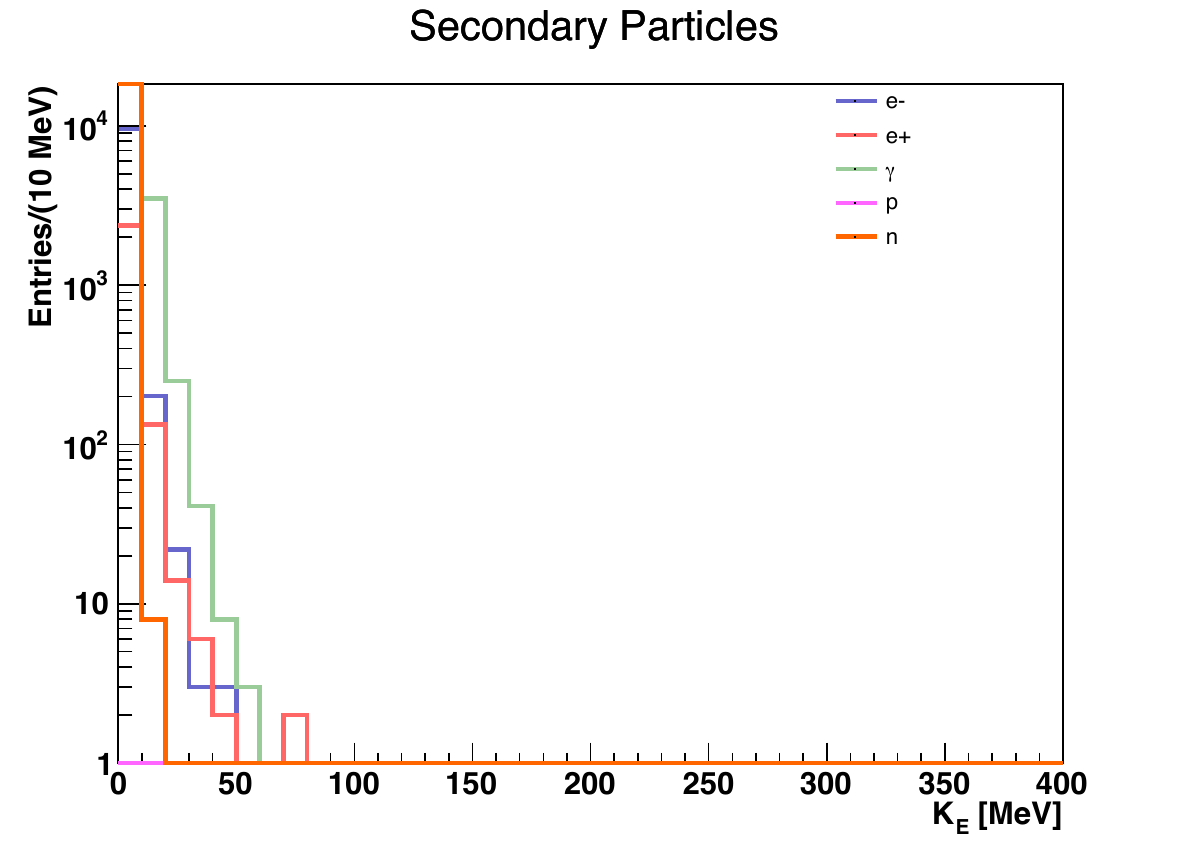}
\caption{Kinetic energy spectra of secondary particles, including electrons, positrons, photons, protons, and neutrons, within the target volume (left) and those escaping the target volume and potentially contributing to backgrounds (right), based on one million generated events.
}
\label{fig:prod}
\end{figure}

The \textsc{GEANT4} simulation uses a simplified geometry but considers all important features of background estimation, properly calculating all secondary particles and their interactions with the experimental structures. The left panel of Fig.~\ref{fig:prod} displays all particles produced inside the target volume, except for the tungsten isotopes. 

The background estimation is based on the energy deposited in the detector volume, including the secondary productions between a vacuum chamber and concrete walls. Neutrons with below 50 MeV may not create signal-like photons, but their background energy deposition in the detector volume degrades the performance of the detector. The right panel of Fig.~\ref{fig:prod} shows the energy distributions of all secondary particles that escaped the target volume. 

\begin{figure}[t]
\centering
\includegraphics[width=0.325\linewidth]{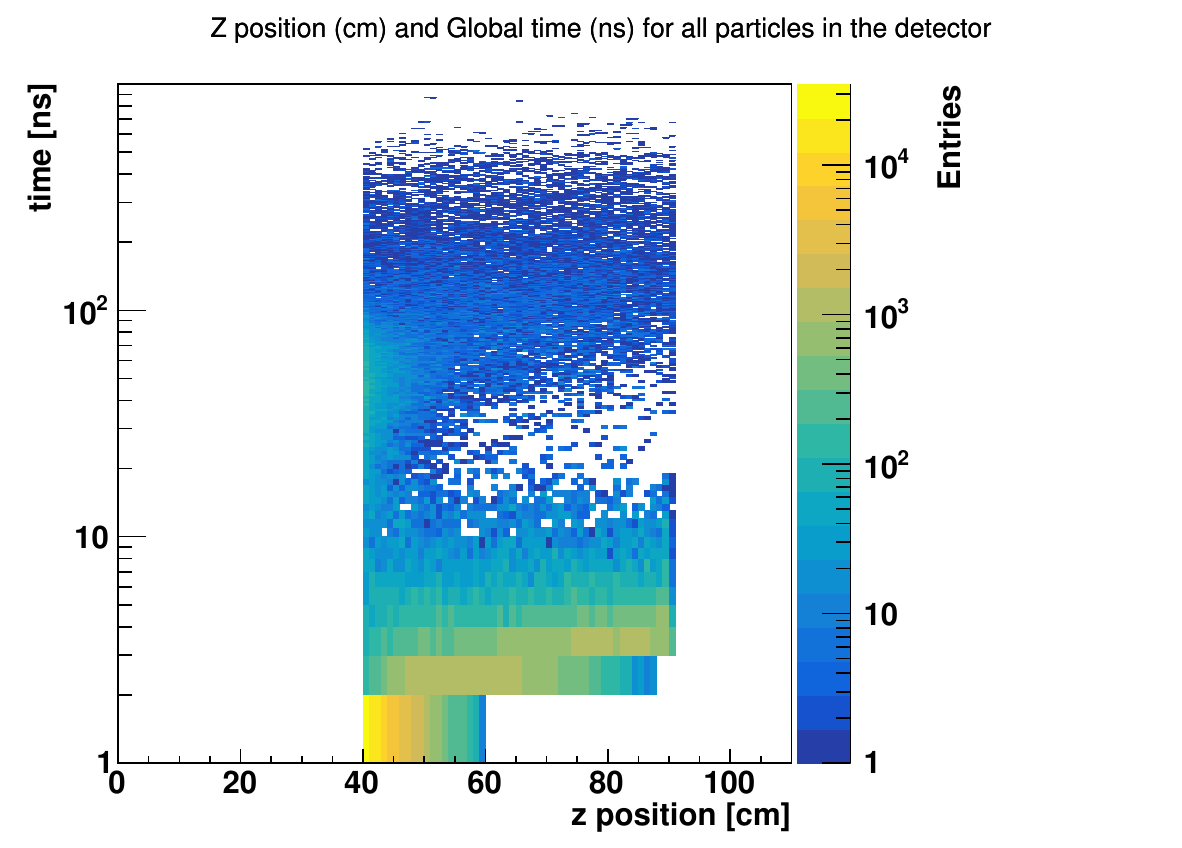}
\includegraphics[width=0.325\linewidth]{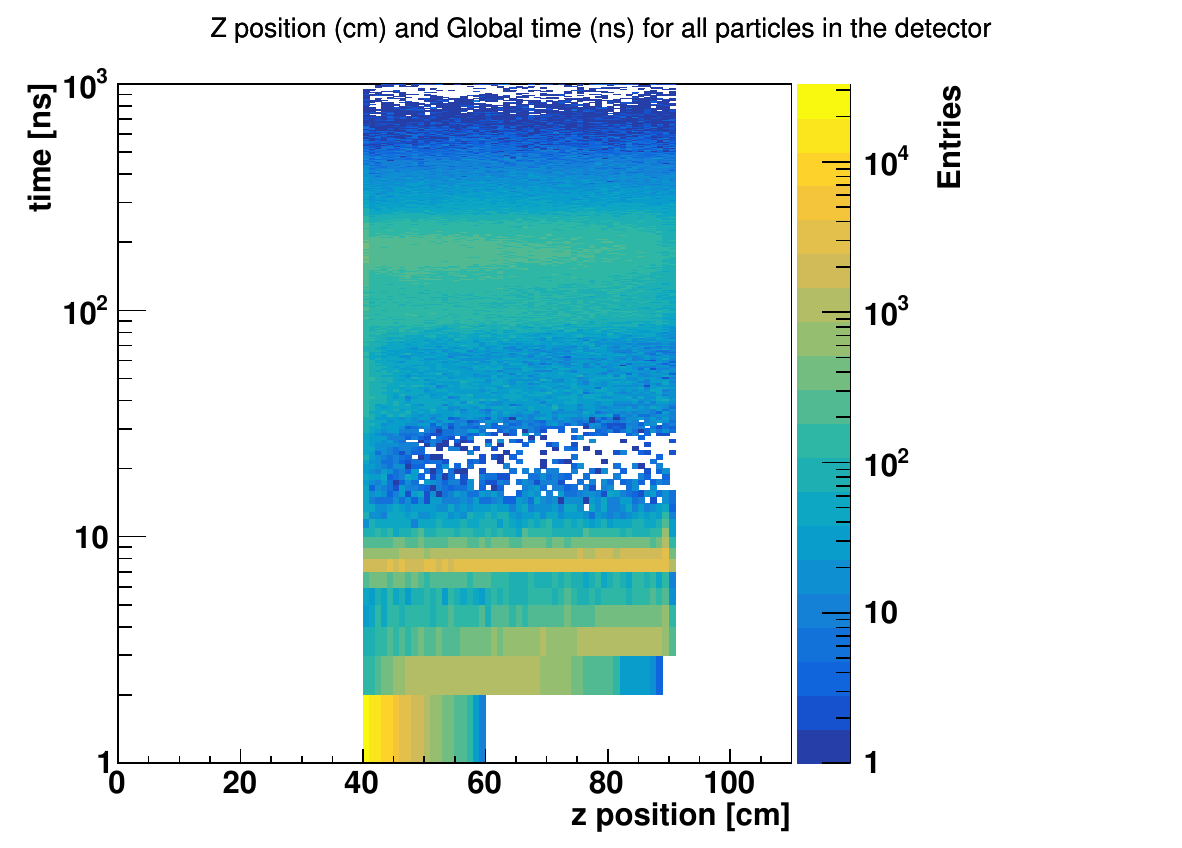}
\includegraphics[width=0.325\linewidth]{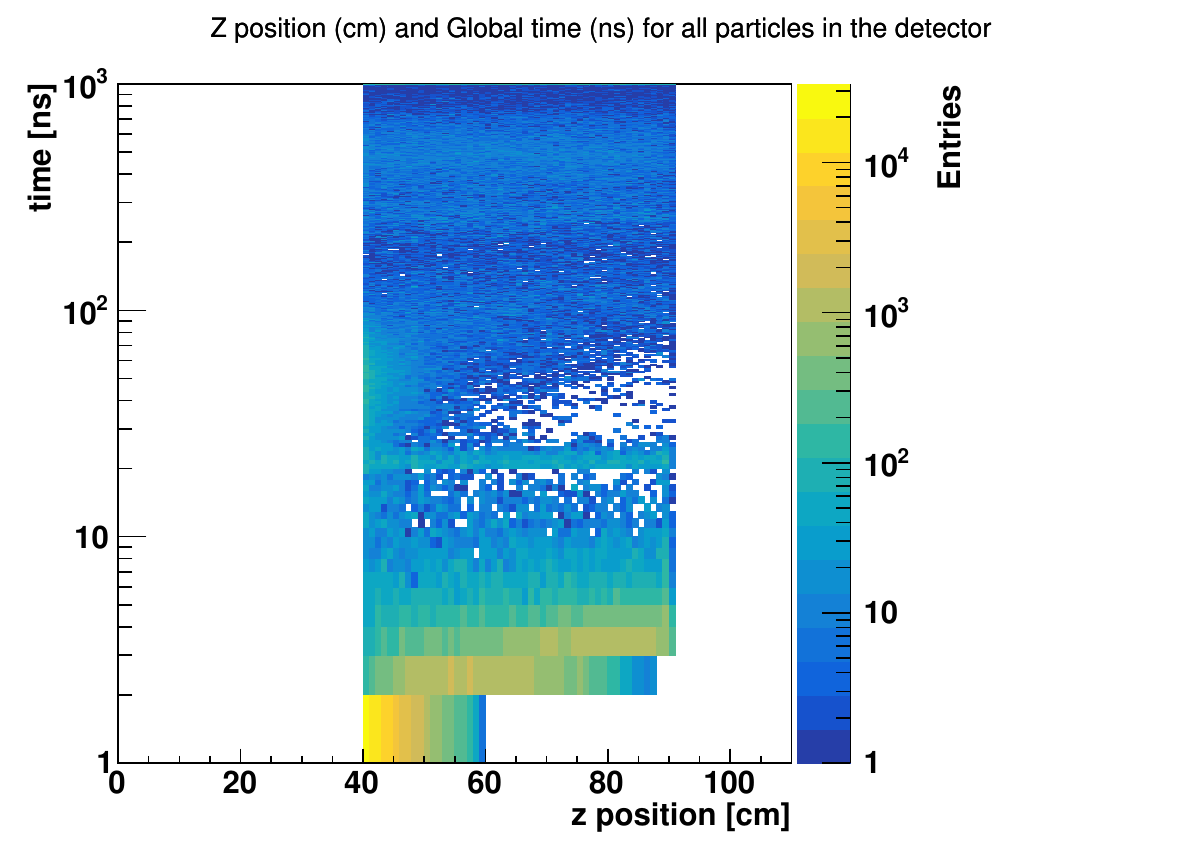}
\includegraphics[width=0.325\linewidth]{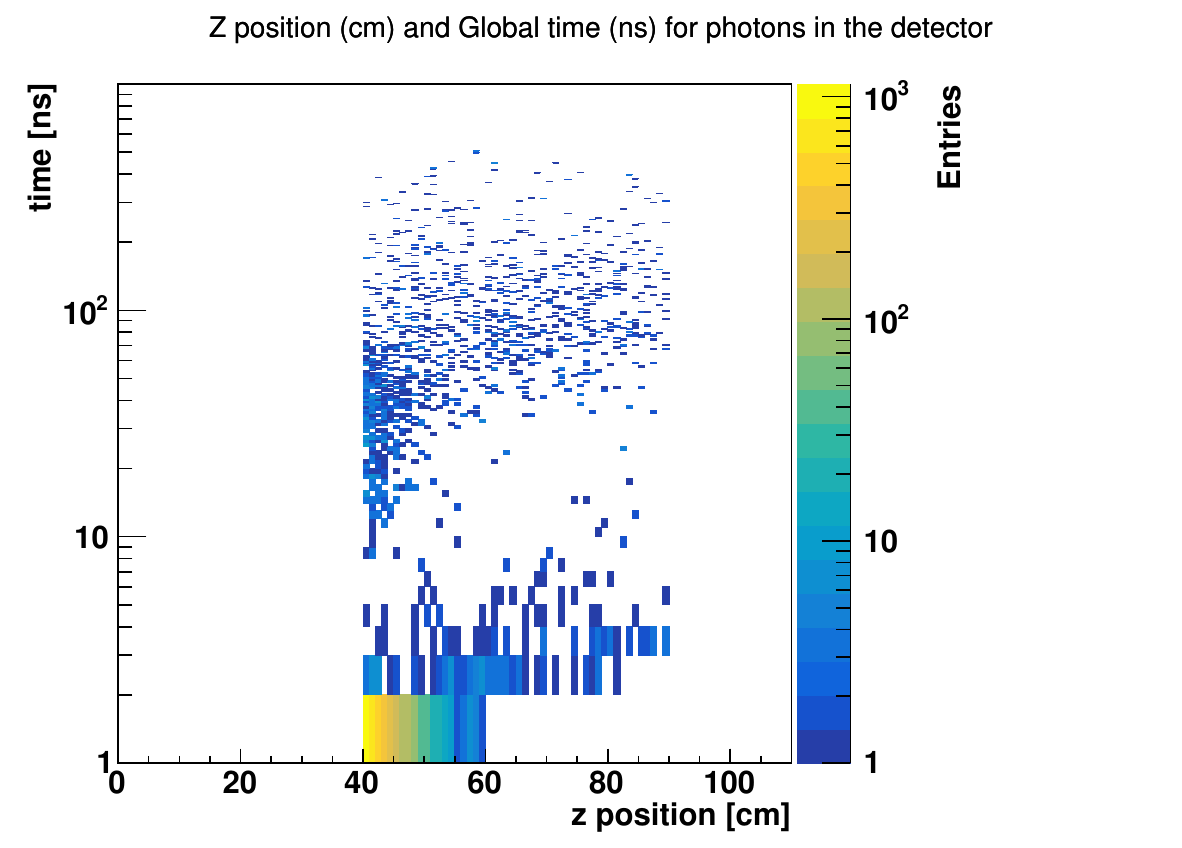}
\includegraphics[width=0.325\linewidth]{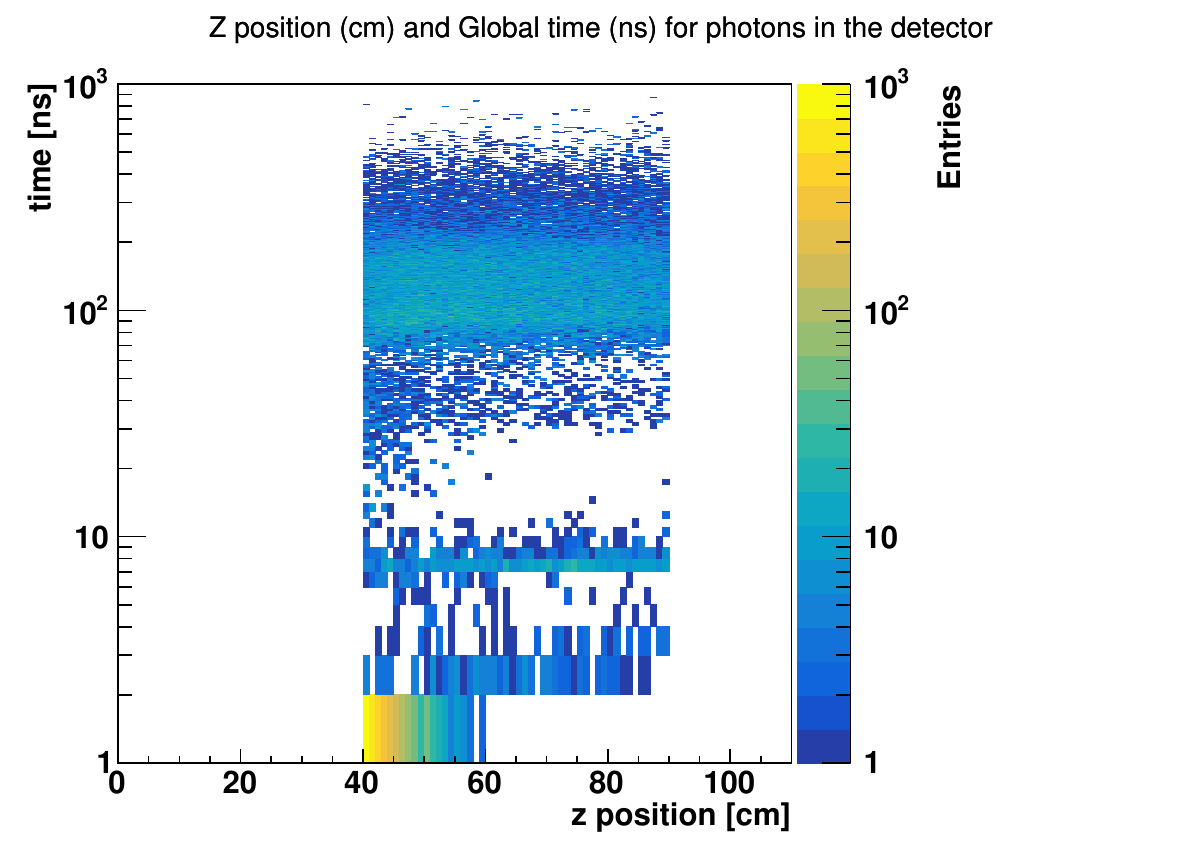}
\includegraphics[width=0.325\linewidth]{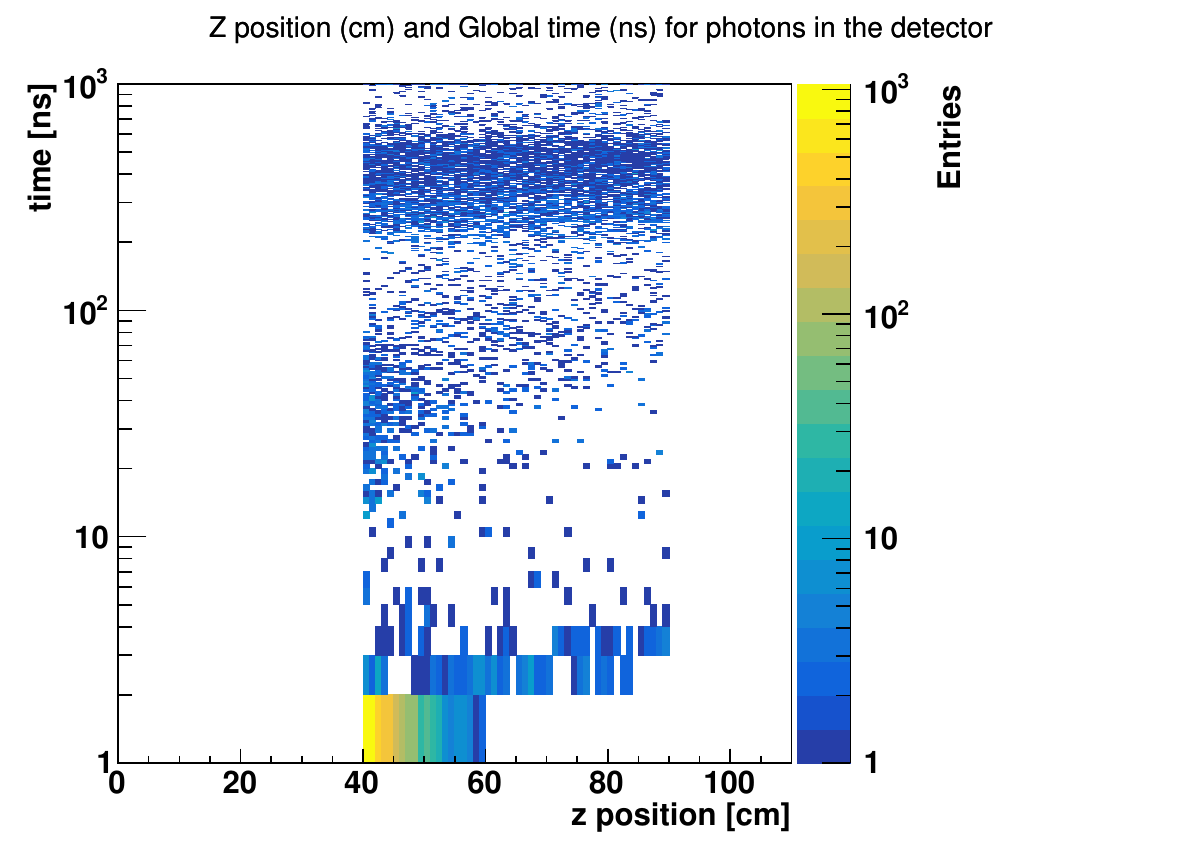}
\caption{All interactions of particles (top) and photons (bottom) escaping from the target and potentially contributing to the backgrounds at the detector are shown in the $z$-position-vs-time plane. Without a concrete wall (left panels), neutron interactions with the wall are not included. However, the 1 m (middle panels) and 2 m (right panels) thick walls show neutron interactions with the concrete wall, clearly illustrating the time difference between the two wall thicknesses.}
\label{fig:allWall}
\end{figure}

To accurately estimate the background, the \textsc{GEANT4} simulation accounts for the time evolution of particles, particularly focusing on the time delay of neutron background interactions with the concrete walls. As a result, the particles (possibly behaving like prompt backgrounds) escaping from the target arrive at the detector within 10 ns, while neutron-induced backgrounds arrive several hundred nanoseconds later. However, most background reaches the detector within 1~ms. Therefore, if the beam separation exceeds 1 ms, there is no significant overlap of backgrounds between beam bunches. The top panel of Fig.~\ref{fig:allWall} displays the timing of interactions between background particles and the detector for configurations without a concrete wall, with a 1~m wall, and a 3~m wall. Additionally, the bottom panel of Fig.~\ref{fig:allWall} highlights the interactions of background photons, both prompt and neutron-induced, which are crucial for background estimation in the detector. The \textsc{GEANT4} simulation does not include thermal neutrons, and the simulation threshold is set to the default value in \textsc{GEANT4}.

\subsection{Stage-0 Measurement at FTBF with 2 GeV Electron Beam and GEANT4 Simulation}
\label{sec:2-GeV-electron-beam-geant4-simulation}
\begin{figure}[htb]
\centering
\includegraphics[width=0.45\linewidth]{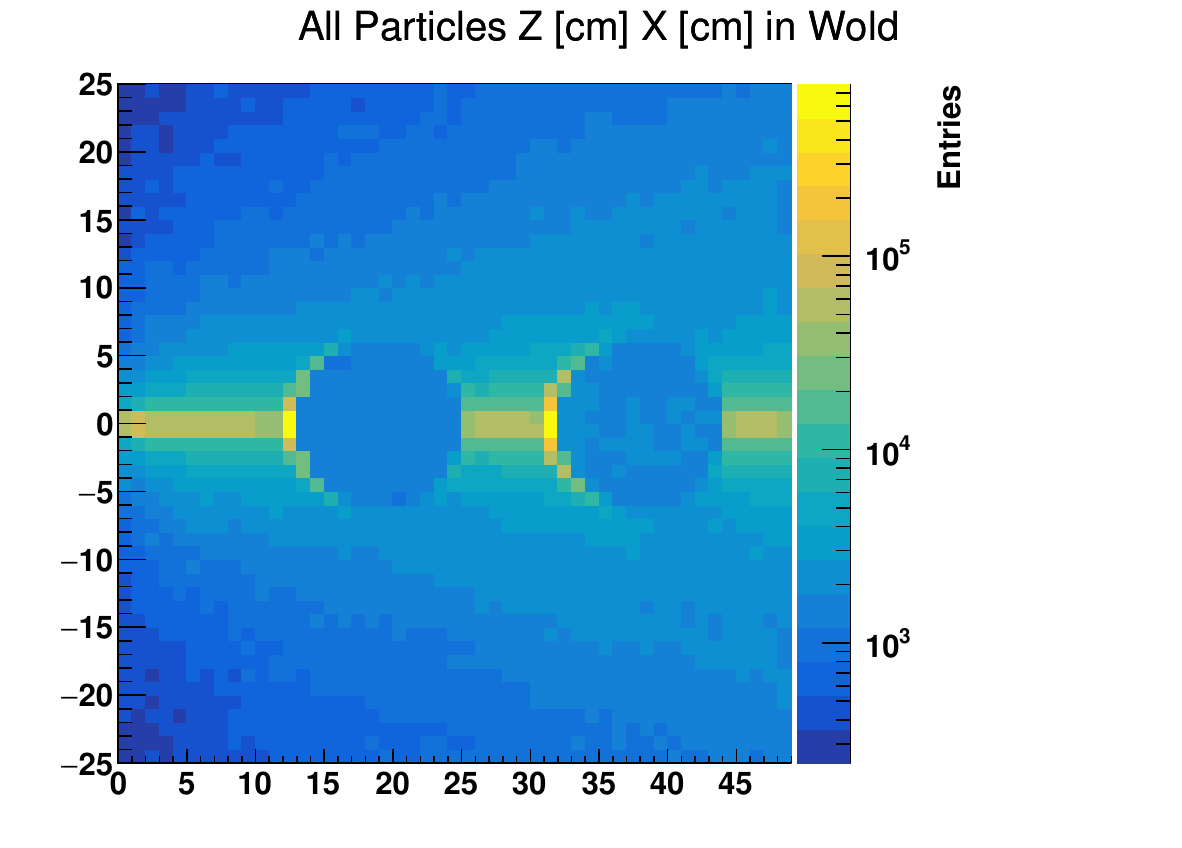}
\includegraphics[width=0.45\linewidth]{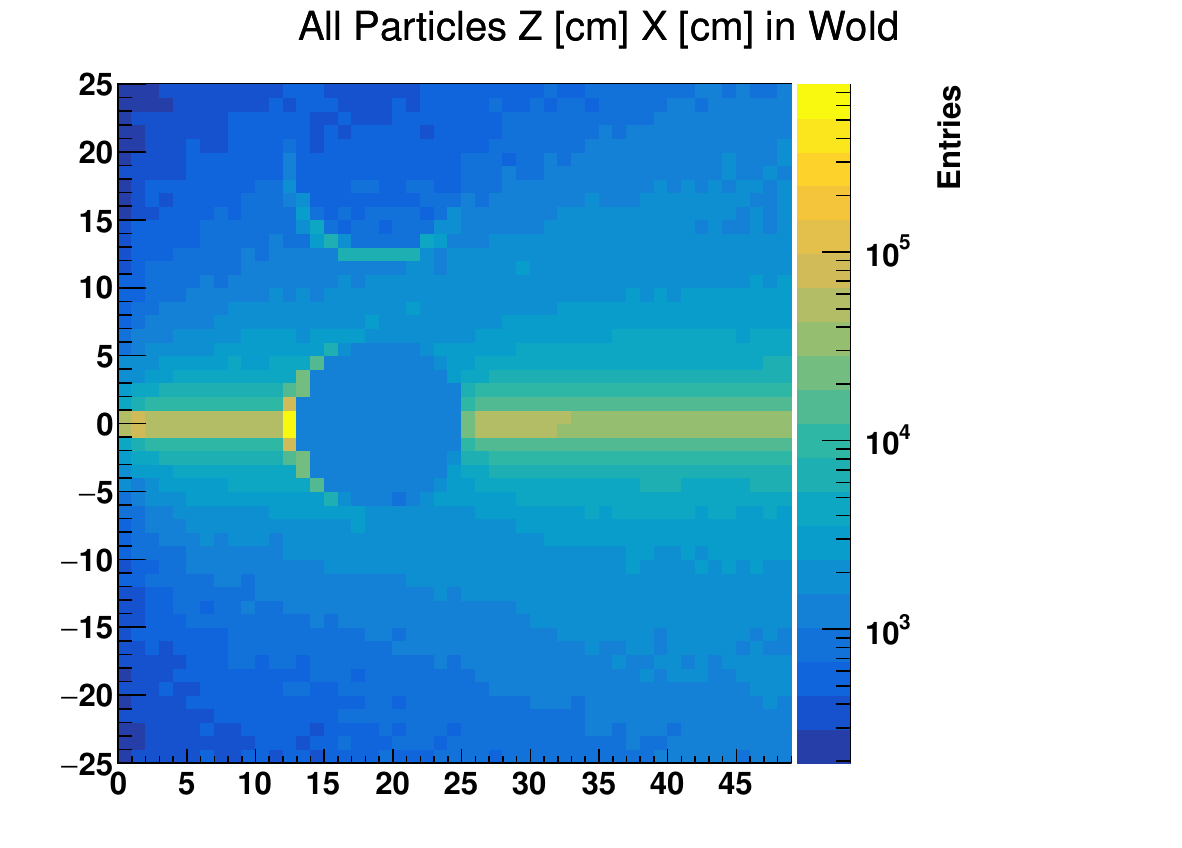}
\caption{The \textsc{GEANT4} simulation results of geometry configurations 1 (left) and 2 (right).} 
\label{fig:geoConf}
\end{figure}

To study the BRN background, we have conducted an experiment at a 2 GeV electron beam facility, irradiating a tungsten target. The experiment involved varying the target length and using two different geometric configurations for the liquid scintillator detectors. \textsc{GEANT4} simulations were performed to evaluate the efficiency of the liquid scintillator trigger, as well as the energy deposition and distribution of particles within the scintillator. Figure~\ref{fig:geoConf} shows the geometry configurations adopted in our \textsc{GEANT4} simulations.

\begin{figure}[htb]
\centering
\includegraphics[width=0.48\linewidth]{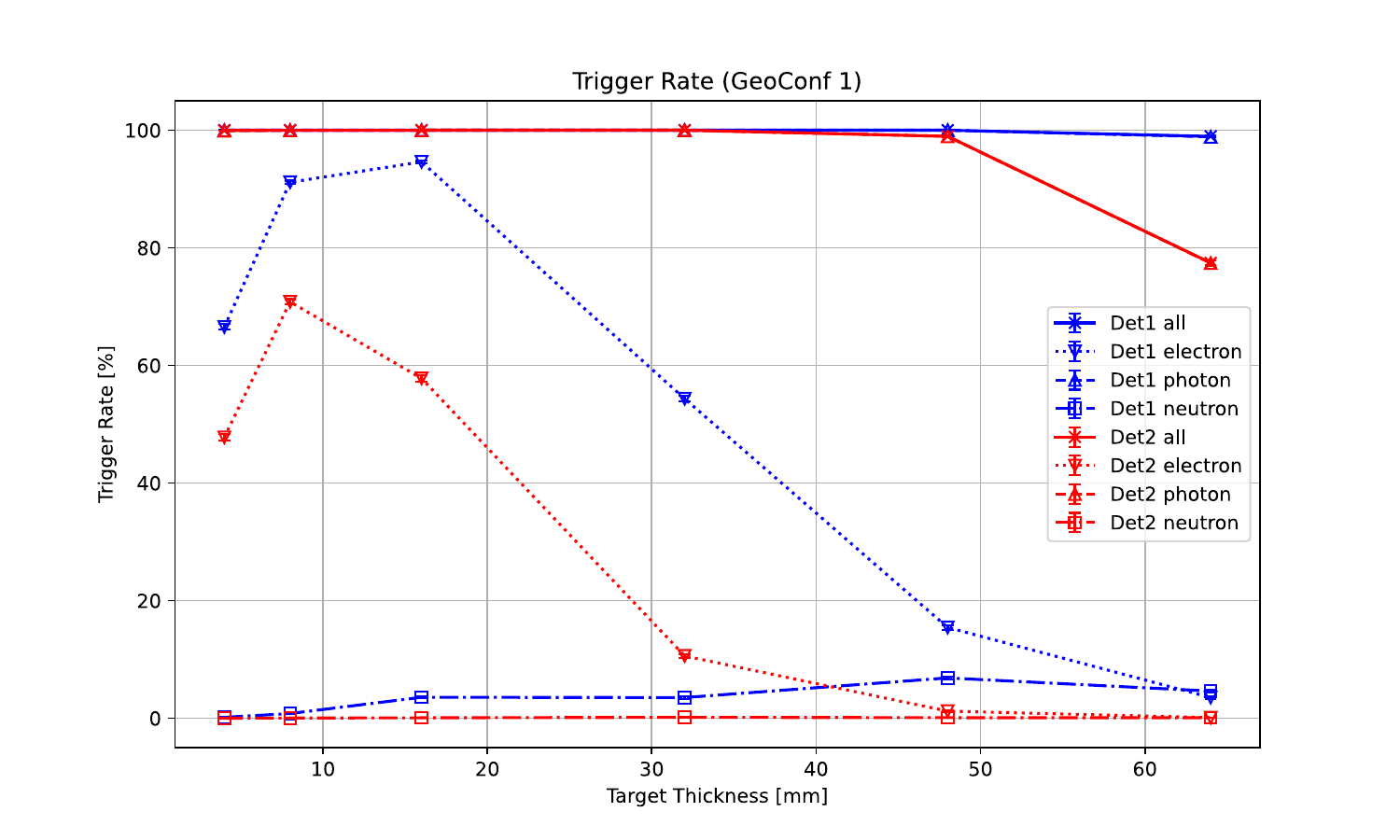}
\includegraphics[width=0.48\linewidth]{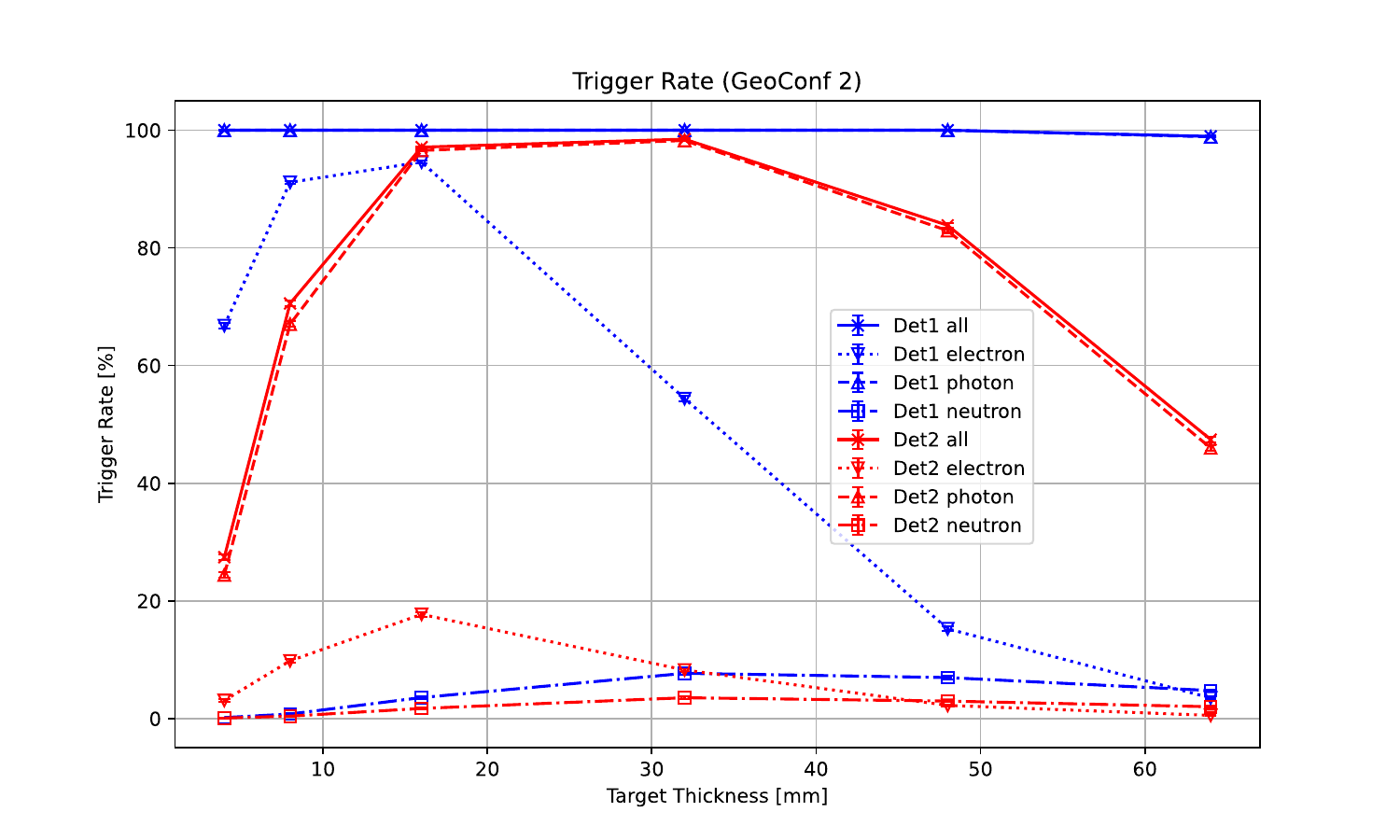}
\caption{The trigger rates of geometry configurations 1 (left) 2 (right). Detector 1 is near the beam target while detector 2 is next to detector 1. The $x$-axis represents the target length, and the $y$-axis represents the trigger rate.} 
\label{fig:trigger}
\end{figure}

The trigger rate is determined by counting how many tracks from the target are produced within the detector. For example, if 100 events are generated and only 10 of them have tracks inside the detector volume, the trigger rate would be 10\%. The threshold for tracking is set to the default value in \textsc{GEANT4}. Our findings indicate that most of the trigger rate contribution comes from photons. The trigger rates for the two geometric configurations are shown in Fig.~\ref{fig:trigger}.

\begin{figure}[t]
\centering
%
\includegraphics[width=0.45\linewidth]{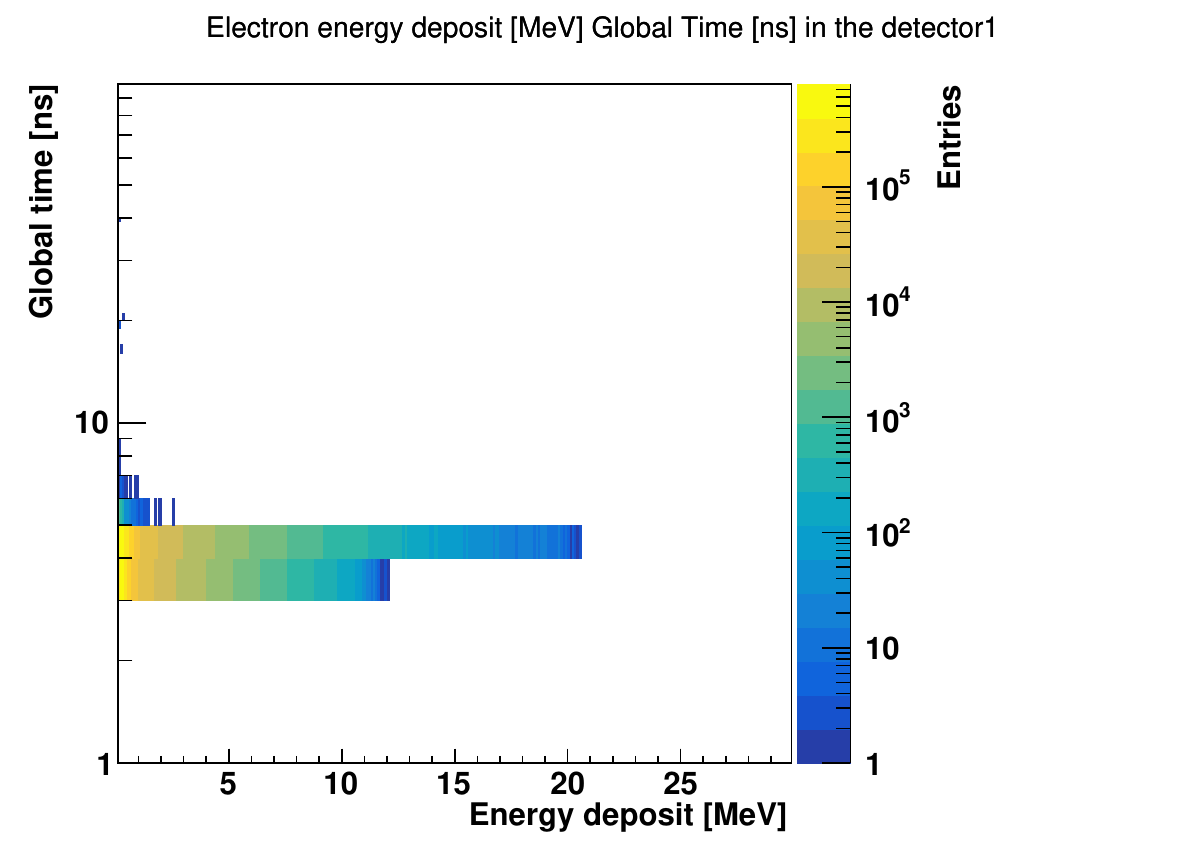}
\includegraphics[width=0.45\linewidth]{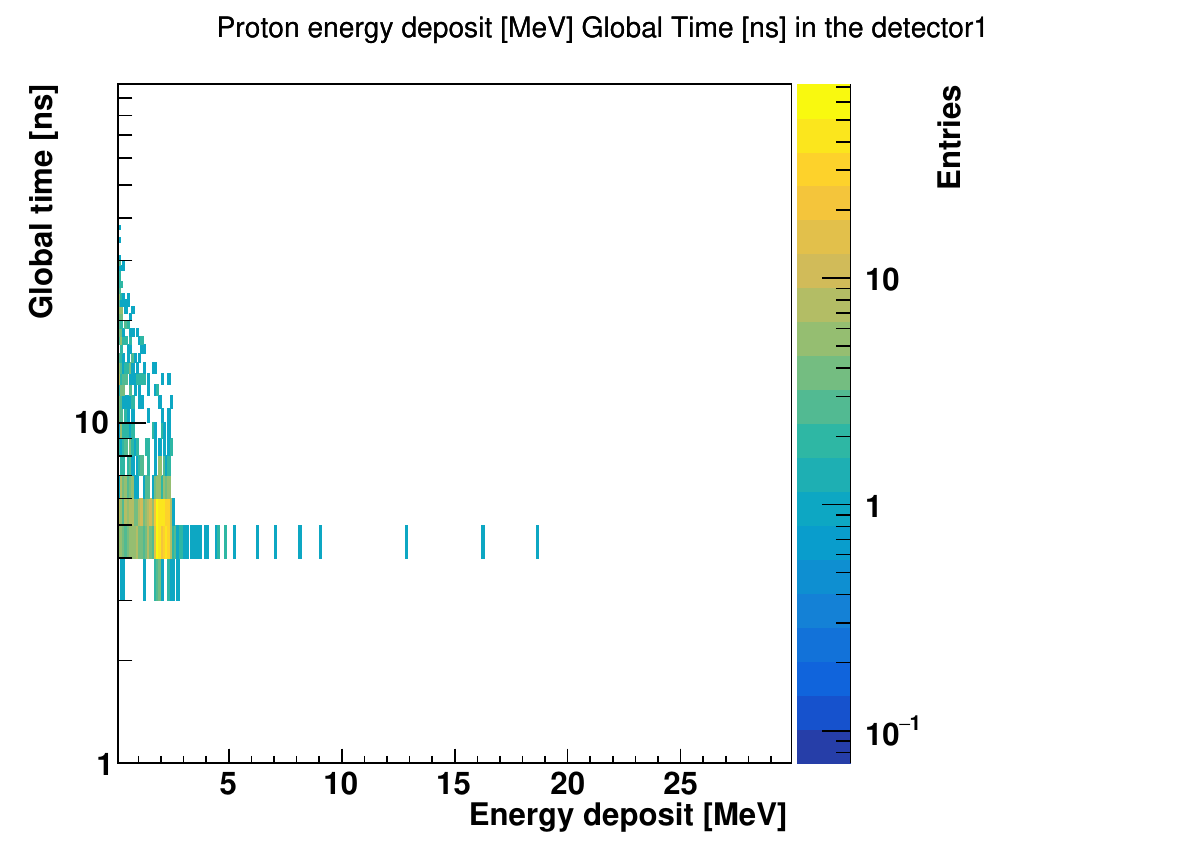}
\includegraphics[width=0.45\linewidth]{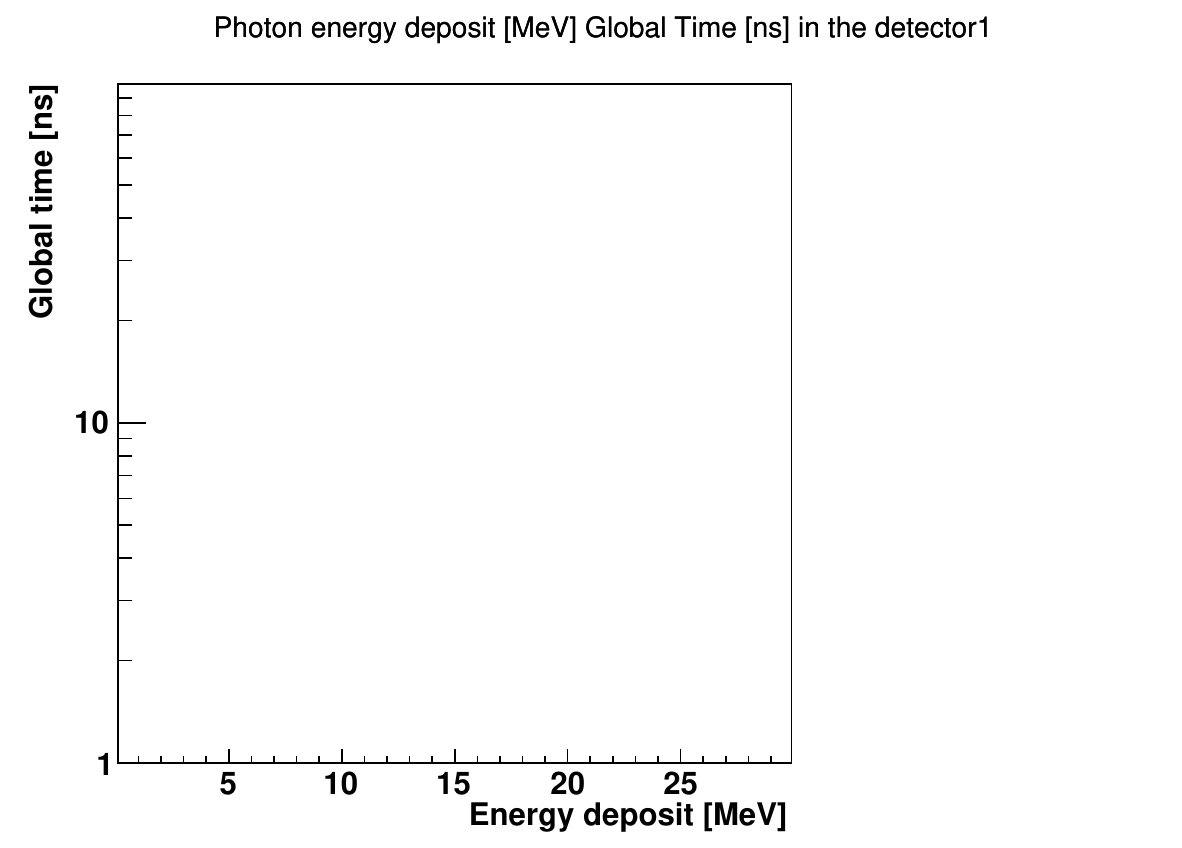}
\includegraphics[width=0.45\linewidth]{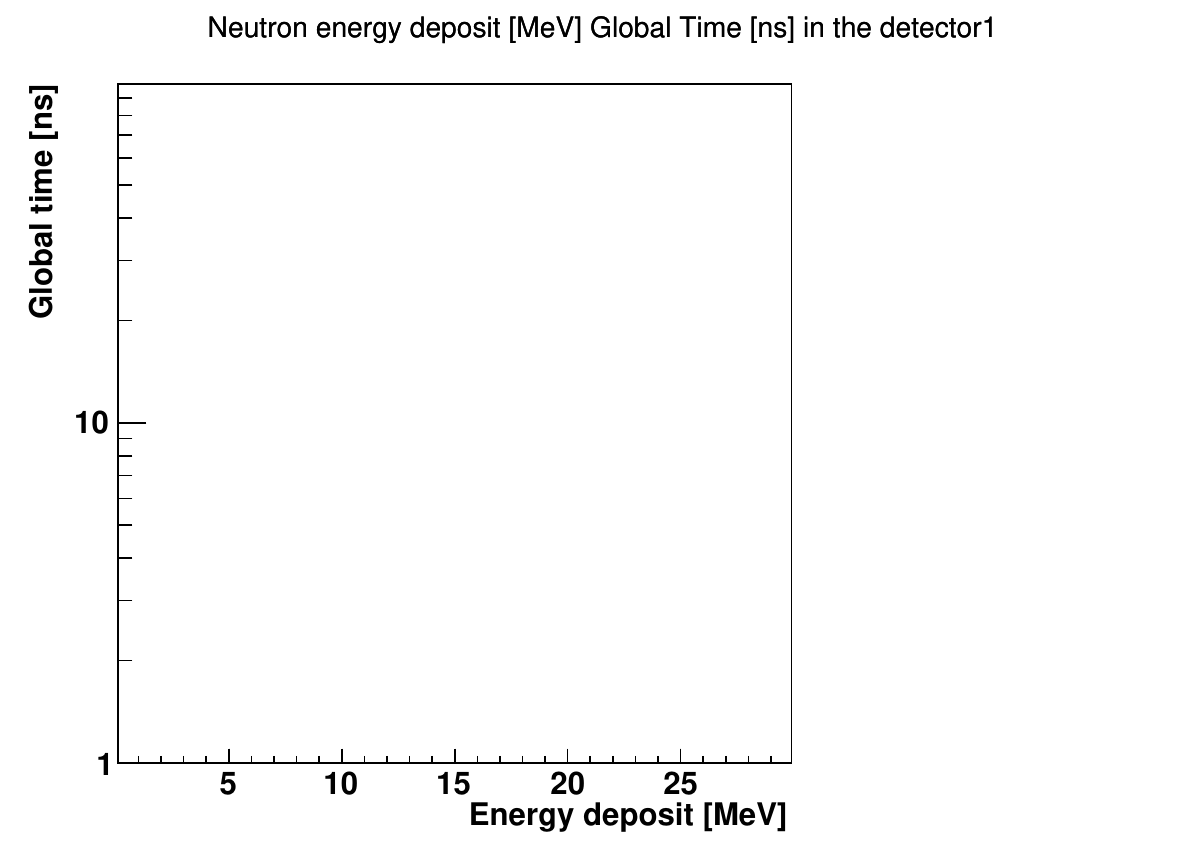}
\caption{Energy depositions by electrons (top left), protons (top right), photons (bottom left), and neutrons (bottom right) in the EJ-301 liquid scintillator, shown in the energy-deposit-vs-time plane. Since the EJ-301 liquid scintillator responds to electrons and protons, the top panels appropriately display energy deposition by these particles, while no energy deposits are recorded for photons and neutrons. For electrons, most prompt particles deposit energy within 6 ns, whereas for protons, the deposition includes neutron-induced protons after 6 ns. These results are based on detector 1 with a 0.4 cm thickness tungsten target and $10^5$ EOTs.} 
\label{fig:lsedep}
\end{figure}

\begin{figure}[t]
\centering
\includegraphics[width=0.65\linewidth]{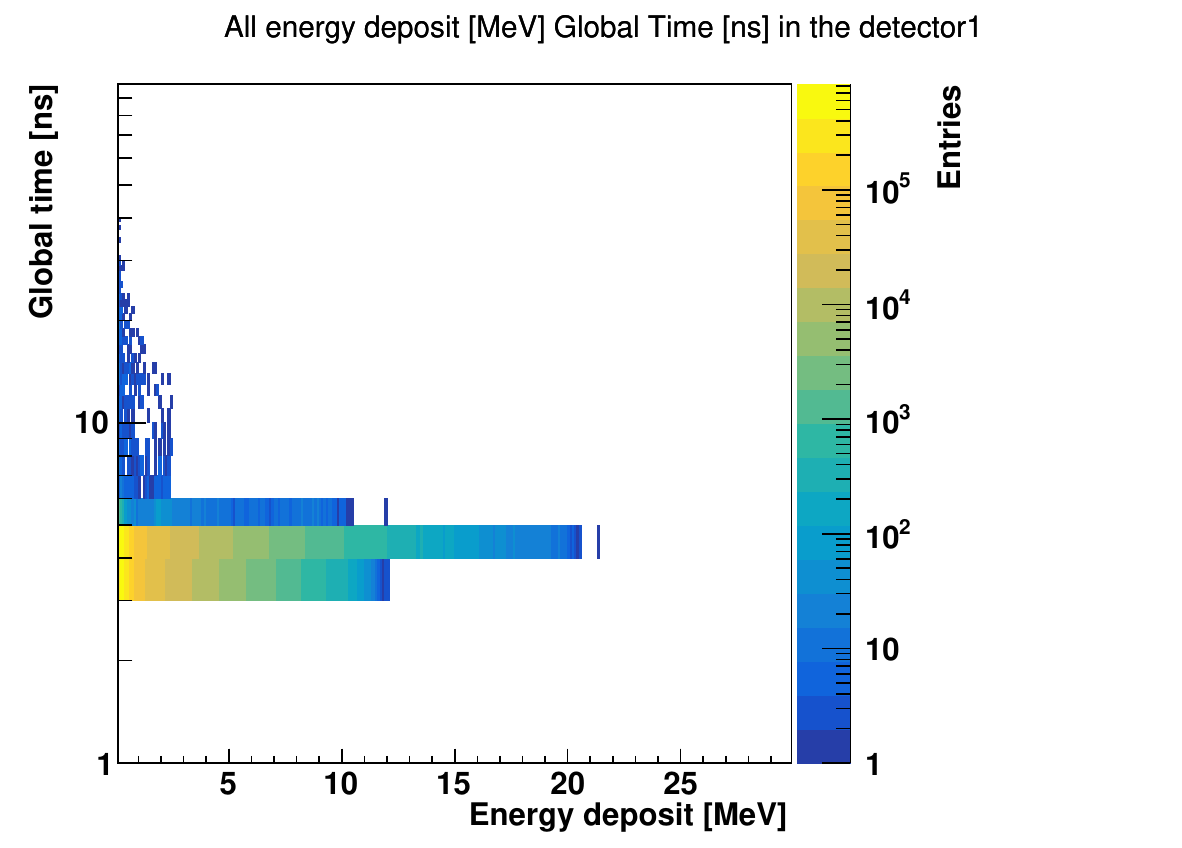}
\caption{Energy depositions by all particles in the EJ-301 liquid scintillator, shown in the energy-deposit-vs-time plane. The results clearly show two distinct groups: prompt particles deposit energy before approximately 6 ns, while neutron-induced particles deposit energy after that, continuing for several tens of nanoseconds. This result is based on the same configuration as in Fig.~\ref{fig:lsedep}.
} 
\label{fig:lsedall}
\end{figure}

The \textsc{GEANT4} simulation models the EJ-301 liquid scintillator and its aluminum housing to calculate the energy distribution and deposition of particles. To compare with experimental results, the energy deposition can be equated to the scintillator light counts observed in the experiment. However, the simulation does not attempt to model the liquid scintillator's response to electrons and protons. Instead, the energy deposition is correlated with signal counts. The \textsc{GEANT4} simulation results specifically depict the energy deposited by electrons and protons in the EJ-301 liquid scintillator (top panel of Fig.~\ref{fig:lsedep}), consistent with the scintillator's specifications. Neutrons and photons, classified as important background particles, 
leave no direct energy deposits as the EJ-301 liquid scintillator responds to electrons and protons (bottom panel of Fig.~\ref{fig:lsedep}). However, they indirectly deposit their energies via induced electrons and protons for which the energy distributions will be displayed later.
Figure~\ref{fig:lsedall} shows the energy deposition from all particles in detector 1 with a 0.4 cm thick tungsten target. The results clearly show two distinct groups: prompt particles depositing energy before approximately 6 ns, and neutron-induced particles depositing energy for several tens of nanoseconds afterward.

\begin{figure}[t]
\centering
\includegraphics[width=0.45\linewidth]{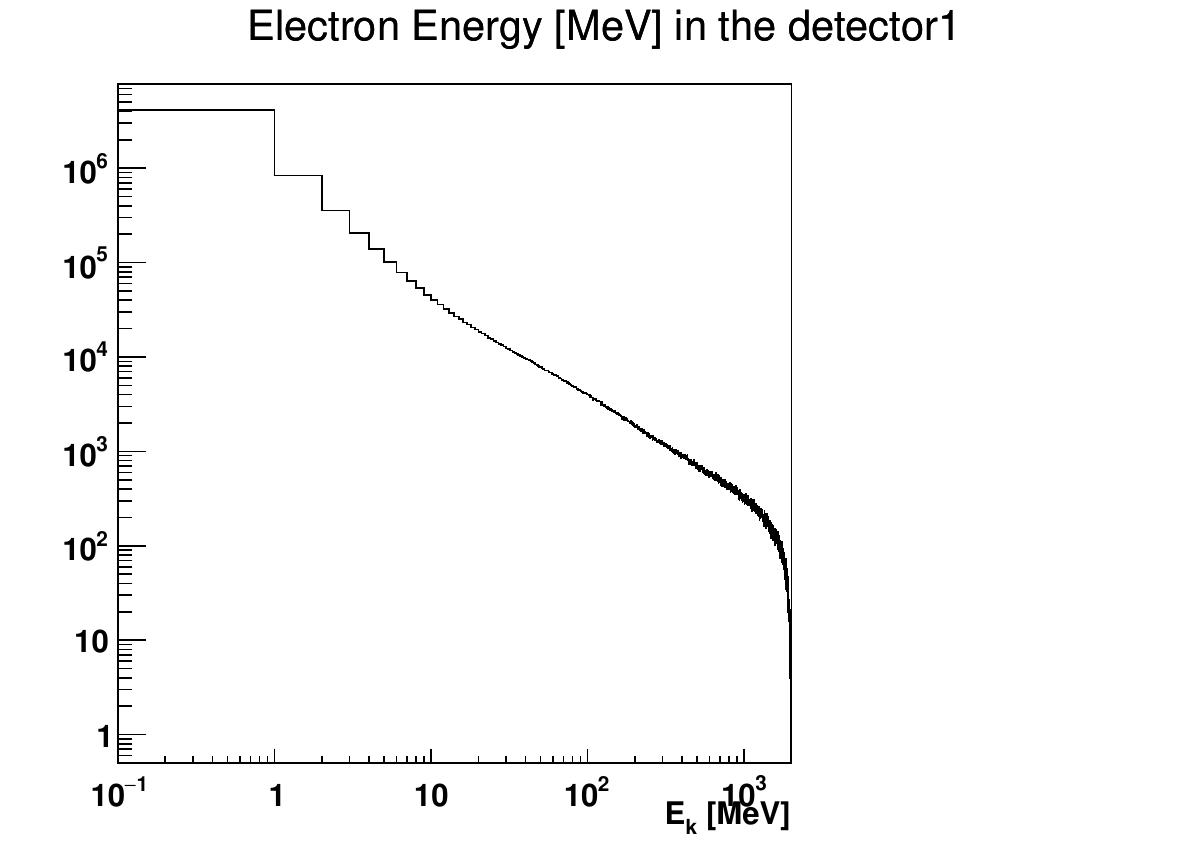}
\includegraphics[width=0.45\linewidth]{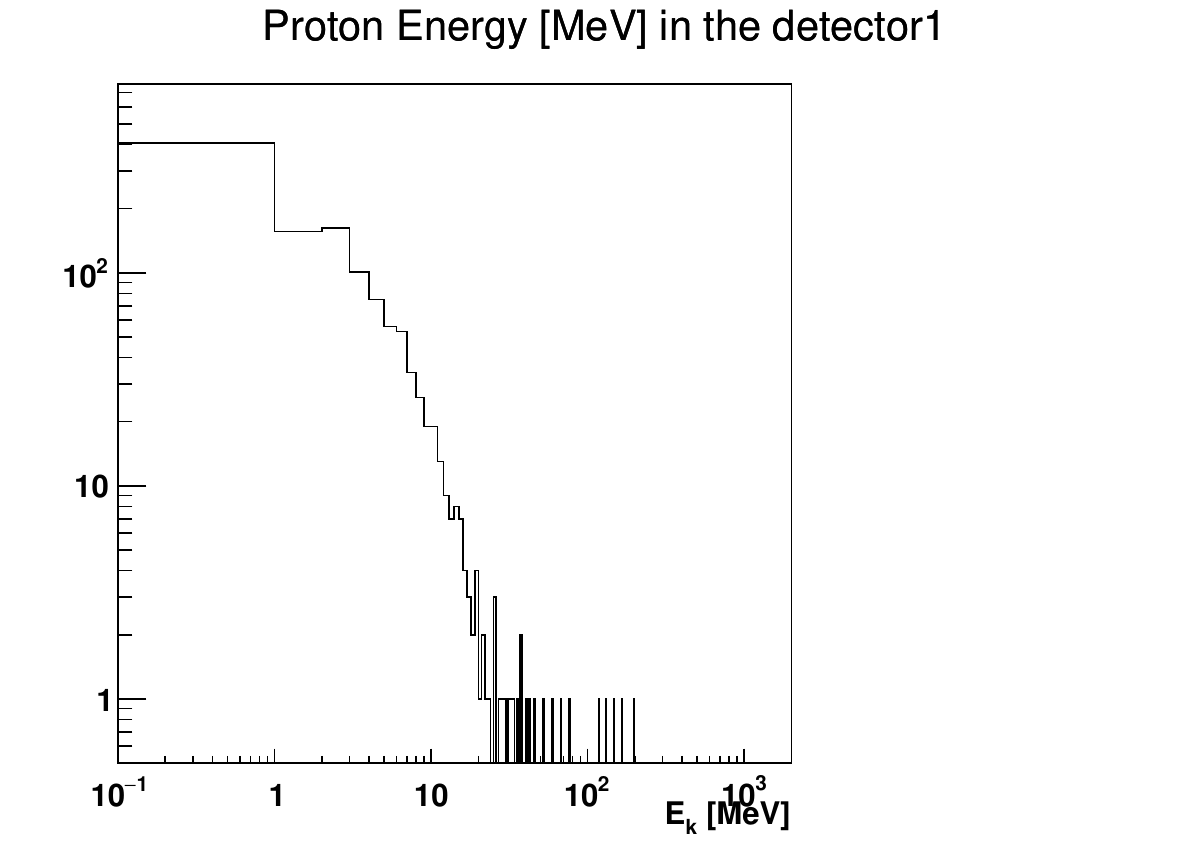}\\
\includegraphics[width=0.45\linewidth]{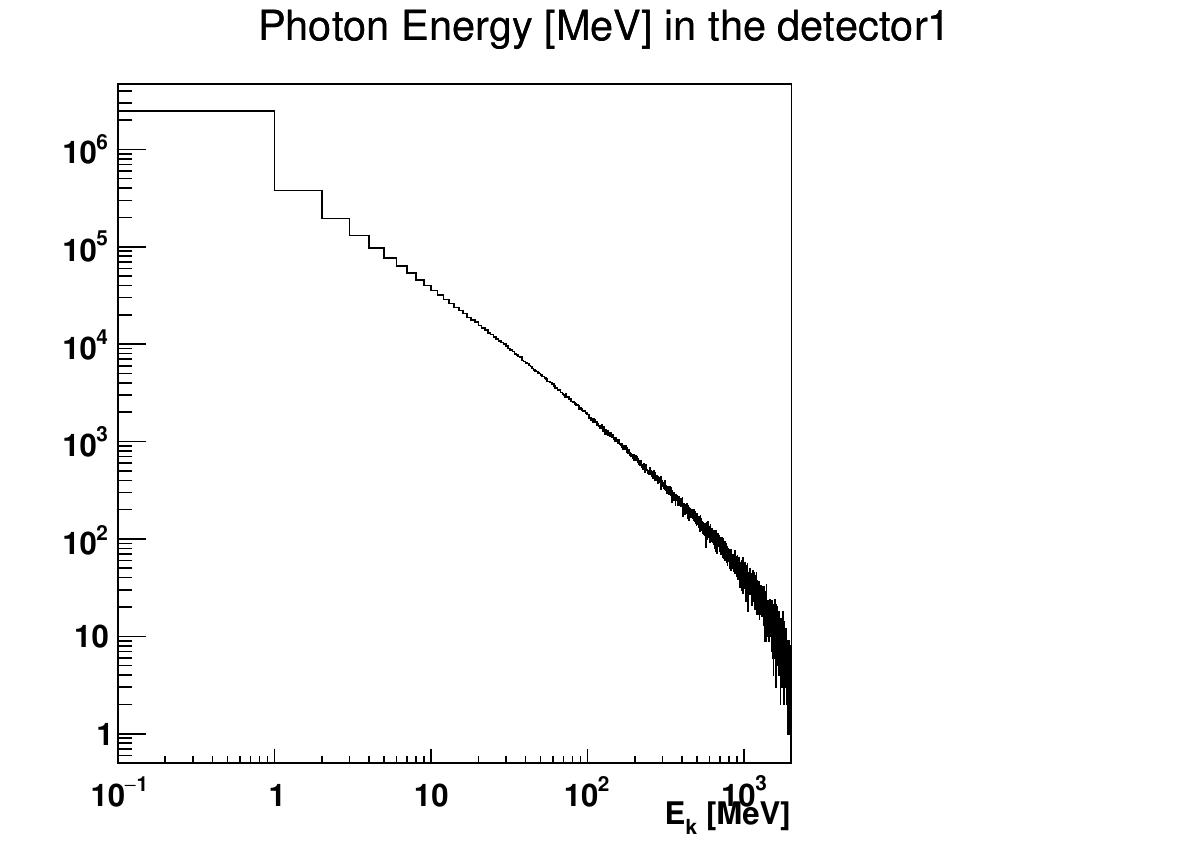}
\includegraphics[width=0.45\linewidth]{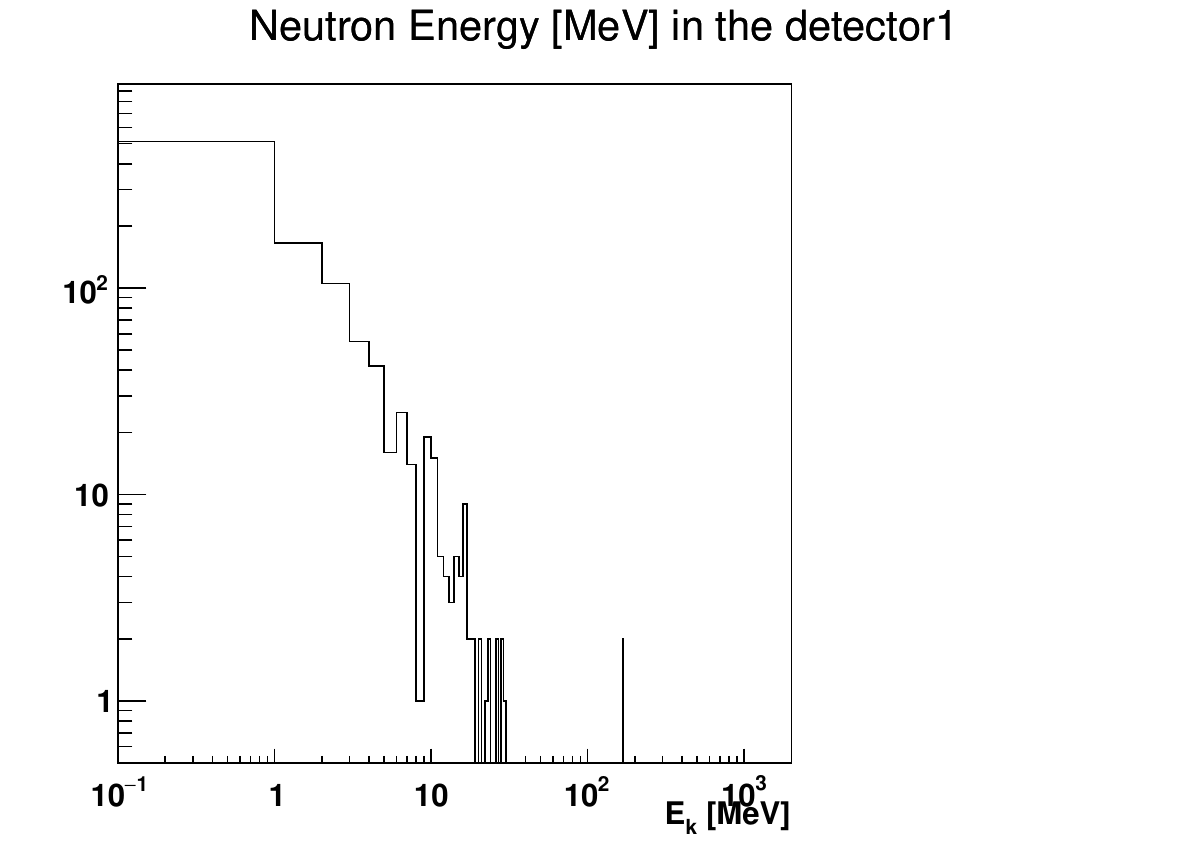}
\caption{Energy distributions of electrons, protons, photons, and neutrons in the EJ-301 liquid scintillator. 
The distributions of photons and neutrons are based on the energy deposited by electrons and protons induced by photon and neutron interactions.
This result is based on the same configuration as in Fig.~\ref{fig:lsedep}. } 
\label{fig:lseall}
\end{figure}

The \textsc{GEANT4} simulation calculated the energy distribution of particles in the detector, accounting for all physics process steps, meaning one particle can create multiple entries. These results will aid in interpreting the experimental data. The EJ-301 liquid scintillator responds only to electrons and protons, although other particles can generate a response indirectly by producing electrons and protons. Furthermore, identifying particles based on the signal shape is crucial for accurate experimental measurements. Figure~\ref{fig:lseall} shows the energy distributions of electrons, protons, photons, and neutrons from the \textsc{GEANT4} simulation. 
As mentioned above, the distributions of photons and neutrons are based on the energy deposited by electrons and protons induced by photon and neutron interactions.
This information helps interpret the measurement result by estimating and distinguishing the particle type and its energy.

\vspace{0.5em}
\section{DAMSA Experiment Staged Plan}
\label{sec:stages}
\begin{figure}[htb]
    \centering
    \includegraphics[width=\textwidth,height=0.7\textwidth]{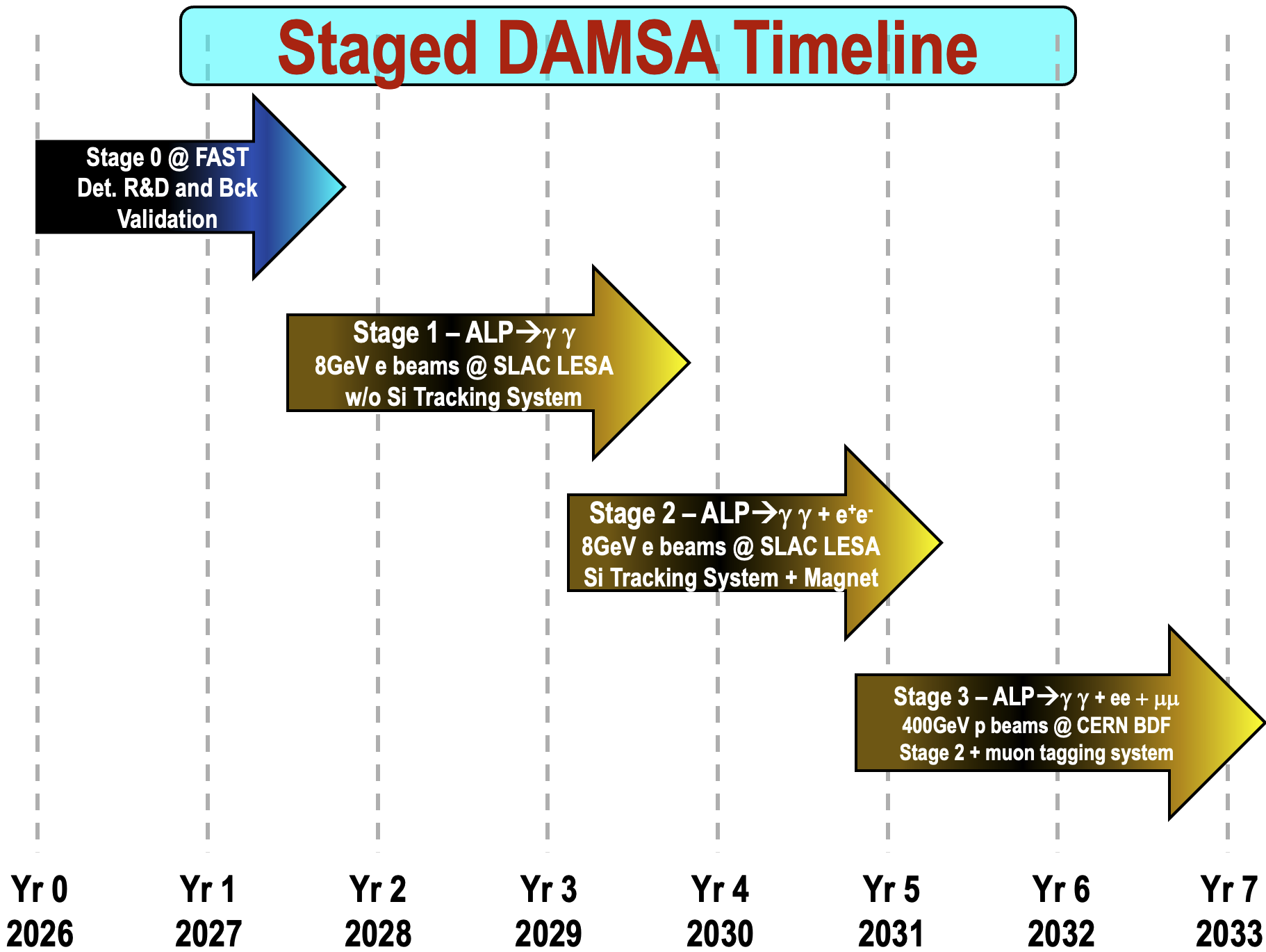}
    \caption{Timeline for the staged DAMSA experiment.}
    \label{fig:damsa-timeline}
\end{figure}

This section describes a staged plan for realizing DAMSA experiment.  This plan ensures each stage to be the bases for the next and to produce meaningful physics outcome in its own rights. 
Figure~\ref{fig:damsa-timeline} shows the overview of the staged DAMSA as a function of the year, in which year 0 is 2026.

\subsection{Stage 0 - Beam Background Validation using 300 MeV electron beams}~\label{sec:phase-0}
As stated in Section~\ref{sec:PhysicsGoals}, DAMSA is a short baseline beam dump experiment that aims to greatly increase the accessible parameter space by minimizing the distance between the target and the detector. 
In this process, as the distance between the detector and the target decreases, managing the beam-induced background becomes important. 
In this stage, we will conduct a background validation run using an existing prototype of the 3DST scintillation counter detector at BNL together with a couple of layers of CsI crystal detectors in the front.
This measurement will be performed using the 300~MeV electron beams available at FAST.
The particle species, their multiplicity and their energies will be measured as a function of the thick of the tungsten target.
These will then be compared to the MC to validate the simulation results.
This will give confidence in our background estimate.

\subsection{Stage 1 - $a\rightarrow \gamma\gamma$ Search at an electron beam facility}
As part of building up the experiment and to expedite search for the ALP, DAMSA will put together the experiment which consists of the tungsten target, the vacuum decay chamber and the 4D total absorption electromagnetic calorimeter.  This stage will have a charged particle tagging system which employs three scintillation counters and occupies the 12~cm gap in between the vacuum decay chamber and the ECAL.  This version of DAMSA will be capable of searching for $a\rightarrow \gamma\gamma$, while potentially providing a capability to tag $a\rightarrow e^{+}e^{-}$ without having the full tracking detector system under magnetic field.  Figure~\ref{fig:stage1} depicts the experiment set up for stage 1 DAMSA.  

We plan on carrying out Stage 1 DAMSA data taking at SLAC LESA facility, described in section~\ref{sec:Facility} to take full advantage of the Dark Current beam configuration flexibility in Table~\ref{tab:LESA_para} and to minimize the neutron background, compared to proton beams.  
This will allow the experiment to focus on the feasibility test of the physics case. 
The expected sensitivity for Stage 1 DAMSA with a $15~cm(L)\times 12~cm(W) \times 12~cm(H)$ tungsten target is presented earlier in Fig.~\ref{fig:alp-slac}, clearly demonstrating its physics reach.
Using the beam intensity of $10^4$ electrons per pulse and the beam repetition rate of 1~kHz, the total integrated electrons on target of $1.5\times10^{14}$ can be accomplished in 3 -- 4 months after data taking.
This beam configuration provides virtually zero background environment, based on our GEANT studies.
In addition, the same study also shows that the neutrons from the beam that lingers around after the initial beam incident to the target most likely be fully dissipated within $1\mu$sec.
This provide an additional factor 100 increase in the repetition rate to 100kHz and enables the experiment to reach electrons on target of $1.5\times10^{16}$ in the same 3 -- 4 months time scale, dramatically expanding the reachable phase space.

The completion of Stage 1 DAMSA will provide a clear demonstration of the physics concept and the fully tested, functioning experiment which includes the most crucial component of the detector, the 4D total absorption ECAL, trigger and data analysis system, ready for the next stage.
We anticipate this stage to take $\sim3$ years. 

\subsection{Stage 2 - $a\rightarrow \gamma\gamma$ and $a\rightarrow e^{+}e^{-}$ Search at an electron beam facility}
\begin{figure}[ht!]
    \centering
    \includegraphics[scale=0.2]{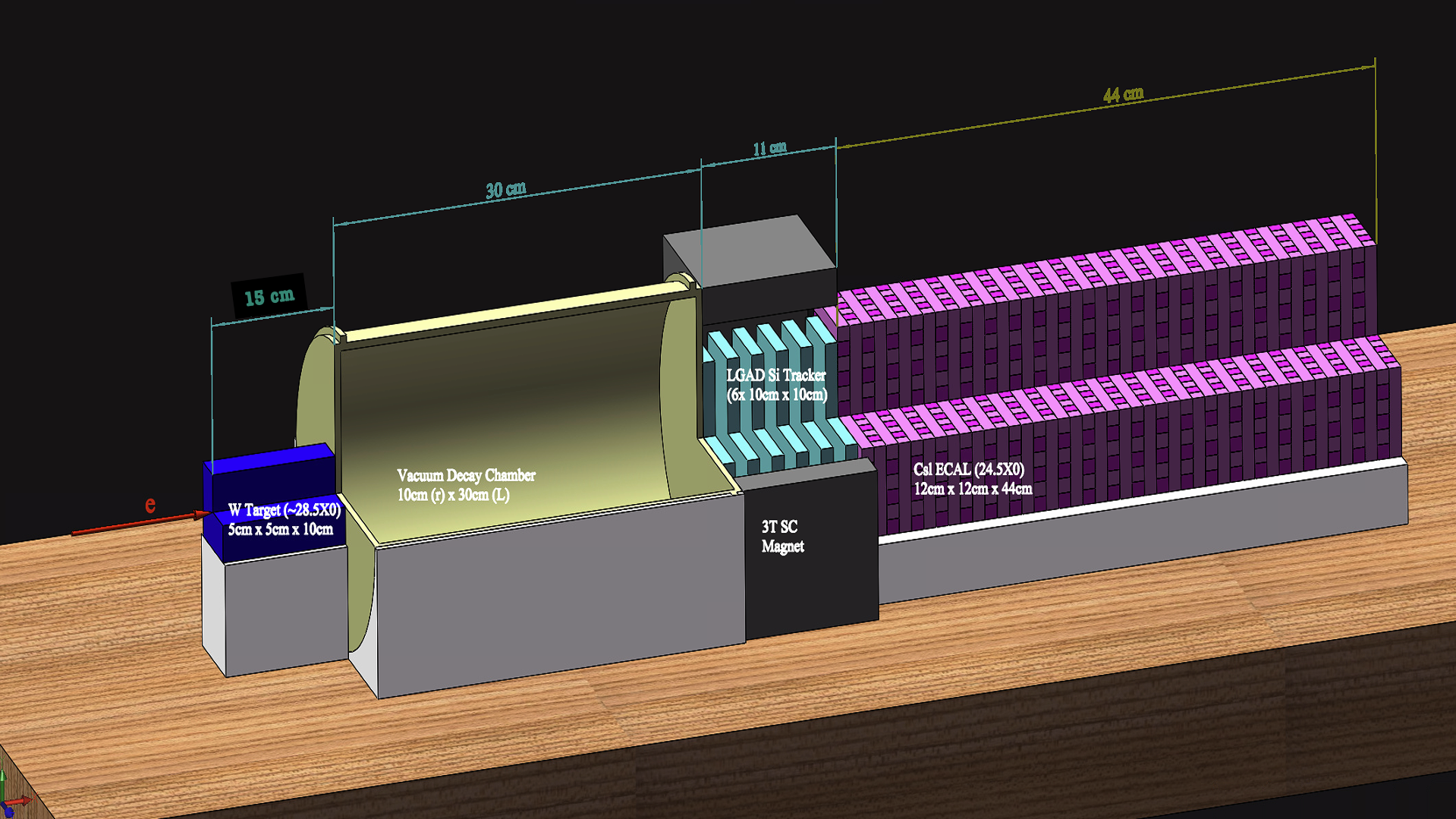}
    \caption{
        Stage 2 DAMSA experiment in which the 3 scintillation counter charged particle ID system is replaced by the full, 6 layer LGAD Si tracking detector under 1~Tesla magnetic field.
    }
    \label{fig:stage2}
\end{figure}

Building on the successful completion of Stage 1 DAMSA, the scintillation counter charged particle ID system will be replaced with the LGAD-based Si tracking system under 1~Tesla magnetic field, as shown in Fig.~\ref{fig:stage2}.
The addition of the tracking system under magnetic field provides the capability to measure the sign of the charged particles as well as the momentum of each charged particles.
The primary goal of Stage 2 is to expand the physics case beyond two photon final states, namely the $a\rightarrow e^{+}e^{-}$ final states.
To continue focus on physics case demonstration and to leverage the existing experimental set up from Stage 1, we plan on carrying out this stage at the same SLAC LESA beam facility.
At the time of writing this report, we do not yet have completed the study for the sensitivity for $a\rightarrow e^{+}e^{-}$ final states.

The completion of Stage 2 DAMSA, however, clearly demonstrates the physics case and provides a great potential for a discovery of an ALP.
It also provides a complete DAMSA detector system for search for any dark portal particle of electromagnetic final stages, ready for stage 3.

\subsection{Stage 3 - ALP Searches at a proton beam facility}
High intensity proton beams, such as those at the CERN Beam Dump Facility described in section~\ref{sec:Facility} provides an opportunity to explore different dark portal particle production mechanism, expanding the DAMSA physics case.
High energy proton beams also provides the potential to explore higher mass portal particles.
The dark red solid line in Figure~\ref{fig:alp-slac} shows DAMSA experiment at CERN's BDF, utilizing the dump and muon magnetic shielding facility under construction at BDF to support SHiP experiment.
While the longer baseline of 20~m due to the thick shield limits the expansion of the "ceiling" beyond that at SLAC electron beams, Stage 3 DAMSA will be able to extend the search of the ALP in a higher mass range at the lower couplings.
In addition, the higher CMS energy available thanks to the 400~GeV proton energy enables the exploration between the two EM particle final states, such as that of $\mu^{+}\mu^{-}$ final states, by adding a muon system behind the 4D ECAL.
Finally, we could also imagine, taking advantage of one of the 7~TeV proton beams at the LHC.

\vspace{0.5em}
\section{Conclusions}      
\label{sec:Conclusions}
DAMSA is a table-top scale dark-sector particle search experiment at an accelerator with excellent physics potential. The pathfinder experiment, Little DAMSA Path-Finder (LDPF), focuses on gaining a deeper understanding of beam-related neutron (BRN) backgrounds in a well-controlled electron beam environment while searching for MeV-scale new physics, including ALPs and dark photons. By leveraging the short baseline between the target and detector, LDPF can probe short-lived particles, testing the experimental concept and serving as the first step toward building a full-scale detector for PIP-II proton beams. The proposed LDPF experiment has significant potential to explore uncharted regions of ALP parameter space, particularly for short-lived ALPs that have eluded previous experiments. With unprecedented sensitivity at FAST beam intensities, this setup could lead to the discovery of ALPs, dark photons, and other unseen dark sector states, or impose stringent new constraints on their interactions with ordinary matter, thereby advancing our understanding of light, weakly interacting particles.


    \section*{Acknowledgments}
    This material is based upon work supported by the U.S. Department of Energy (DOE) Office of Science and the National Research Foundation of Korea (NRF). We are grateful to the Center for Theoretical Underground Physics and Related Areas (CETUP*), The Institute for Underground Science at Sanford Underground Research Facility (SURF), and the South Dakota Science and Technology Authority for their hospitality and financial support.

    \appendix
    
    \bibliography{main,detector}
    
    \newpage

  

\end{document}